\documentclass[11pt,a4paper]{article}
\pdfoutput=1
\usepackage{jcappub}
\usepackage[T1]{fontenc}

\usepackage{slashed}
\usepackage{epstopdf}
\usepackage{color}
\usepackage{subfigure}
\newcommand{\dm}{\psi^0_1}
\title{When Freeze-out Precedes Freeze-in: Sub-TeV Fermion
Triplet Dark Matter with Radiative Neutrino Mass}
\author[a]{Anirban Biswas,}
\emailAdd{anirban.biswas.sinp@gmail.com}
\affiliation[a]{Department of Physics, Indian Institute of Technology
Guwahati, Assam 781039, India}
\author[a,1]{Debasish Borah, \note{Corresponding author}}
\emailAdd{dborah@iitg.ac.in}
\author[a]{Dibyendu Nanda}
\emailAdd{dibyendu.nanda@iitg.ac.in}
\abstract{
We propose a minimal predictive scenario for dark matter and radiative neutrino mass where the relic abundance of dark matter is generated from a hybrid setup comprising of both thermal freeze-out as well as non-thermal freeze-in mechanisms. Considering three copies of fermion triplets and one scalar doublet, odd under an unbroken $\mathbb{Z}_2$ symmetry, to be responsible for radiative origin of neutrino mass, we consider the lightest fermion triplet as a dark matter candidate which remains under-abundant in the sub-TeV regime from usual thermal freeze-out. Late decay of the $\mathbb{Z}_2$-odd scalar doublet into dark matter serves as the non-thermal (freeze-in) contribution which not only fills the thermal dark matter deficit, but also constrains the mother particle's parameter space so that the correct relic abundance of dark matter is generated. Apart from showing interesting differences from the purely freeze-out and purely freeze-in dark matter scenarios, the model remains testable through disappearing charge track signatures at colliders, observable direct and indirect detection rates for dark matter and prediction of almost vanishing lightest neutrino mass.}

\begin{document}
\maketitle
%\flushbottom
%%%%%%%%%%%%%%%%%%%%%%%%%%%%%%%%%%%%%%%%%%%%%%%%%%%%%%%%%%%%%%%%%%%%%%%%%
\section{Introduction}
\label{sec:intro}
%%%%%%%%%%%%%%%%%%%%%%%%%%%%%%%%%%%%%%%%%%%%%%%%%%%%%%%%%%%%%%%%%%%%%%%%%%
There have been a convincing number of evidences from astrophysics and cosmology related experiments in the last several decades which suggest the presence of some non-baryonic and non-luminous form of matter, called dark matter (DM) in large amount distributed across the Universe. From the galaxy cluster observations by Fritz Zwicky \cite{Zwicky:1933gu} back in 1933, observations of galaxy rotation curves in 1970's \cite{Rubin:1970zza}, the more recent observation of the bullet cluster \cite{Clowe:2006eq} and the latest cosmology data provided by the Planck satellite \cite{Ade:2015xua}, it is now certain that around $27\%$ of the present Universe is composed of such dark matter which is approximately five times more than the ordinary baryonic matter. The present abundance of DM is often quoted in terms of the density parameter $\Omega$ as \cite{Ade:2015xua}
\begin{equation}
\Omega_{\text{DM}} h^2 =\frac{\rho_{\text{DM}}}{\rho_{\text{c}}}h^2= 0.1186 \pm 0.0020
\label{dm_relic}
\end{equation}
where $h = \dfrac{H_0}{100\,\,{\rm km\,s^{-1}\,Mpc^{-1}}}$ is a parameter of order
unity while $H_0=(67.8\pm 0.9)\,{\rm km\,s^{-1}\,Mpc^{-1}}$ \cite{Ade:2015xua},
is the present value of the Hubble parameter
and $\rho_{\text{c}}=\frac{3H^2_0}{8\pi G}$ is the critical density
of the Universe with $G$ being the universal constant of gravity.
In spite of all these evidences, the particle nature of DM is not yet
known leaving a wide range of possibilities.
It is however, certain that none of the standard model (SM) particles can
satisfy the requirements \cite{Taoso:2007qk} of a typical DM candidate.
This has led to several beyond standard model (BSM) proposals in the last few decades,
the most popular of which being the so called weakly interacting massive particle (WIMP) paradigm. In this framework, a dark matter candidate typically with electroweak scale mass and interaction rate similar to electroweak interactions can give rise to the correct dark matter relic abundance, a remarkable coincidence often referred to as the \textit{WIMP Miracle}.

One interesting aspect of WIMP dark matter paradigm is their masses and interactions falling around the ballpark of the electroweak scale, which lead to their thermal production in the early Universe. The relic of such a particle arise when its interactions fall below the rate of expansion of the Universe, known as the freeze-out mechanism. These interactions between DM and SM particles leading to its thermal production in the early Universe can also lead to observable DM nucleon scattering cross sections. However, no such observations have so far been made which could have been a direct detection of particle DM. The most recent direct detection experiments like LUX, PandaX-II and XENON1T have also reported their null results \cite{Akerib:2016vxi, Tan:2016zwf, Aprile:2017iyp, Cui:2017nnn, Aprile:2018dbl}. The absence of such direct detection signals have progressively lowered the exclusion curve in DM-nucleon cross section versus DM mass plane. With such high precision measurements, the WIMP-nucleon cross section will soon overlap with the neutrino-nucleon cross section, known as the neutrino floor, making it difficult to distinguish a probable DM signal from the neutrino ones. Although such null results at direct detection frontier could indicate a very constrained region of WIMP parameter space, they have also motivated the particle physics community to seek for a paradigm shift. One such scenario that has drawn attention in the last few years is the non-thermal dark matter scenarios \cite{McDonald:2001vt, Hall:2009bx}. In this case, the DM particles have so feeble interactions with the remaining thermal bath that it never attains thermal equilibrium at any epoch in the early Universe. However, it can be produced from decay of some heavy particles or scattering processes, popularly known as the freeze-in mechanism \cite{Hall:2009bx, Konig:2016dzg, Biswas:2016bfo, Biswas:2016iyh}, leading to a new paradigm called freeze-in (or feebly interacting) massive particle (FIMP). For a recent review of this DM paradigm, please see \cite{Bernal:2017kxu}. 

Instead of completely giving up on the WIMP framework due to the null results at direct search experiments, here we consider a hybrid scenario where both thermal (freeze-out) and non-thermal (freeze-in) contributions to DM relic abundance can be important. To be more specific, DM can have sizeable interactions to be thermally produced in the early Universe but also needs a non-thermal source to satisfy the correct relic abundance in the present epoch. This idea was explored in several earlier works including \cite{Fairbairn:2008fb, Cheung:2010gj, Medina:2014bga, Gherghetta:2015ysa} and references therein. Recently, such a work was performed for the inert Higgs doublet dark matter model \cite{Borah:2017dfn} where the SM is extended by an additional scalar doublet odd under an in-built $\mathbb{Z}_2$ symmetry so that the lightest $\mathbb{Z}_2$-odd component is a stable dark matter candidate. In that work the thermally under-abundant DM parameter space was revisited and it was shown that a non-thermal contribution from an additional $\mathbb{Z}_2$-odd singlet neutral fermion can fill this deficit while the other two cousins of this singlet neutral fermion along with the inert scalar doublet can play a role in generating one loop radiative neutrino masses in scotogenic fashion \cite{Ma:2006km}. It is worth mentioning at this point that the observation of non-zero neutrino masses and large leptonic mixing \cite{Olive:2016xmw} has also been another motivation for BSM physics for last few decades. While the addition of singlet right handed neutrinos to the SM content can give rise to the usual seesaw mechanism \cite{Minkowski:1977sc, Mohapatra:1979ia, Schechter:1980gr} for neutrino mass at tree level, the scotogenic framework can explain the origin of neutrino mass and dark matter in a unified manner. In the earlier work \cite{Borah:2017dfn} where both thermal and non-thermal contributions to scalar dark matter abundance were studied, the initial abundance of the heavy singlet fermion was not explained, but was fixed suitably in order to generate the required non-thermal contribution.

In the present work, we consider a more general study of this hybrid scenario within the framework of another minimal model for dark matter and neutrino masses but with more observable consequences. Here the singlet neutral fermions of the minimal scotogenic model \cite{Ma:2006km} are replaced by fermion triplets having zero hypercharge, but odd under the $\mathbb{Z}_2$ symmetry. As known from the discussions of fermion triplet dark matter (FTDM) model proposed by Ma and Suematsu \cite{Ma:2008cu}, the fermion triplet dark matter is usually under-abundant below 2.2 TeV mass due to its large annihilation cross section. Inclusion of non-perturbative effects on dark matter annihilations \cite{Hisano:2004ds, Hisano:2006nn} which are more dominant for heavier dark matter masses, pushes this bound on fermion triplet DM mass to around 2.7 TeV \cite{Garcia-Cely:2015quu}. In this work, we particularly focus on the fermion triplet dark matter mass around 1 TeV and check if a non-thermal contribution can fill the deficit coming from thermal freeze-out. In a scotogenic model with fermion triplet, the only non-thermal contribution for such 1 TeV DM can come from the $\mathbb{Z}_2$ odd scalar doublet. Unlike the earlier work \cite{Borah:2017dfn} where the mother particle was a fermion singlet with tiny Yukawa coupling with the dark mater candidate, here the mother particle has gauge interactions and hence can have thermal abundance due to the usual freeze-out scenario. We solve the coupled Boltzmann equations corresponding to the comoving number densities of the inert scalar doublet as well as fermion triplets to determine the final relic abundance of fermion triplet DM. The inert scalar doublet, being heavier, freezes out first followed by the freeze-out of the fermion triplet. Then a non-thermal contribution from the inert scalar doublet fills the deficit in fermion triplet relic abundance from thermal freeze-out. The requirement for correct fermion triplet DM abundance not only constrains its coupling with the inert scalar doublet, but also the freeze-out abundance and hence the parameter space of the latter. The other two cousins of the fermion triplet dark matter can have sizeable coupling with inert scalar doublet and SM leptons to generate tiny neutrino masses at one loop. Since the lightest fermion triplet almost decouples from neutrino mass generation due to tiny Yukawa couplings (required for non-thermal production or freeze-in), the lightest neutrino remains massless in our scenario. We also check the testability of the model and find that such TeV scale fermion triplet DM can have observable consequences and direct, indirect dark matter detection experiments as well as collider experiments like the large hadron collider (LHC).

This paper is organised as follows. In section \ref{sec:ftdm} we briefly discuss the fermion triplet dark matter followed by its scotogenic extension and details of the model we adopt for our present work in section \ref{sec:scto_ftdm}. In section \ref{sec:neutrino}, we briefly discuss the generation of neutrino mass at one loop and then move on to discussing the details of dark matter relic calculation in section \ref{sec:relic_density}. We then discuss different constraints or observational aspects of our dark matter scenario at LHC, direct detection and indirect detection experiments in section \ref{sec:lhc_cons}, \ref{sec:dd}, \ref{sec:id} respectively. We finally conclude in section \ref{sec:conclu}.

%%%%%%%%%%%%%%%%%%%%%%%%%%%%%%%%%%%%%%%%%%%%%%%%%%%%%%%%%%%%%%%%%%%%
%\newpage
\section{Fermion Triplet Dark Matter (FTDM)}
\label{sec:ftdm}
%%%%%%%%%%%%%%%%%%%%%%%%%%%%%%%%%%%%%%%%%%%%%%%%%%%%%%%%%%%%%%%%%%%%%
In this section, we discuss a stable fermion triplet of zero hypercharge as a dark matter candidate, stabilised by an in-built $\mathbb{Z}_2$ symmetry. It is straightforward to realise the need of additional $\mathbb{Z}_2$ symmetry as otherwise the triplet can decay into the SM Higgs and lepton due to renormalizable couplings among them. A SU(2)$_{\rm L}$ multiplet of higher dimensions can however be naturally stable without any need of additional symmetries, along the minimal dark matter spirit \cite{Cirelli:2005uq}. Fermion triplet dark matter was also studied by several other groups, the most recent of which can be found in \cite{Garcia-Cely:2015quu} within the framework of another class of models. To be more technical, in a minimal setup of FTDM, the fermionic sector of the SM is extended by a
SU(2)$_{\rm L}$ triplet $\Sigma_R=\left(\Sigma^{1}_R\,~\Sigma^{2}_R\,~
\Sigma^{3}_R \right)^T$ with zero hypercharge. The triplet $\Sigma_R$
which is in the adjoint representation of SU(2)$_{\rm L}$,
can also be expressed in the fundamental representation as
\begin{eqnarray}
\Sigma_R~=~\dfrac{\sigma^i\,\,\Sigma^{i}_R}{\sqrt{2}}
~=~\left(\begin{array}{cc}
   {\Sigma_R^0}/{\sqrt{2}} & \Sigma^{+}_R \\
   \Sigma^{-}_R  & -\,{\Sigma_R^0}/{\sqrt{2}} \\
  \end{array} \right)\,,
\label{sigmar}  
\end{eqnarray}
where $\sigma_i$'s ($i=1$ to 3) are the Pauli spin matrices, the
generators of the fundamental representation of SU(2)$_{\rm L}$ while
$\Sigma_R^{\pm}=\left(\Sigma_R^{1}{\mp}\,i\,\Sigma^{2}_R\right)/\sqrt{2}$,
$\Sigma_R^{0}=\Sigma_R^{3}$. 
%Moreover, an additional
%discrete $\mathbb{Z}_2$ symmetry is also imposed on the triplet
%$\Sigma_R$, as result the neutral component $\Sigma^0_R$ of $\Sigma_R$
%becomes absolutely stable and can be a dark matter candidate.
Let us define $\Sigma=\Sigma_R + {\Sigma_R}^c$,
where ${\Sigma_R}^c = C\,{\overline{\Sigma_R}}^{T}$ is
the CP conjugate of $\Sigma_R$ and $C$
being the charge conjugation operator. 
Note that, by construction $\Sigma$ is a Majorana
fermion ($\Sigma^c=\Sigma$) however, not all the
components of $\Sigma$ are Majorana fermions.
We have shown in the Appendix \ref{App:lag}
that only neutral component of $\Sigma$
is a Majorana fermion while the charged one is
as usual a Dirac fermion. The Lagrangian of the
triplet $\Sigma$ is given by
\begin{eqnarray}
\mathcal{L}_{triplet} &=& \dfrac{i}{2}{\rm Tr}
[\overline{\Sigma}\,\slashed{D} \Sigma] -
\dfrac{1}{2}\,{\rm Tr}[\overline{{\Sigma}}\,M_{\Sigma}\,\Sigma] \,,\\
%\dfrac{1}{2}\,{\rm Tr}[\overline{\Sigma^c}\, M^{\star}_{\Sigma}\,\Sigma]\,,
&&=\dfrac{i}{2}{\rm Tr}[\overline{\Sigma_R}\,\slashed{D}\,{\Sigma_R}]
+\dfrac{i}{2}{\rm Tr}[\overline{{\Sigma_R}^c}\,\slashed{D}\,{\Sigma_R}^c]-
\left(\dfrac{1}{2}\,{\rm Tr}[\overline{{\Sigma_R}^c}\,
M_{\Sigma}\,\Sigma_R]+h.c.\right) \,,
\label{tripletLag}
\end{eqnarray}
where, $D_{\mu}$ is the covariant derivative of $\Sigma_R$ and its
expression is given in Eq.\,\,\eqref{Eq:dmu} of Appendix \ref{App:lag}. 
Now, inserting Eq. \eqref{sigmar}
into Eq. \eqref{tripletLag}, and using the relation
${\Sigma_R}^c = C\,{\overline{\Sigma_R}}^{T}$, we get
\begin{eqnarray}
\mathcal{L}_{triplet} &=& \overline{\psi^{-}}\,i\,\slashed{\partial} \psi^{-} +
\dfrac{1}{2}\overline{\psi^0}i\,\slashed{\partial} \psi^0 - 
M_{\Sigma}\,\overline{\psi^{-}}\,\psi^{-} 
- \dfrac{M_{\Sigma}}{2}\,\overline{{\psi^0}}\,{\psi^0}  
-\,g\left(\overline{\psi^{-}}\gamma^{\mu}\,\psi^{0}\,W^{-}_{\mu} + h.c.\right)
\nonumber \\
&&+\,g\,\cos\theta_w\,\overline{\psi^{-}}\gamma_{\mu}\psi^{-}\,Z^{\mu}
+ \,g\,\sin\theta_w\,\overline{\psi^{-}}\gamma_{\mu}\psi^{-}\,A^{\mu}\,,
\label{lag_psi0_psi-}
\end{eqnarray}
where, we have defined $\psi^{-} = \Sigma^{-}_R + {\Sigma^{+}_R}^c$, a four
component Dirac spinor while $\psi^{0}$ is a Majorana fermion
in four component notation as $\psi^{0} = \Sigma^{0}_R + {\Sigma^{0}_R}^c$
and $\theta_{w}$ is the usual weak mixing angle known as the Weinberg angle.
In Appendix \ref{App:lag}, we have explicitly derived Eq.\,\,\eqref{lag_psi0_psi-}
from Eq.\,\,\eqref{tripletLag}.
%From the above equation, it is clearly evident that the
%neutral component of the triplet $\Sigma_R$ is a Majorana fermion
%with a Majorana type mass term $\dfrac{M_{\Sigma}}{2}\overline{{\Sigma^0_R}^c}
%\,{\Sigma^0_R}$ while the charged components are Dirac fermion. 
Now, if one introduces a $\mathbb{Z}_2$ parity on $\Sigma_R$
then the neutral component $\psi^0$, which is a Majorana fermion,
can be a viable thermal dark matter candidate (WIMP).
As the Yukawa interaction terms among $\Sigma_R$, SM leptons and
Higgs boson are forbidden by the $\mathbb{Z}_2$ symmetry hence the
DM candidate $\psi^0$ interacts with the SM particle only through
gauge interactions. Moreover, from the Lagrangian (Eqs.\,\,\eqref{tripletLag}, \eqref{lag_psi0_psi-})
it is evident that the bare masses for all members of the fermionic
triplet $\Sigma_R$ are identical to $M_{\Sigma}$. However, one can make the
charged components $\psi^{\pm}$ heavier by considering one loop
electroweak radiative corrections \cite{Cirelli:2005uq, Ma:2008cu}, which result in
a mass splitting $\sim 166$ MeV between $M_{\psi^{\pm}}$ and
$M_{\psi^0}$ for $M_{\Sigma}\gtrsim1\,{\rm TeV}$.
In this framework, the relic density of the lightest $\mathbb{Z}_2$-odd
particle $\psi^{0}$ is determined by its pair annihilation
into $W^{+}W^{-}$ final state via t-channel exchange of $\psi^{\pm}$. Besides,
the co-annihilations among $\psi^0$ and $\psi^{\pm}$ into SM
particles also play a crucial role in relic density calculation
as the mass splitting between the members of the fermionic triplet
is extremely small ($\sim\mathcal{O}(100\,{\rm MeV})$). The co-annihilations
include several process like $\psi^0\,\psi^{\pm}\rightarrow W^{\pm}Z$
(via t-channel exchange of $\psi^{\mp}$), $\psi^+\,\psi^-
\rightarrow f\bar{f}$, $W^+W^-$ (via s-channel exchange of $Z$),
$\psi^{\pm}\,\psi^{\pm} (\psi^{\mp})\rightarrow
W^{\pm}W^{\pm}(W^{\mp})$ (via t-channel exchange of $\psi^0$) etc.
After including all these annihilation and co-annihilation channels,
the relic density of $\psi^0$ satisfies the Planck limit \cite{Ade:2015xua}
for $M_{\psi^0}\sim$ 2.2 TeV \cite{Ma:2008cu} (see green dashed line
in Fig.\,\,\ref{Fig:omega-mdm} in Section \ref{sec:relic_density}). Therefore,
the dark matter mass allowed by the observational data from Planck satellite
may not be produced efficiently in LHC at the present centre of mass energy and luminosity. This has also led to some recent discussions on fermion triplet dark matter with a singlet admixture, in order to lower the bound on fermion triplet from relic abundance point of view and to enhance its production at colliders, see for example, \cite{Choubey:2017yyn}. It should be noted that such admixture of singlet fermion with the neutral component of fermion triplet as dark matter candidate was proposed long back by the authors of \cite{Chardonnet:1993wd}, motivated from its search prospects at LEP collider. We consider another minimal modifications of the present
model which is also motivated from neutrino mass point of view,
so that one can still have a much lighter triplet fermionic
dark matter which at the same time satisfies all the existing direct
and indirect bounds and can have tantalising detection prospects at the LHC. We will present a detailed discussion on this topic in Section \ref{sec:lhc_cons}.
%%%%%%%%%%%%%%%%%%%%%%%%%%%%%%%%%%%%%%%%%%%%%%%%%%%%%%%%%%%%%%%%%%%%%%%%%%%%%%%
\section{Scotogenic Extension of FTDM}
\label{sec:scto_ftdm}
%%%%%%%%%%%%%%%%%%%%%%%%%%%%%%%%%%%%%%%%%%%%%%%%%%%%%%%%%%%%%%%%%%%%%%%%%%%%%%%
In order to have nonzero neutrino masses and a light triplet
fermionic dark matter, we have extended the minimal model,
described in the previous section, by two more triplets and
one inert doublet. Therefore, besides the usual SM fields, the present model
contains three fermionic SU(2)$_{\rm L}$ triplets and one extra
SU$(2)_{\rm L}$ doublet $\Phi=\left(\phi^+~~~~\dfrac{\phi^0 + i\,A^0}
{\sqrt{2}}\right)^{T}$. All these extra BSM fields are odd under
the discrete symmetry $\mathbb{Z}_2$.\,\,As a result, the Yukawa interaction
terms involving fermionic triplet, SM Higgs doublet and lepton doublet
are forbidden. Hence, one cannot generate tiny neutrino masses following
usual Type-III seesaw mechanism \cite{Foot:1988aq, Ma:1998dn,
Ma:2002pf}, where the above mentioned Yukawa
terms play a pivotal role. Instead, in the present model, we can write
the Yukawa terms using the inert Higgs doublet $\Phi$. However, the
extra doublet $\Phi$, being an inert one (does not interact with SM fermions),
does not have any vacuum expectation value (VEV) (in order to maintain $\mathbb{Z}_2$ symmetry unbroken).
Hence, there are no neutrino masses at tree level
(Type-III seesaw is not possible) and instead, the light neutrino
masses can be generated radiatively at one loop level following
the scotogenic model \cite{Ma:2006km}. This scotogenic version of Type-III seesaw model is also known as radiative Type-III seesaw model, studied in different contexts by several authors \cite{Ma:2008cu, Chao:2012sz, vonderPahlen:2016cbw}.

Further, the Majorana type bare mass terms
${\rm Tr}\left[\overline{{{\Sigma_{\alpha}}_{R}}^c}\,M^{\alpha\beta}_{\Sigma}\,
{\Sigma_{\beta}}_{R}\right]$ are although invariant under $\mathbb{Z}_2$
symmetry, the origin of such terms is not obvious in the present
scenario. Therefore, to understand a possible origin
of bare mass term of fermionic triplet we introduce a
scalar field $S^{\prime}$, which is a singlet under SM gauge group.
Moreover, we also impose an additional $\mathbb{Z}_3$ charges
to ${\Sigma_{\beta}}_{R}$ ($\beta =$ 1 to 3) and $S^{\prime}$ such
that the bare mass term of ${\Sigma_{\beta}}_{R}$ is
forbidden by $\mathbb{Z}_3$ symmetry while at the same time
${\rm Tr}\left[\overline{{{\Sigma_{\alpha}}_R}^c}
\,y_s^{\alpha\beta}\,{\Sigma_{\beta}}_{R}\right]{S^{\prime}}^{\dagger}$
term is allowed. Thus, when $S^{\prime}$ gets a VEV,
$\langle S^{\prime}\rangle=\dfrac{v_s}{\sqrt{2}}$,
this term will generate the bare mass term
$M^{\alpha\beta}_{\Sigma}\sim {y_s^{\alpha\beta}\,v_s}$.
Apart from that the singlet scalar $S^{\prime}$ also helps
the present model to evade the indirect detection bounds
on $W^{+}W^{-}$ annihilation channel \cite{Ahnen:2016qkx}.\,\,This
limit is obtained from the non-observation of
excess gamma-rays, over the known backgrounds, due to dark
matter annihilation from dwarf spheroidal galaxies. Another interesting aspect of including this singlet scalar is to enhance the dark matter direct detection rate by introducing a tree level scattering while the minimal model had only radiative direct detection scattering at one loop level. We will have a detailed
discussion on this topic in Section \ref{sec:id}.
Furthermore, we have to impose appropriate $\mathbb{Z}_3$ charges
to the SM leptons as well so that the Yukawa interaction terms
involving both inert doublet $\Phi$ and SM Higgs doublet $H$
remain invariant under $\mathbb{Z}_3$ symmetry. The $\mathbb{Z}_3$ charges of
${\Sigma_{\beta}}_{R}$ and SM leptons can be $\omega^2$ while that
of $S^{\prime}$ can be $\omega$, where $\omega^2$, $\omega$ are the
cube roots of unity.

The Lagrangian involving ${\Sigma_{\beta}}_{R}$, $\Phi$ and $S^{\prime}$ fields,
which is invariant under SU(2)$_{\rm L}\times {\rm U}(1)
_{\rm Y}\times {\mathbb{Z}_2} \times \mathbb{Z}_3$ symmetry,
is given by
\begin{eqnarray}
\mathcal{L}_{\rm BSM} &\supset& \dfrac{i}{2}\left(\sum_{\beta =1}^3{\rm Tr}
[\overline{{{\Sigma}_{\beta}}_{R}}\,\slashed{D} {{\Sigma}_{\beta}}_{R}]
+{\rm Tr}[\overline{{{{\Sigma}_{\beta}}_R}^c}\,\slashed{D} {{{\Sigma}_{\beta}}_{R}}^c]\right) -
\left(\sum_{\alpha,\,\beta=1}^3\dfrac{1}{2}\,{\rm Tr}[\overline{{{{\Sigma}_{\alpha}}_R}^c}\,
\sqrt{2}\,y_s^{\alpha\beta}\,{{\Sigma}_{\beta}}_R]\,{S^{\prime}}^{\dagger}+h.c.\right)
 \nonumber\\
&-&
\left(\sum_{\alpha,\,\beta=1}^3y^{\alpha\beta}_{\Sigma}\,\overline{{l_{\alpha}}_{L}}
{\Sigma_{\beta}}_{R} \tilde{\Phi} + h.c.\right)\, + 
\mathcal{L}_{\rm scalar}(H,\,\Phi,\,S^{\prime}) \,,
\label{lag_bsm}
\end{eqnarray}
where $\tilde{\Phi} = i\sigma_2 \Phi^\star$ and the Lagrangian
for the scalar fields are given by
\begin{eqnarray}
\mathcal{L}_{\rm scalar}(H,\,\Phi,\,S^{\prime}) &=& 
\left(D_{\mu}H\right)^{\dagger}\left(D^{\mu}H\right)
+\left(D_{\mu}\Phi\right)^{\dagger}\left(D^{\mu}\Phi\right) 
+\partial_\mu {{S^{\prime}}^\dagger}\,\partial^\mu S^{\prime}
+\mu^2_H\,H^\dagger H - \lambda_H\,\left(H^\dagger H\right)^2
\nonumber \\ 
&-&\mu_2^2\, \Phi^\dagger \Phi - \lambda_2 (\Phi^\dagger \Phi)^2
+\mu_s^2\,{S^{\prime}}^\dagger S^{\prime} 
-\mu_3 \left({S^\prime}^3+{{S^\prime}^\dagger}^3\right)
-\lambda_s \left({S^{\prime}}^\dagger S^{\prime}\right)^2 
\nonumber\\ &-&
\lambda_3 \left(H^\dagger H\right)\left(\Phi^\dagger \Phi\right)
-\lambda_4 \left(H^\dagger \Phi\right)\left(\Phi^\dagger H\right)
%\nonumber \\ &-&
-\dfrac{\lambda_5}{2}\left((H^\dagger \Phi)^2+h.c.\right)
\nonumber\\ &-&
\lambda_6 \left(H^\dagger H\right)\left({S^{\prime}}^\dagger S^{\prime}\right)
-\lambda_7 \left(\Phi^\dagger \Phi\right)\left({S^{\prime}}^\dagger S^{\prime}\right)\,.
\label{lscalar}
\end{eqnarray}
where $S^{\prime} = \dfrac{h_2 + i\, \zeta}{\sqrt{2}}$,
$h_2$ and $\zeta$ are the real and imaginary parts of
$S^{\prime}$ respectively while $H=\left(0\,\,\,\,\,\frac{h_1+v}{\sqrt{2}}\right)^T$
is the SM Higgs doublet in the unitary gauge. 
Following the procedure given in the Appendix \ref{App:lag}, we
can write the triplet Lagrangian in terms of $\psi^{-}_{\beta}=
{\Sigma^{-}_{\beta}}_R + {{\Sigma^{+}_{\beta}}_R}^c$ and
$\psi^{0}_{\beta}={\Sigma^{0}_{\beta}}_R + {{\Sigma^{0}_{\beta}}_R}^c$ as 
\begin{eqnarray}
\mathcal{L}_{\rm BSM} &\supset& 
\sum_{\beta=1}^3\overline{\psi^{-}_{\beta}}\,i\,\slashed{\partial} \psi^{-}_{\beta} +
\dfrac{1}{2}\overline{\psi^0_{\beta}}i\,\slashed{\partial} \psi^0_{\beta} - 
\left(\sum_{\alpha,\,\beta=1}^3y_{s}^{\alpha\beta}\,\overline{\psi^{-}_{\alpha}}\,\psi^{-}_{\beta} 
+\dfrac{y_{s}^{\alpha\beta}}{2}\,\overline{\psi^0_{\alpha}}\,\psi^0_{\beta}\right)
h_2
\nonumber \\ &&
-i\left(\sum_{\alpha,\,\beta=1}^3y_{s}^{\alpha\beta}\,\overline{\psi^{-}_{\alpha}}
\gamma_5\,\psi^{-}_{\beta} 
+\dfrac{y_{s}^{\alpha\beta}}{2}\,\overline{\psi^0_{\alpha}}\gamma_5\,\psi^0_{\beta}\right)
\zeta
-\,g\left(\overline{\psi^{-}_{\beta}}\gamma^{\mu}\,\psi^{0}_{\beta}\,W^{-}_{\mu} + h.c.\right)
\nonumber \\&&
+\,g\,\cos\theta_w\,\overline{\psi^{-}_{\beta}}\gamma_{\mu}\psi^{-}_{\beta}\,Z^{\mu}
+ \,g\,\sin\theta_w\,\overline{\psi^{-}_{\beta}}\gamma_{\mu}\psi^{-}_{\beta}\,A^{\mu}
\nonumber \\ &&
-\Bigg[\sum_{\alpha,\,\beta=1}^3 y^{\alpha\beta}_{\Sigma}
\left(\dfrac{1}{\sqrt{2}}\overline{{\nu_{\alpha}}_L}\,P_R \psi^0_{\beta} +
\overline{{e_{\alpha}}_L}\,P_R\psi^{-}_{\beta}\right)
\left(\dfrac{\phi^0 -i\,A^0}{\sqrt{2}}\right) 
-y^{\alpha\beta}_{\Sigma}\,\overline{P_L\,{\psi^{-}_{\beta}}}\,{{\nu_{\alpha}}_{L}}^c\,\phi^{-}
\nonumber \\ &&  
+ \dfrac{y^{\alpha\beta}_{\Sigma}}{\sqrt{2}}\overline{{e_{\alpha}}_L}P_R\psi^0_{\beta}\,\phi^{-}
+ h.c.\Bigg] +\mathcal{L}_{\rm scalar}(H,\,\Phi,\,S^{\prime})\,.
\label{lag_bsm_psi0_psi-}
\end{eqnarray}
As mentioned earlier, when
the real part of $S^{\prime}$ gets VEV $v_s$, the mass terms for both $\psi^{-}_{\alpha}$
and $\psi^{0}_{\alpha}$ are generated from the third and fourth terms of the above
Lagrangian as $M^{\alpha\beta}_{\psi} = {y^{\alpha\beta}_s\,v_s}$.
For simplicity, we have assumed that $M^{\alpha\beta}_{\psi}$ is a diagonal matrix with
real and nonzero elements. Therefore, both $\psi^{0}_{\alpha}$
and $\psi^{-}_{\alpha}$ are representing the physical states
and the lightest neutral fermion $\psi^{0}_1$ can be our
dark matter candidate with mass $M_{\dm}={y_s\,v_s}$
\footnote{For notational simplicity we choose $y^{11}_s=y_s$.}.
Furthermore, the first term within
the square bracket is responsible for tiny neutrino
mass generation at one loop level. We will have
a detailed discussion on this topic in the next section (Section \ref{sec:neutrino}). 
After SU(2)$_{\rm L}\times{\rm U}(1)_{\rm Y}\times\mathbb{Z}_3$ symmetry
breaking, the mass terms for the components of inert Higgs doublet
are
\begin{eqnarray}
M^2_{\phi^{\pm}} &=& \mu_2^2 + \dfrac{1}{2}\left(\lambda_3 v^2 
+ \lambda_7 v^2_s\right)\,\,,
\label{Eq:idm_mphic}\\
M^2_{A^0} &=& \mu_2^2 + \dfrac{1}{2}\left[(\lambda_3+\lambda_4-\lambda_5)
v^2 + \lambda_7 v^2_s\right]\,\,,
\label{Eq:idm_mA}\\
M^2_{\phi^0} &=& \mu_2^2 + \dfrac{1}{2}\left[(\lambda_3+\lambda_4+\lambda_5)
v^2 + \lambda_7 v^2_s\right]\,\,.
\label{Eq:idm_mphi0}
\end{eqnarray} 
On the other hand, after symmetry breaking there will be a
mixing between the real scalar fields of $H$ and $S^{\prime}$.
The mixing matrix with respect to the basis ($h_1$, $h_2$)
is given by
\begin{eqnarray}
\mathcal{M}^2_{\rm scalar} =
\begin{pmatrix}
2\,\lambda_H v^2 &\,\,\,\lambda_6\,v\,v_s\\
\lambda_6\,v\,v_s &\,\,\,2\,\lambda_s v_s^2 + 
\dfrac{3\,\mu_3\,v_s}{\sqrt{2}}
\end{pmatrix}\,.
\label{scalarmmatrix}
\end{eqnarray}
Clearly, $h_1$ and $h_2$ do not represent physical fields. However,
after diagonalising the above mass matrix, one can have two physical
scalar fields $h$ and $S$, which are two orthogonal linear combinations
of $h_1$, $h_2$ and the corresponding mixing angle is given by
\begin{eqnarray}
\tan 2\xi = \dfrac{\lambda_6\,v\,v_s}{\lambda_H\,v^2-\lambda_s\,v_s^2-
\frac{3\,\mu_3\,v_s}{2\,\sqrt{2}}}\,\,.
\label{Eq:scalar_mixing_angle}
\end{eqnarray}
We consider $h$ is the SM-Like Higgs boson which was
discovered at the LHC \cite{Aad:2012tfa, Chatrchyan:2012xdj}.
The mass of the pseudo scalar $\zeta$ is given by
\begin{eqnarray}
M^2_{\zeta} = -\dfrac{3\,\mu_3\,v_s}{\sqrt{2}}\,,
\label{mzeta}
\end{eqnarray}
where we need $\mu_3<0$ for $M^2_{\zeta}>0$. The trilinear terms of
$S^\prime$ and ${S^\prime}^\dagger$ allowed by $\mathbb{Z}_3$
symmetry play a crucial role in generating the mass term for
the imaginary part of $S^{\prime}$ after spontaneous symmetry breaking
of $\mathbb{Z}_3$. As a result, we can choose the mass of the
pseudo scalar ($\zeta$) different than $S$ and also
heavier compared to the mass of our dark matter
candidate $\dm$, so that it will not affect the
relic density of $\dm$. From Eqs.\,\,(\ref{mzeta}),
one can easily find that $M_{\zeta}$ will be
heavier than our dark matter candidate $\dm$ when $\mu_3<-
\dfrac{\sqrt{2}\,M^2_{\dm}}{3\,v_s}$. Now, using $M_{\dm}=
y_s v_s$, we can rewrite the above limit on $\mu_3$ as $\mu_3 <
-\dfrac{\sqrt{2}\,y^2_s}{3} v_s$. On the other hand,
since $\mu_3$ acts oppositely to the expression
of $M_{s}$, therefore one can also find a lower bound
on self coupling $\lambda_s$ from the requirement of $M_s>0$.
Therefore, under the approximation of small mixing angle $\xi$,
the quartic coupling $\lambda_s$ must be greater than $\dfrac{y^2_s}{2}$.
In Sections \ref{sec:relic_density} and \ref{sec:id},
we have considered $y_s=1.5$ for $M_{\dm}=1$ TeV
and hence we need $\mu_3<-707.11$ GeV and $\lambda_s>1.13$.
One thing we want to note here that the considered
hierarchy $M_{\zeta}>M_{\dm}>M_s$ is solely for the
calculational simplification. There may be a situation
when $M_{\dm}>M_{\zeta}, M_s$, where our dark matter candidate
can annihilate to both $\zeta$ as well as $S$ and in that case
due to the extra annihilation modes, the thermal contribution
to dark matter relic abundance will be even less compared
to the present scenario.
The main difference between the minimal
model discussed in the previous section and the present model
is that here, our dark matter candidate ${\psi^0_1}$, after
its thermal freeze-out, gets some non-thermal contribution to the
relic abundance mainly from the decay of $\phi^0$. This eventually 
compensates the under abundance of ${\psi^0_1}$ in
low mass regime $M_{\psi_1}\lesssim2$ TeV. We also consider the mass of $\phi^0$ to be smaller than the heavier cousins of $\Sigma_{1R}$ so that $\phi^0$ decays preferentially to the dark matter candidate only. This is also important from the point of view of neutrino mass (as we discuss in the next section) because the couplings of $\Sigma_{2R, 3R}$ with $\Phi$ and SM leptons have to be sizeable in order to generate the correct one loop neutrino mass at sub-eV. Therefore, if $\Sigma_{2R, 3R}$ are lighter than $\Phi$, then $\phi^0$ will also decay into these triplets reducing the branching ratio into the dark matter candidate $\Sigma_{1R}$, falling short of providing the required non-thermal contribution. We will discuss
in great detail about the thermal as well as non-thermal contributions
to the relic density of our dark matter candidate in Section \ref{sec:relic_density}.
%%%%%%%%%%%%%%%%%%%%%%%%%%%%%%%%%%%%%%%%%%%%%%%
\section{Neutrino mass generation at one loop}
\label{sec:neutrino}
\begin{figure}[h!]
\centering
\includegraphics[scale=0.5]{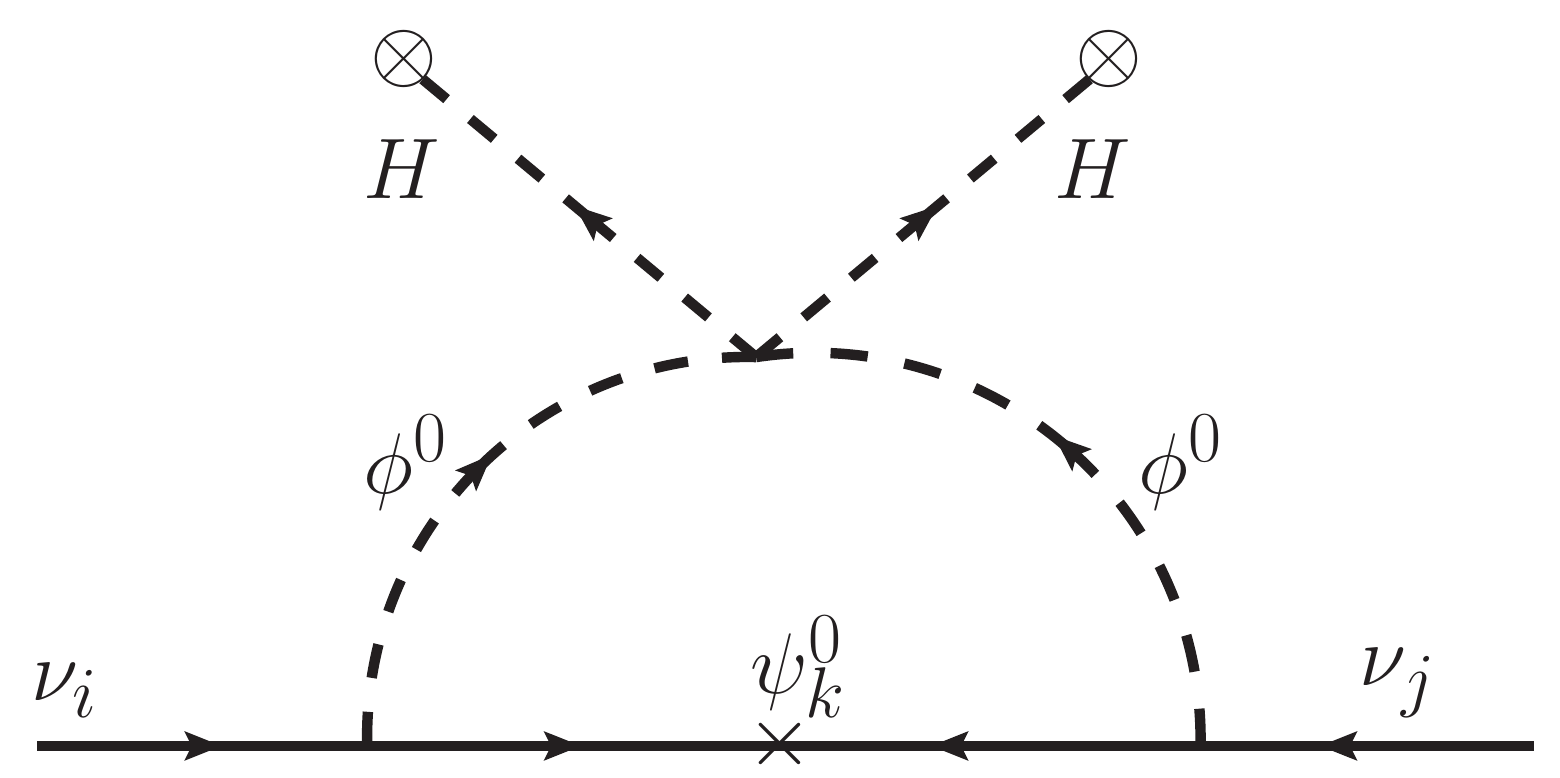}
\caption{Feynman diagram of neutrino mass generation at one loop.}
\label{Fig:neutrino-mass}
\end{figure}

%%%%%%%%%%%%%%%%%%%%%%%%%%%%%%%%%%%%%%%%%%%%%%%
In the present model since we have imposed both
$\mathbb{Z}_2$ as well as $\mathbb{Z}_3$ charges
to all three fermion triplets ${\Sigma_{\beta}}_R$,
neutrino masses can not be generated via tree level Type-III
seesaw mechanism. However, in this
model we have an inert doublet $\Phi$ which
is also odd under $\mathbb{Z}_2$ symmetry.
Hence we can write a Yukawa interaction
term involving fermion triplet ${\Sigma_{\beta}}_R$
inert doublet $\Phi$ and SM lepton doublet
${l_{\alpha}}_L$. Such term will be automatically
$\mathbb{Z}_2$ invariant. However, $\mathbb{Z}_3$
invariance requires same $\mathbb{Z}_3$
charge to SM leptons as well since $\Phi$ does not have any
$\mathbb{Z}_3$ charge. Because, any nonzero $\mathbb{Z}_3$
charge of $\Phi$, other than unity, will forbid the mass
splitting between $A^0$ and $\phi^0$, which is $\lambda_5 v^2$
(see Eqs.\,\,(\ref{Eq:idm_mA} and \ref{Eq:idm_mphi0})).
Later in this section, we will see that the light
neutrino masses are proportional to the mass
splitting between $A^0$ and $\phi^0$, hence we
cannot impose any non-trivial $\mathbb{Z}_3$
charge to $\Phi$.
Therefore, in this work
as mentioned in the previous section, we choose $\mathbb{Z}_3$
charge of ${\Sigma_{\beta}}_R$ and ${l_{\alpha}}_L$ is $\omega^2$.
The Yukawa interaction terms invariant under both continuous
and discrete symmetries of the model are given as
\begin{eqnarray}
\mathcal{L}_{\rm Yukawa} \supset &-&
\sum_{\alpha,\,\beta=1}^3y^{\alpha\beta}_{\Sigma}\,\overline{{l_{\alpha}}_{L}}
{\Sigma_{\beta}}_{R} \tilde{\Phi} + h.c.\,\,,\nonumber \\
&-&\sum_{\alpha,\,\beta=1}^3 y^{\alpha\beta}_{\Sigma}
\left(\dfrac{1}{\sqrt{2}}\overline{{\nu_{\alpha}}_L}\,P_R \psi^0_{\beta} +
\overline{{e_{\alpha}}_L}\,P_R\psi^{-}_{\beta}\right)
\left(\dfrac{\phi^0 -i\,A^0}{\sqrt{2}}\right) 
-y^{\alpha\beta}_{\Sigma}\,\overline{P_L\,{\psi^{-}_{\beta}}}\,{{\nu_{\alpha}}_{L}}^c\,\phi^{-}
\nonumber \\ 
&+& \dfrac{y^{\alpha\beta}_{\Sigma}}{\sqrt{2}}\overline{{e_{\alpha}}_L}P_R\psi^0_{\beta}\,\phi^{-}
+ h.c.\,\,,
\end{eqnarray}
where we have used the definition of $\psi_{\beta}^{-}$
and $\psi_{\beta}^0$ give in Appendix \ref{App:lag}. As
discussed in previous section, the first term in the above
Lagrangian is responsible for neutrino mass generation
in one loop level following the scotogenic model \cite{Ma:2006km}.
The Feynman diagram for neutrino mass generation
at one loop level is shown in Fig.\,\,\ref{Fig:neutrino-mass}.
In Fig.\,\,\ref{Fig:neutrino-mass}, we have drawn
the diagram with $\phi^0$ as an intermediate
loop particle. However, the same diagram with $\phi^0$
replaced by pseudo scalar $A^0$ is also possible. These
two diagrams will differ by a -ve sign which actually 
helps us to cancel the divergences of both diagrams.
The expression of neutrino mass is then given by \cite{Ma:2006km}
\begin{eqnarray}
{\mathcal{M}_{\nu}}^{\alpha\beta} = \sum_{\rho} \dfrac{y^{\alpha \rho}_{\Sigma}
\,y^{\beta \rho}_{\Sigma}\,M_{\psi^0_\rho}}{64\pi^2}\,
\left(\dfrac{M^2_{\phi^0}}{M^2_{\phi^0}-M^2_{\psi^0_\rho}}{\rm ln}
\dfrac{M^2_{\phi^0}}{M^2_{\psi^0_\rho}}
-\dfrac{M^2_{A^0}}{M^2_{A^0}-M^2_{\psi^0_\rho}}{\rm ln}
\dfrac{M^2_{A^0}}{M^2_{\psi^0_\rho}}\right)\,.
\end{eqnarray} 
Now, if we consider the mass splitting between $M_A^0$
and $M_{\phi^0}$ is small compared to $(M_A^0 + M_{\phi^0})/2$,
which is indeed true for our work (see Section \ref{sec:relic_density}),
the expression for $\alpha \beta$ element of neutrino mass
matrix simplifies to
\begin{eqnarray}
{\mathcal{M}_{\nu}}^{\alpha\beta} = \dfrac{\lambda_5\,v^2}{64\pi^2}
\sum_{\rho} \dfrac{y^{\alpha \rho}_{\Sigma}
\,y^{\beta \rho}_{\Sigma}\,M_{\psi^0_\rho}}{M^2_{0}-M^2_{\psi^0_\rho}}\,
\left(1-\dfrac{M^2_{\psi_\rho^0}}{M^2_{0}-M^2_{\psi^0_\rho}}{\rm ln}
\dfrac{M^2_{0}}{M^2_{\psi^0_\rho}}\right)\,,
\end{eqnarray} 
where $M^2_0=(M_A^0 + M_{\phi^0})/2$ and
$M^2_{A^0}-M^2_{\phi^0}=\lambda_5\,v^2$ using
Eqs.\,\,(\ref{Eq:idm_mA} and \ref{Eq:idm_mphi0}). It should be noted that the requirement for non-thermal contribution to dark matter candidate in our model constrains the Yukawa couplings of the lightest fermion triplet with the SM leptons to be very small $y^{11}_{\Sigma} \approx y^{21}_{\Sigma} \approx y^{31}_{\Sigma} \leq 10^{-9}$, as we discuss in details below. As can be seen from the one loop neutrino mass formula given above, such tiny couplings will practically have a negligible contribution to light neutrino masses. This makes the lightest fermion triplet to effectively decouple from the neutrino mass generation mechanism giving rise to one almost massless and two massive light neutrinos which can be tested at experiments which are sensitive to the absolute mass scale of neutrinos.
%%%%%%%%%%%%%%%%%%%%%%%%%%%%%%%%%%%%%%%%%%%%%%%
\section{Relic Abundance of dark matter candidate ${\Sigma^0_1}_R$
in Scotogenic FTDM}
\label{sec:relic_density}
\begin{figure}[h!]
\centering
\includegraphics[height=2cm,width=4cm]{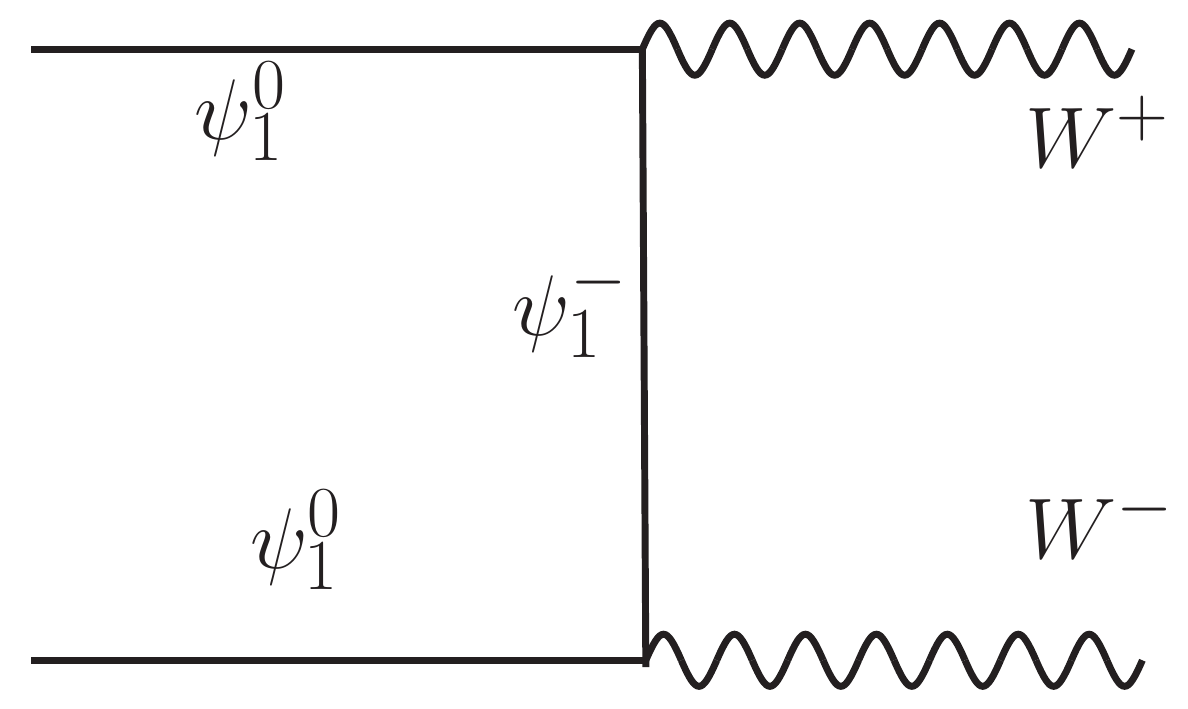}
\includegraphics[height=2cm,width=4cm]{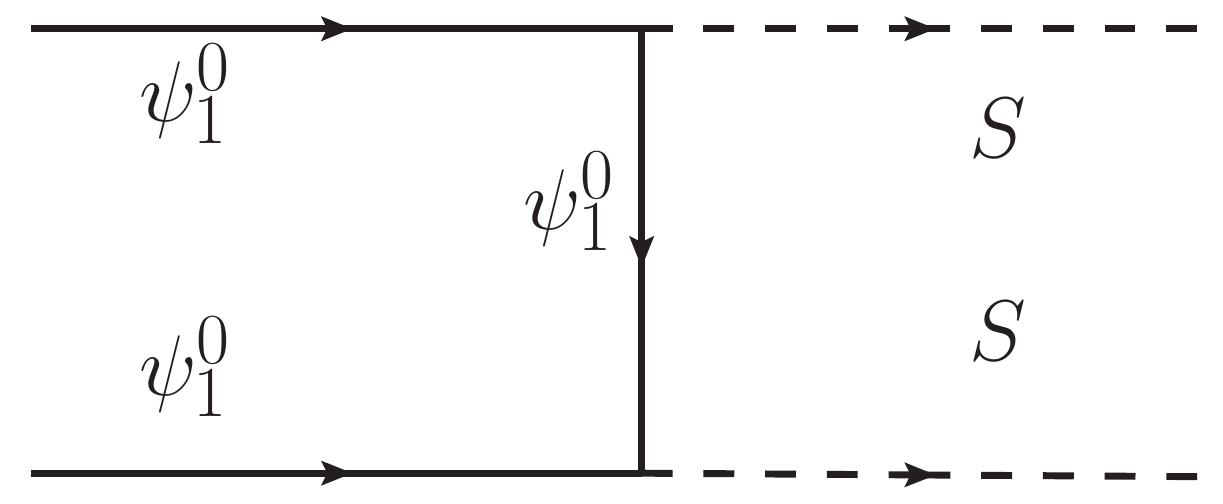}\\
\vskip 0.1in
\includegraphics[height=2cm,width=4cm]{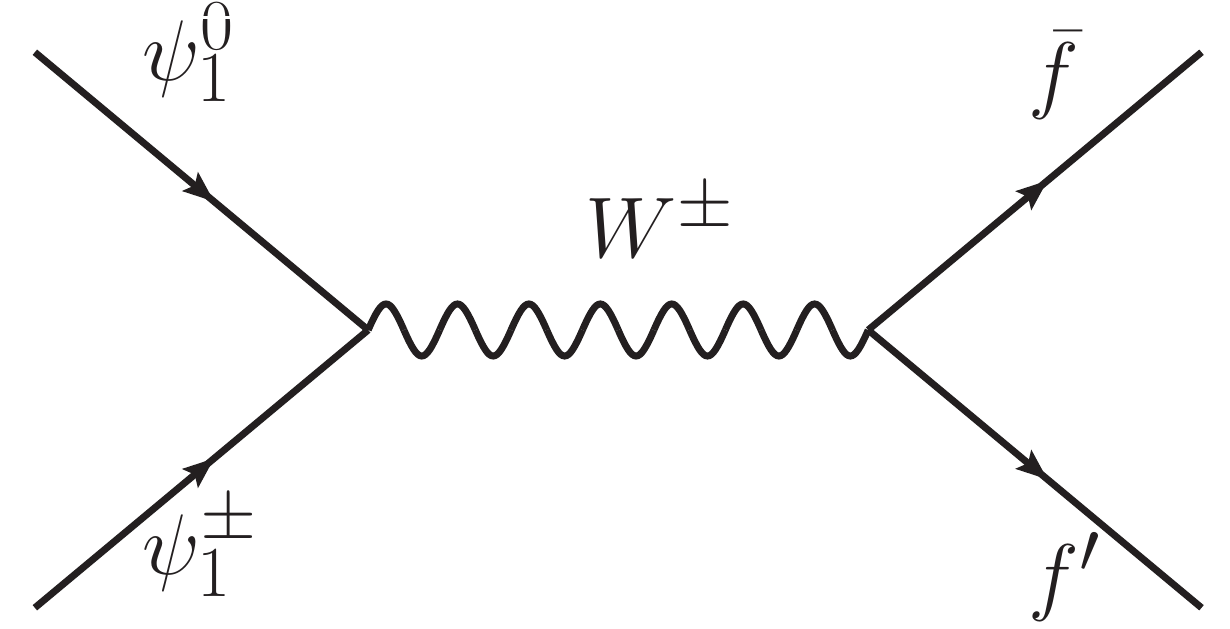}
\includegraphics[height=2cm,width=4cm]{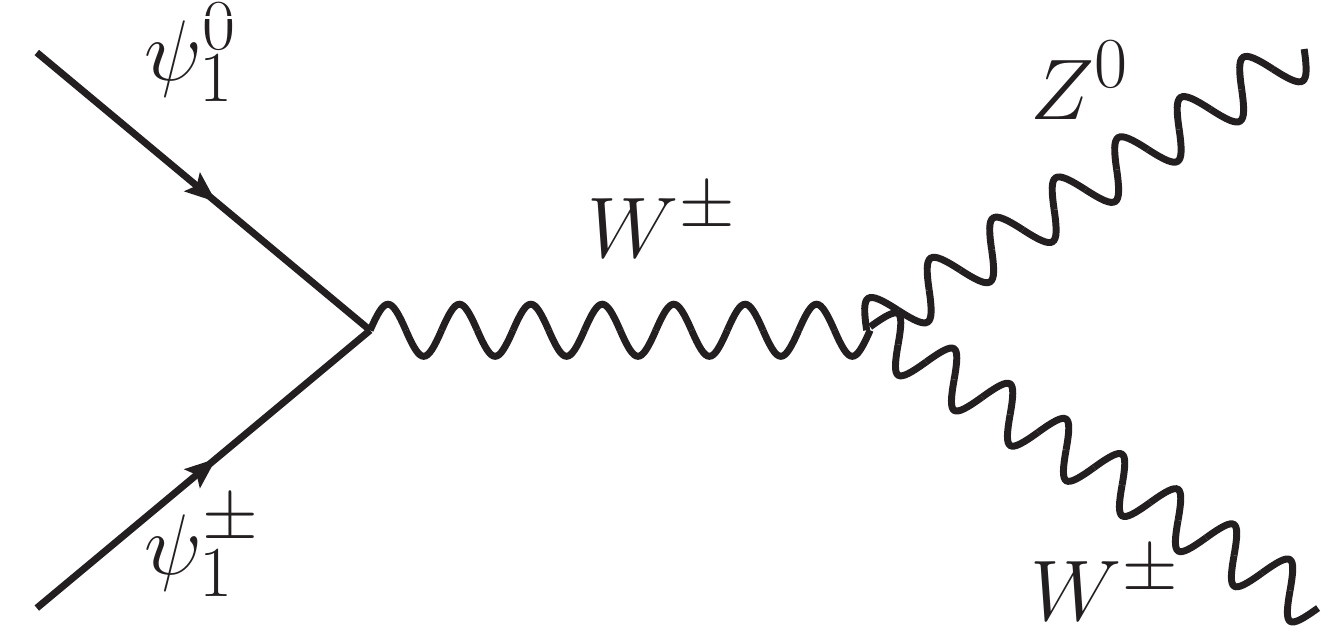}
\includegraphics[height=2cm,width=4cm]{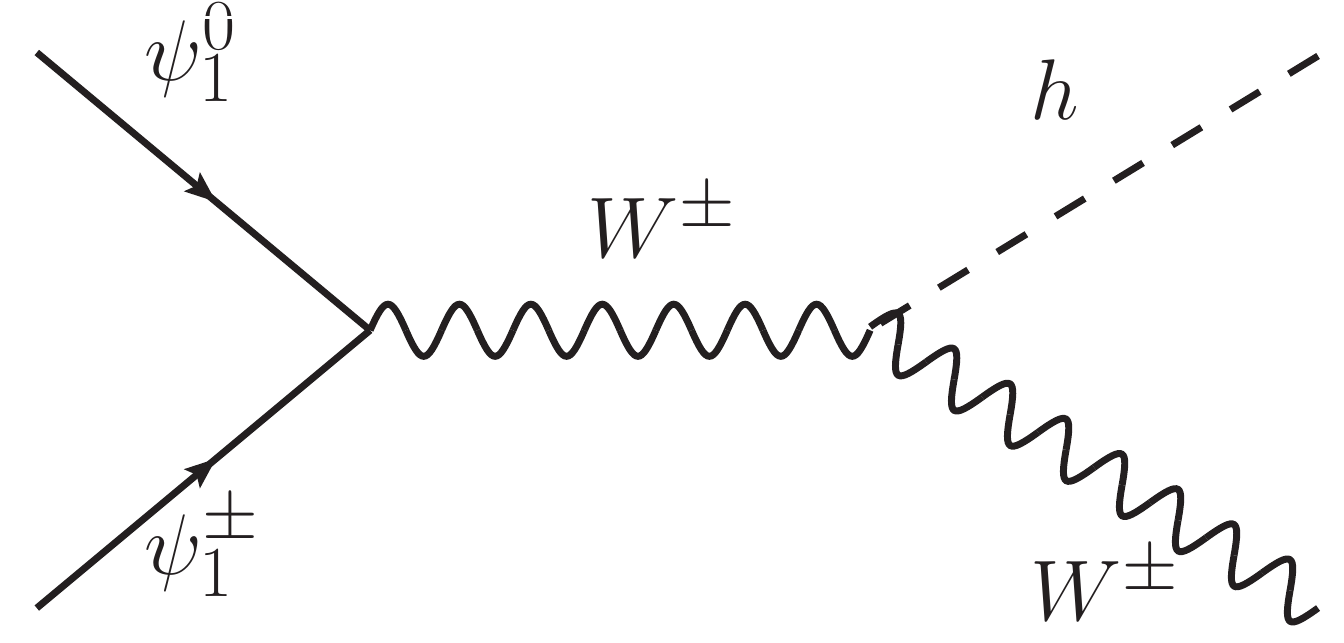}
\includegraphics[height=2cm,width=4cm]{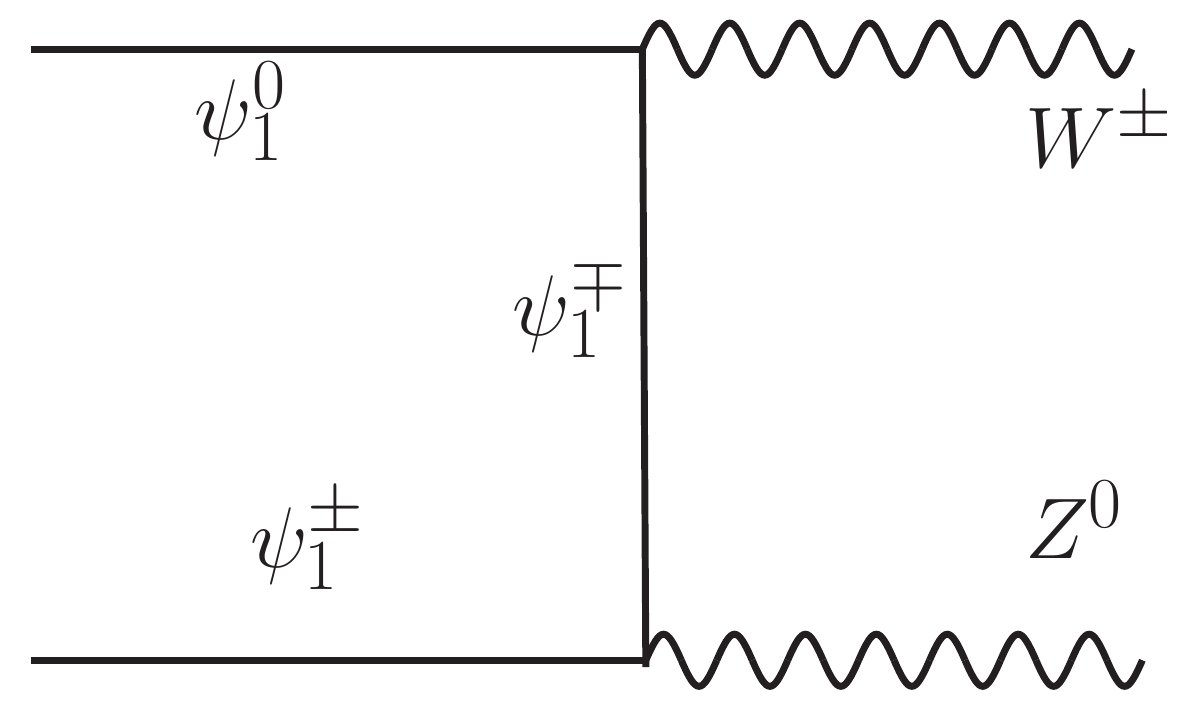}\\
\vskip 0.1in
\includegraphics[height=2cm,width=4cm]{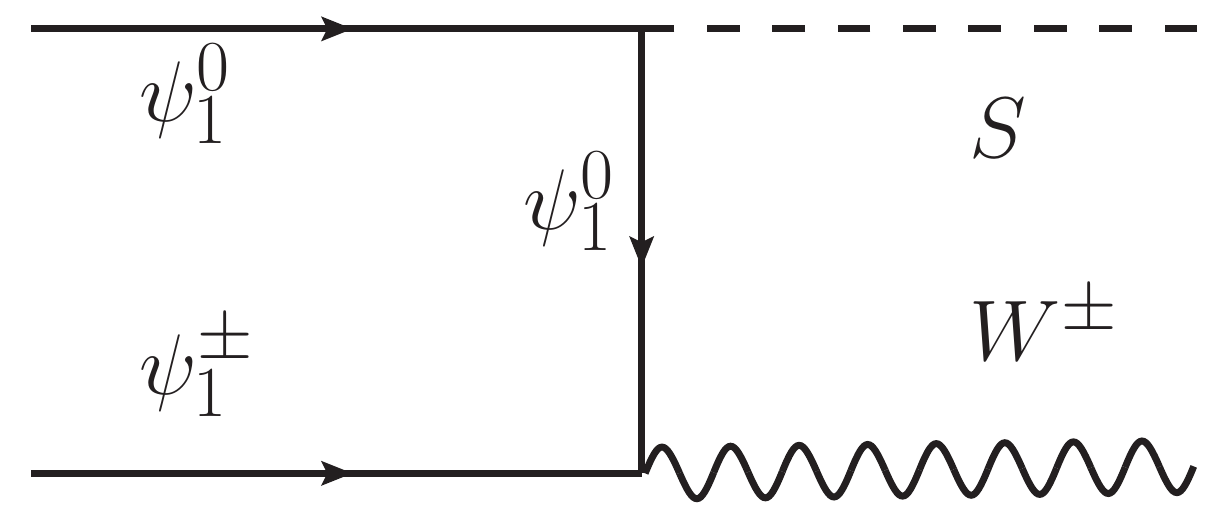}
\includegraphics[height=2cm,width=4cm]{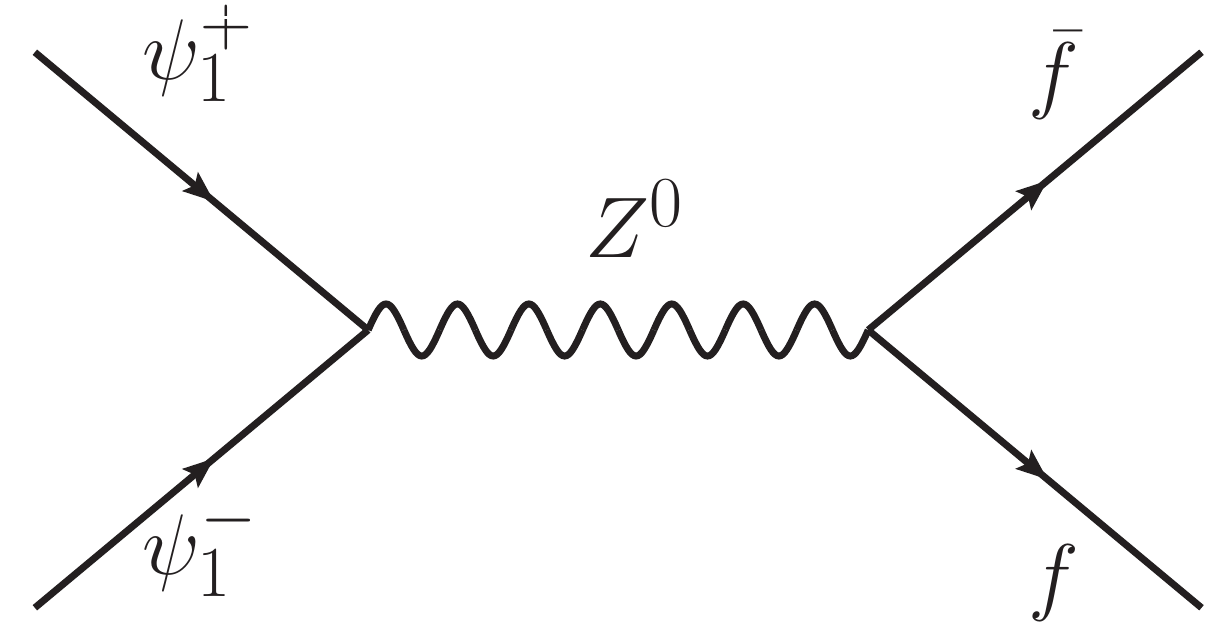}
\includegraphics[height=2cm,width=4cm]{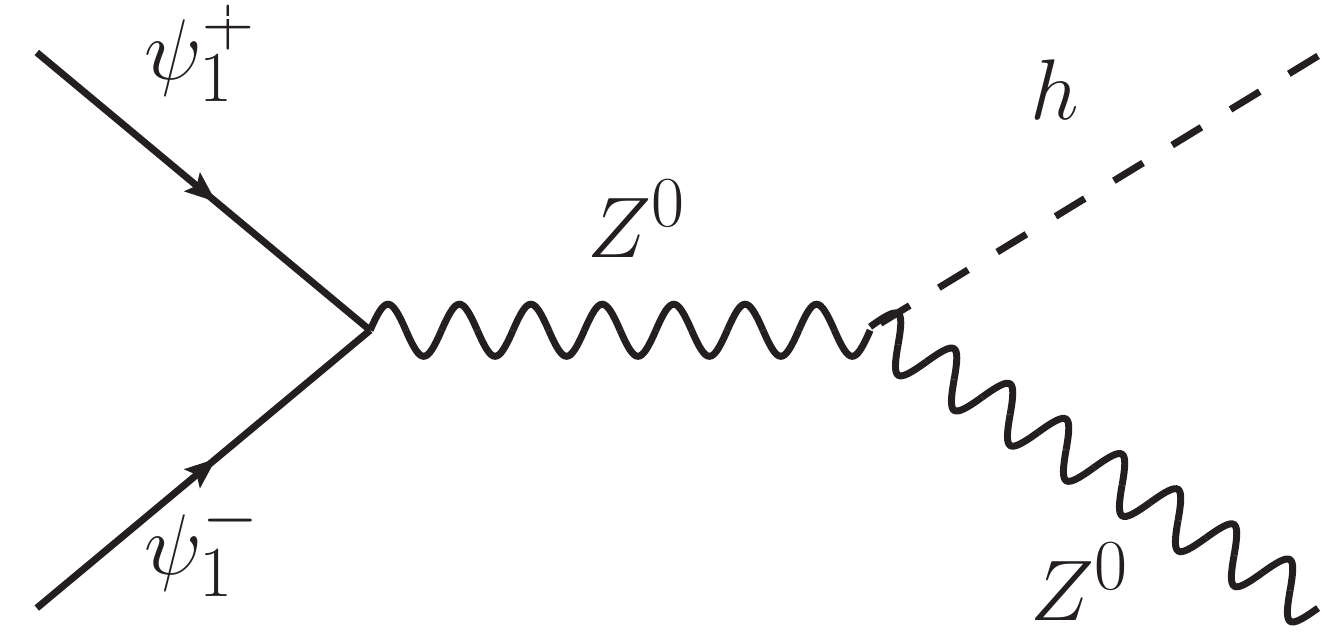}
\includegraphics[height=2cm,width=4cm]{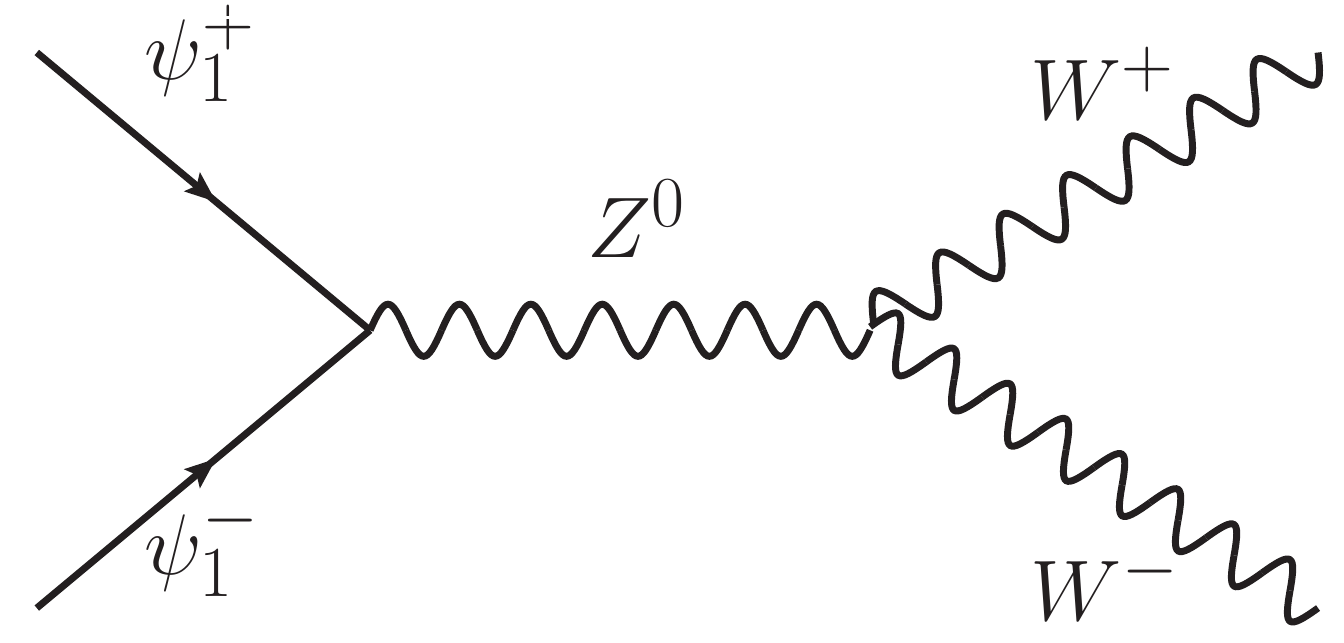}\\
\vskip 0.1in
\includegraphics[height=2cm,width=4cm]{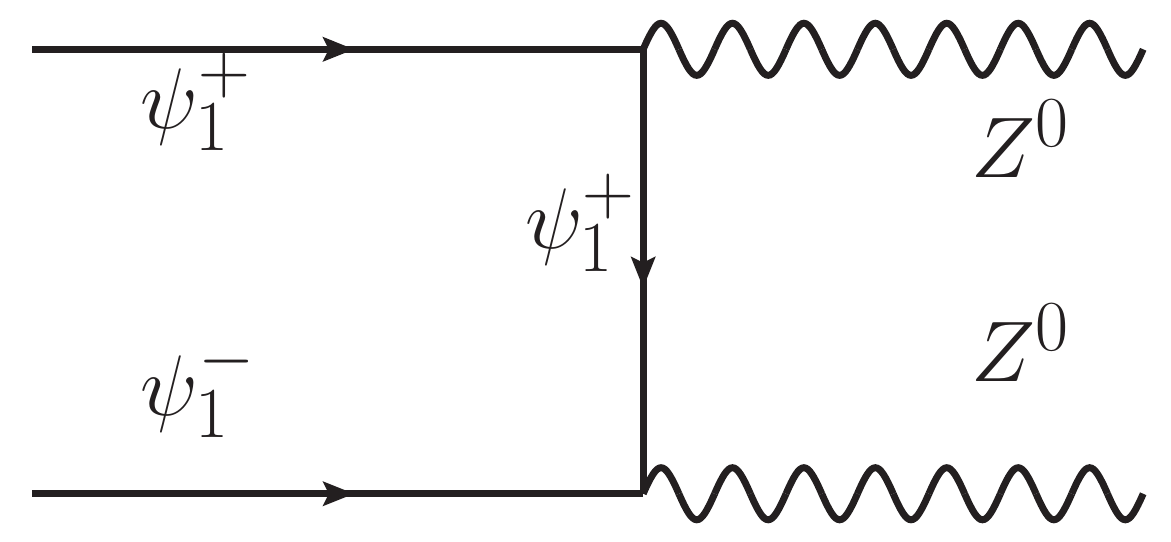}
\includegraphics[height=2cm,width=4cm]{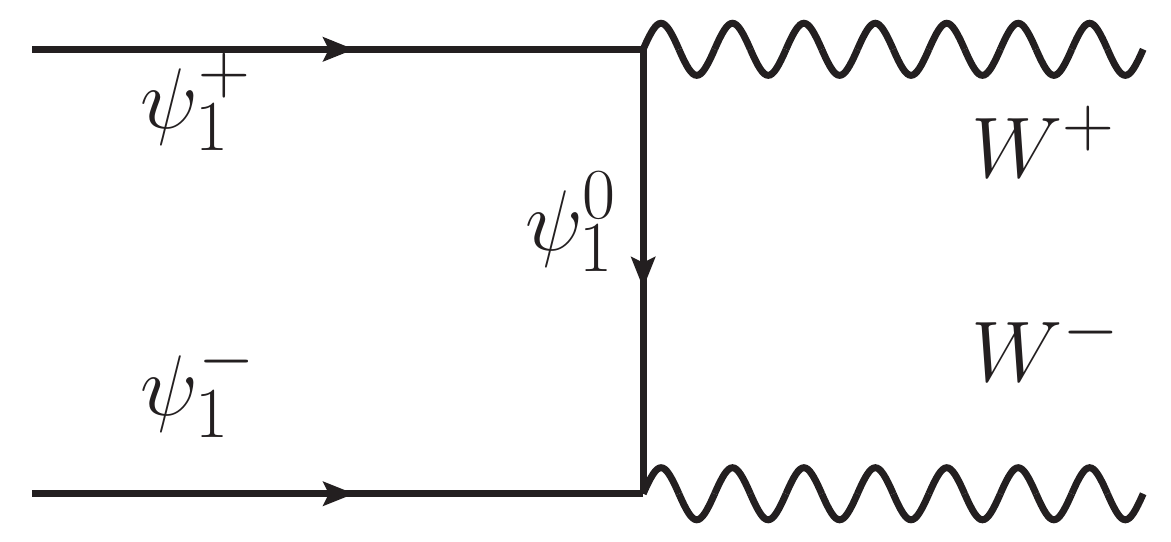}
\includegraphics[height=2cm,width=4cm]{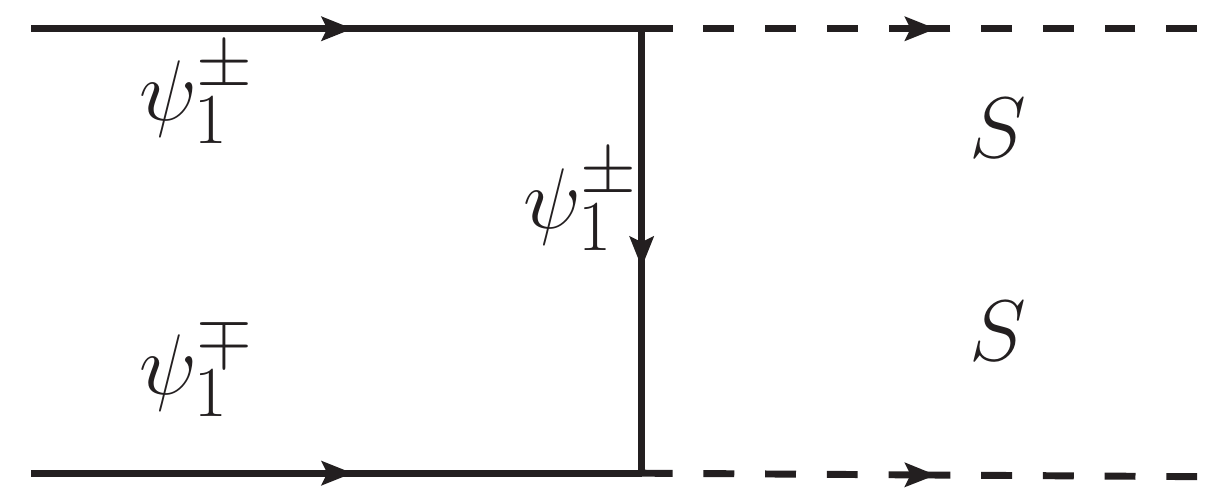}
\includegraphics[height=2cm,width=4cm]{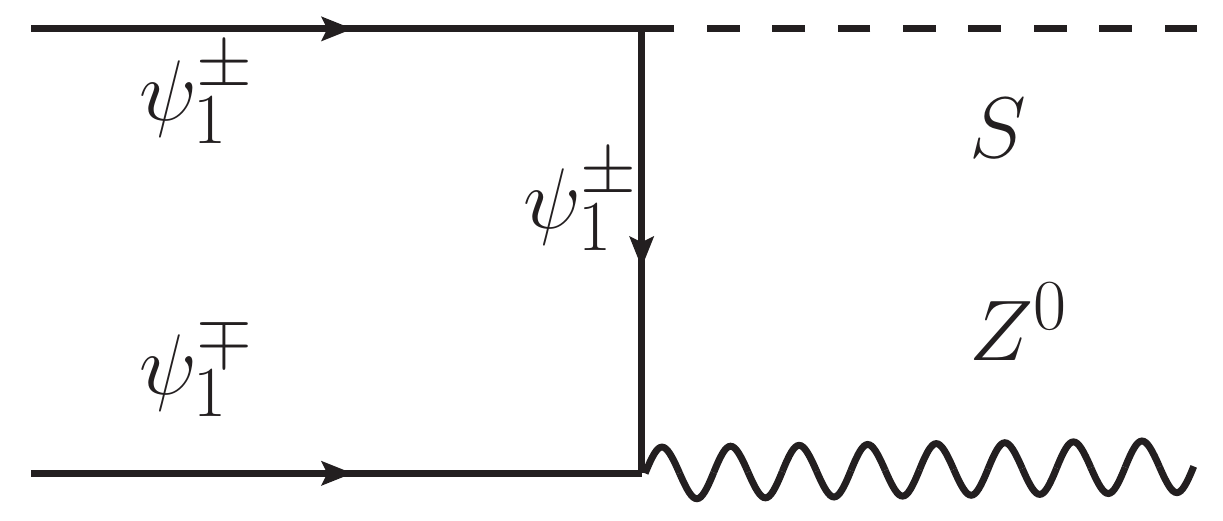}\\
\vskip 0.1in
\includegraphics[height=2cm,width=4cm]{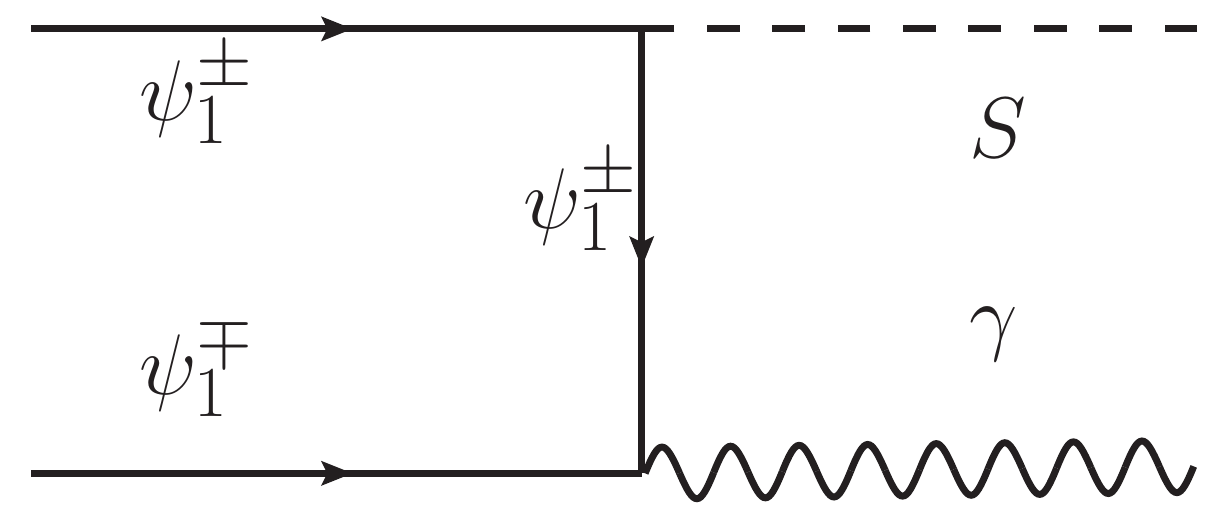}
\includegraphics[height=2cm,width=4cm]{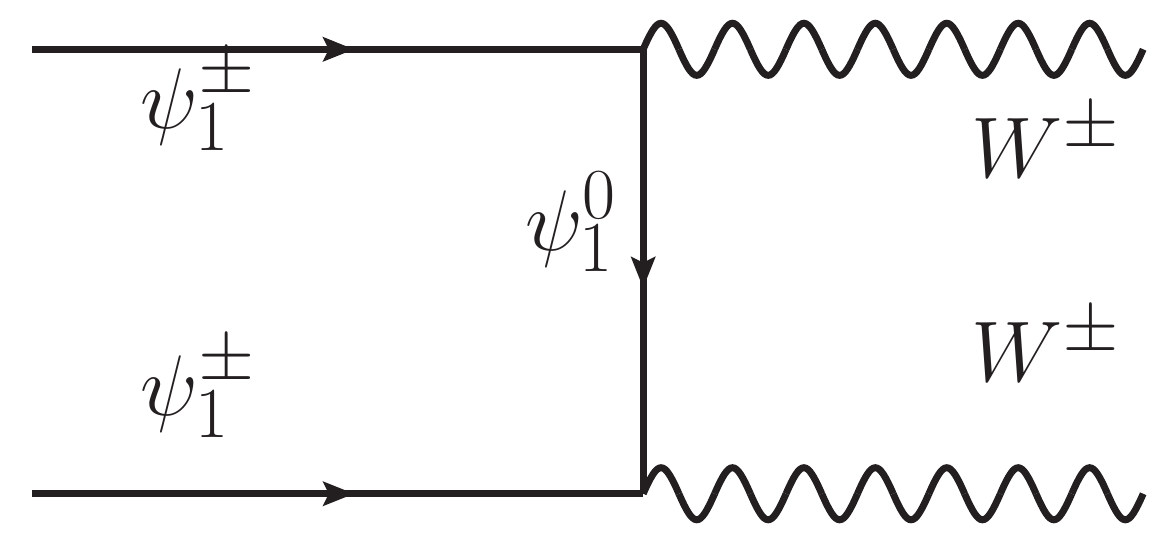}
\caption{Feynman diagrams for all possible annihilation
and co-annhilation channels. Two diagrams in first row
are annihilation of $\dm$ while rest are co-annihilations.}
\label{Fig:feyn_dia}
\end{figure}

%%%%%%%%%%%%%%%%%%%%%%%%%%%%%%%%%%%%%%%%%%%%%%%%
To compute the relic abundance of a dark matter candidate,
one has to find the value of comoving number density of
dark matter species at the present epoch, which can be
found by solving the corresponding Boltzmann equation for
the dark matter candidate we are considering i.e.\,\,$\dm$. Moreover,
while solving the Boltzmann equation for $\dm$, we need
to have the information about the comoving number densities
of other particles which are not in thermal equilibrium
but have significant impact on the production 
processes of $\dm$. In other words, we need to solve
a system of coupled Boltzmann equations for all the
out of equilibrium particles. The coupled Boltzmann
equations for our present scenario are given by
\begin{eqnarray}
\frac{dY_{\dm}}{dx} &=& -\left(\frac{45\,G}{\pi}\right)^{-1/2}
\frac{M_{sc}}{x^2}\,\sqrt{g_{\star}}\,\langle\sigma {\rm v}\rangle_{\rm Triplet}
\left( Y_{\dm}^2-(Y_{\dm}^{eq})^2\right) \nonumber \\
&& + \left(\frac{4 \pi^3 G}{45}\right)^{-1/2} \frac{x}{M_{sc}^2}
\frac{\sqrt{g_{\star}}}{g_s} \left( \langle \Gamma_{\phi^0 \rightarrow \dm} \rangle
\ Y_{\phi^0} + 2\,\langle \Gamma_{\psi_1^{\pm} \rightarrow \dm} \rangle  \ Y_{\psi_1^{-}}
\right)\,, 
\label{Eq:BEpsi0}\\
\nonumber\\
%\end{eqnarray}
%\begin{eqnarray}
\frac{dY_{\phi^0}}{dx} &=& -\left(\frac{45\,G}{\pi}\right)^{-1/2}
\frac{M_{sc}}{x^2}\,\sqrt{g_{\star}}\,\langle\sigma {\rm v}\rangle_{\rm IDM}
\left( Y_{\phi^0}^2-(Y_{\phi^0}^{eq})^2\right) \nonumber \\
&& - \left(\frac{4 \pi^3 G}{45}\right)^{-1/2} \frac{x}{M_{sc}^2}
\frac{\sqrt{g_{\star}}}{g_s} \langle \Gamma_{\phi^0}^{\rm Total} \rangle 
\ Y_{\phi^0}\,,
\label{Eq:BEphi0}
\\
%\end{eqnarray}
%\begin{eqnarray}
\nonumber\\
\frac{dY_{\psi_1^-}}{dx} &=& \left(\frac{4 \pi^3 G}{45}\right)^{-1/2}
\frac{x}{M_{sc}^2}\frac{\sqrt{g_{\star}}}{g_s} 
\left(\langle \Gamma_{\phi^0 \rightarrow \psi_1^-} \rangle\,
Y_{\phi^0} - \langle \Gamma_{\psi_1^- \rightarrow \dm} \rangle
\, Y_{\psi_1^-} \right)\,,
\label{Eq:BEpsi-}
\end{eqnarray}
where, $M_{sc}$ is some arbitrary mass scale and 
here we have considered $M_{sc}=M_{\dm}$. The first
Boltzmann equation describes the evolution of comoving
number density of $\dm$ where the first term in the R.H.S.
represents the thermal WIMP contribution to $Y_{\dm}$ coming
from the annihilations and co-annihilations among the $\dm$ and
$\psi_1^{\pm}$. As mentioned earlier, since the mass splitting between $\psi_1^{\pm}$ and
$\dm$ is $\sim \mathcal{O}(100\,{\rm MeV})$, the effect of
co-annihilations \cite{Griest:1990kh, Edsjo:1997bg} is significant to $Y_{\dm}$.
In Eq.\,\,\eqref{Eq:BEpsi0}, ${\langle \sigma {\rm v}\rangle}_{\rm Triplet}$
is the thermally averaged cross section for the annihilation and co-annihilation channels
which are contributing significantly to the freeze-out process of $\dm$.
The Feynman diagrams of these processes are shown in Fig.\,\,\ref{Fig:feyn_dia}. It is worth mentioning that we are not taking any non-perturbative effects on DM annihilations into account here which is justified by the fact that such effects are small for DM masses in the sub TeV regime \cite{Hisano:2004ds, Hisano:2006nn}. This is in fact, another motivation for confining our discussion to the sub-TeV mass regime of dark matter. Here, we have considered the mass difference between $\dm$ and
other heavier triplet fermions ($\psi^0_{\beta}$, $\psi^{\pm}_{\beta}$, $\beta=2,\,3$)
to be large so that the co-annihilations of these heavier triplet
fermions do not affect the freeze-out process of $\dm$.

The second term in the R.H.S. of Eq.\,\,\eqref{Eq:BEpsi0},
which appears with an opposite sign to the first
one, is the non-thermal contributions to $Y_{\dm}$ coming from
the decays of $\phi^0$ and $\psi_1^{\pm}$ and the corresponding
thermal averaged decay widths are indicated by
$\langle\Gamma_{\phi^0 \rightarrow \dm}\rangle$
and $\langle\Gamma_{\psi_1^{\pm} \rightarrow \dm}\rangle$
respectively.\,\,The general expression of thermal averaged
decay width for a decay process $A\rightarrow B\, C$ is
$\langle \Gamma_{A\rightarrow B\, C} \rangle =
\Gamma_{A\rightarrow B\, C}\dfrac{K_1 (\frac{M_A}{T})}{K_2(\frac{M_A}{T})}$, where
$M_A$ is the mass of the mother particle $A$ and $K_n$ is the $n$th order Modified
Bessel function of second kind.\,\,These non-thermal contributions proportional
to the thermal averaged decay widths $\langle\Gamma_{\phi^0 \rightarrow \dm}\rangle$
and $\langle\Gamma_{\psi_1^{\pm} \rightarrow \dm}\rangle$ respectively,
are effective only after the freeze-out of $\dm$ before which $\dm$ has equilibrium number
density governed by the Maxwell-Boltzmann distribution
function.\,\,The decay of $\phi^0\rightarrow \psi_1^0\,\nu$
is due to the Yukawa interaction (first term within the square bracket of
Eq.\,\,\eqref{lag_bsm_psi0_psi-}), which is also responsible for the
neutrino mass generation radiatively, while the decay of $\psi_1^{\pm}\rightarrow
\psi^0_1\,\pi^\pm$ is possible due to an $\mathcal{O}(100\,{\rm MeV})$ mass splitting
between $\psi^{\pm}_1$ and $\psi^{0}_1$. Although, $\psi_1^\pm$ has some three-body
decay modes like $\psi_1^{+(-)}\rightarrow\dm+\bar{l}({l}) + {\nu}(\bar{\nu})$ and
$\psi_1^{+(-)}\rightarrow\dm + u(\bar{u}) + \bar{d}(d)$, $\psi_1^{\pm}\rightarrow
\psi^0_1\,\pi^\pm$ is the dominant decay mode of charged fermion $\psi_1^{\pm}$
with nearly 97\% branching ratio \cite{Cirelli:2005uq}. The expression of decay width
for $\psi_1^{\pm}\rightarrow \psi^0_1\,\pi^\pm$ is given by
\begin{eqnarray}
\Gamma_{\psi_1^\pm \rightarrow \dm}&=&\dfrac{g^4\,f^2_{\pi}\,\,V^2_{ud}}
{128\pi\,M^4_W\,M_{\psi^\pm_1}} \Delta{M^2_{\text{triplet}}}
\left((M_{\dm}+M_{\psi_1^\pm})^2
-M^2_{\pi}\right)\times \nonumber \\ &&
\sqrt{1-\dfrac{(M_{\dm}-M_{\pi})^2}{M^2_{\psi_1^\pm}}}
\sqrt{1-\dfrac{(M_{\dm}+M_{\pi})^2}{M^2_{\psi_1^\pm}}}\,\,,
\label{psidecay}
\end{eqnarray}
where $\Delta{M_{\text{triplet}}}=M_{\psi_1^\pm}-M_{\dm}\simeq$ 166 MeV,
$g$ is the SU(2)$_{\rm L}$ gauge coupling, the pion decay constant
$f_{\pi}=131$ MeV \cite{Cirelli:2005uq} and 
the first diagonal element of the CKM matrix $V_{ud}\simeq0.974$
respectively.\,\,Since we have assumed that there is no asymmetry
between the comoving number densities of $\psi_1^{+}$ and $\psi_1^{-}$
and decay widths $\Gamma_{\psi_1^{-} \rightarrow \dm} =
\Gamma_{\psi_1^{+} \rightarrow \dm} = \Gamma_{\psi_1^\pm \rightarrow \dm}$,
the non-thermal contribution arising from the decays of
$\psi_1^{+}$ and $\psi_1^{-}$ are equal. Therefore, without explicitly
showing the contribution of $\psi_1^{+}$ in $Y_{\psi_1^0}$, we have
multiplied the term for $\psi_1^{-}$ in Eq.\,\,(\ref{Eq:BEpsi0}) by a
factor of 2.   

The second Boltzmann equation
in Eq.\,\,\eqref{Eq:BEphi0}, is the evolution equation for $\phi_0$.
In the R.H.S. of this equation, the first term is the collision
term due to the annihilation and co-annihilation processes among
the components of inert doublet $\Phi$. 
These processes were
in both thermal as well as chemical equilibrium in the early
Universe and the lightest neutral component $\phi^0$ of $\Phi$
freezes-out when these interaction rates become less than
the expansion rate of the Universe, which is governed by the Hubble
parameter $\mathbf{H}(T)$. The effective
annihilation cross section, which we have denoted by
$\langle {\sigma {\rm v}}\rangle_{\rm IDM}$ for $\phi^0$
and $\langle {\sigma {\rm v}}\rangle_{\rm Triplet}$ for $\psi_1^0$ respectively,
can be expressed as a weighted sum of individual
annihilation and co-annihilation cross sections \cite{Griest:1990kh}.
We have calculated $\langle {\sigma {\rm v}}\rangle_{\rm IDM}$
using \texttt{micrOMEGAs} package \cite{Belanger:2013oya}.
In the present scenario, although
$\phi^0$ has a $\mathbb{Z}_2$ charge, it is not the lightest $\mathbb{Z}_2$-odd
particle in the model. Hence, it can decay to other lighter
$\mathbb{Z}_2$-odd particles such as $\psi^{\pm}_1$, $\dm$ \footnote{Here
we have assumed other triplet fermions are heavier than $\phi^0$.}.
These decay modes further decrease the number density of
$\phi^0$ after its thermal freeze-out
and this effect has been included in the Boltzmann equation of $\phi^0$
by the second term in the R.H.S. of Eq.\,\,\eqref{Eq:BEphi0}, which
is, as expected, proportional to the total decay width $\Gamma_{\phi^0}^{\rm Total}$
of $\phi^0$. Finally, Eq.\,\,\eqref{Eq:BEpsi-} is the
Boltzmann Equation for $\psi_1^{-}$. As we know that after the freeze-out
of $\dm$, the abundance of $\psi_1^{\pm}$ is zero. However, $\psi_1^{\pm}$
can again be produced from the late decay of $\phi^0$ and those $\psi_1^{\pm}$
eventually decay to $\dm$ and contribute to the non-thermal
abundance of $\dm$. In the R.H.S. of Eq.\,\,\eqref{Eq:BEpsi-}, the first term is the
production term of $\psi_1^{-}$ from the decay of $\phi^0$ while
the second one with an opposite sign is the depletion term of
$\psi_1^{-}$. Like $\psi_1^{-}$, one can also write a same Boltzmann
equation for $\psi_1^{+}$ as well, however since both $\psi_1^{+}$
and $\psi_1^{-}$ have identical interactions with other
particles and also we have assumed that there is no asymmetry
in the initial number densities of $\psi_1^{+}$ and $\psi_1^{-}$,
hence we do not need to solve an extra Boltzmann equation
similar to Eq.\,\,(\ref{Eq:BEpsi-}) for $\psi_1^{+}$.
%This is taken into account by multiplying Eq.\,\,\eqref{Eq:BEpsi-} by an overall 2 factor. 
Ultimately, we have solved three coupled Boltzmann equations
given in Eqs.\,\,(\ref{Eq:BEpsi0}-\ref{Eq:BEpsi-}) numerically
to find the comoving number density $Y_{\dm}(T_0)$ of $\dm$
at the present epoch, which contains both thermal as well
as non-thermal contributions.
Finally, the relic density of our dark matter candidate $\dm$
can be computed from the value of $Y_{\dm}(T_0)$ using the
following relation \cite{Edsjo:1997bg}
\begin{eqnarray}
\Omega_{\dm} h^2 = 2.755 \times 10^8 \left(\dfrac{M_{\dm}}{\rm GeV}\right)
\,Y_{\dm}(T_0)\,\,,
\end{eqnarray}
where $T_0\simeq2.73$ K, the present temperature of the Universe.
\begin{figure}[h!]
\centering
\includegraphics[height=5cm,width=7cm]{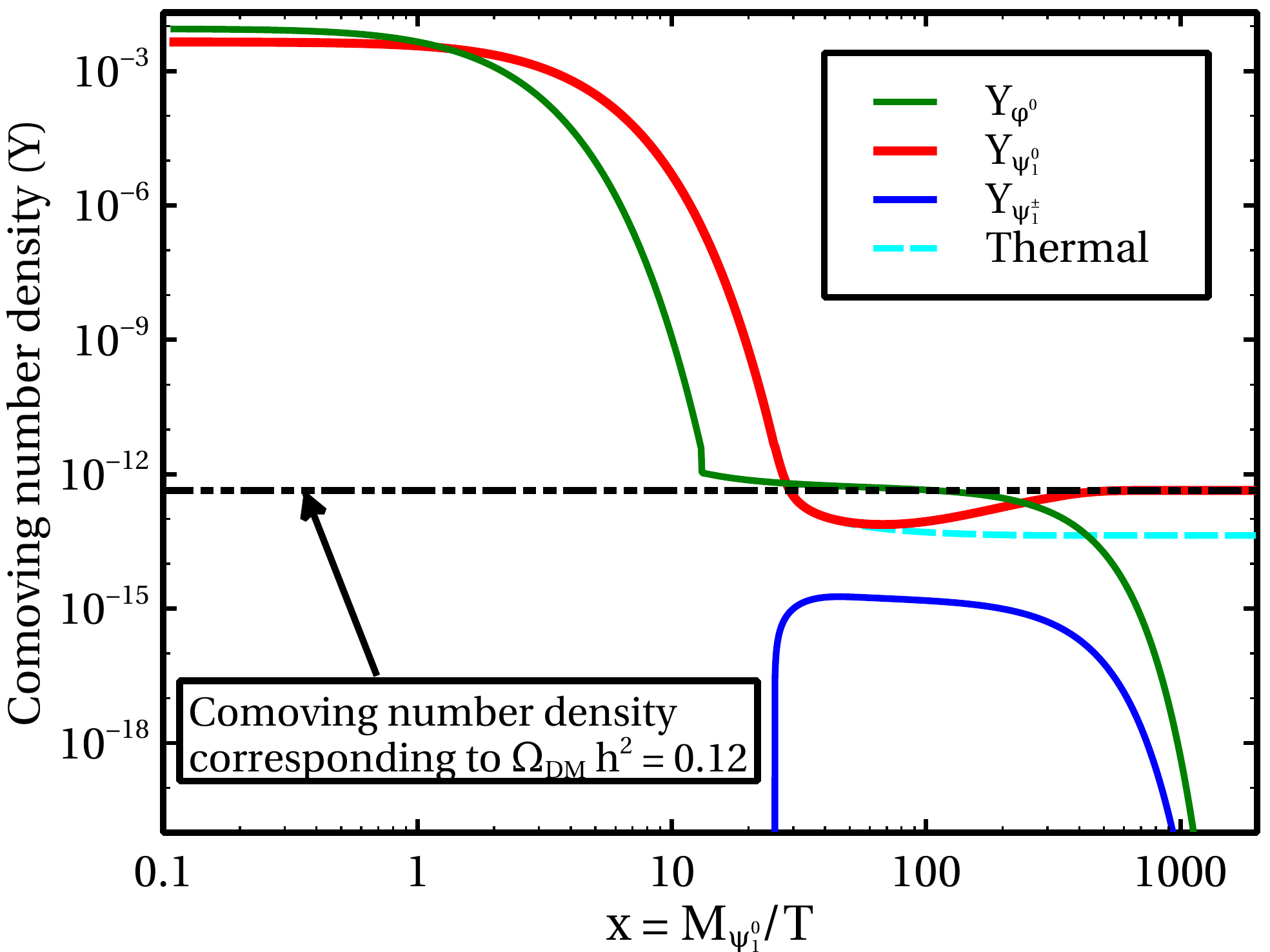}
\,\,
\includegraphics[height=5cm,width=7cm]{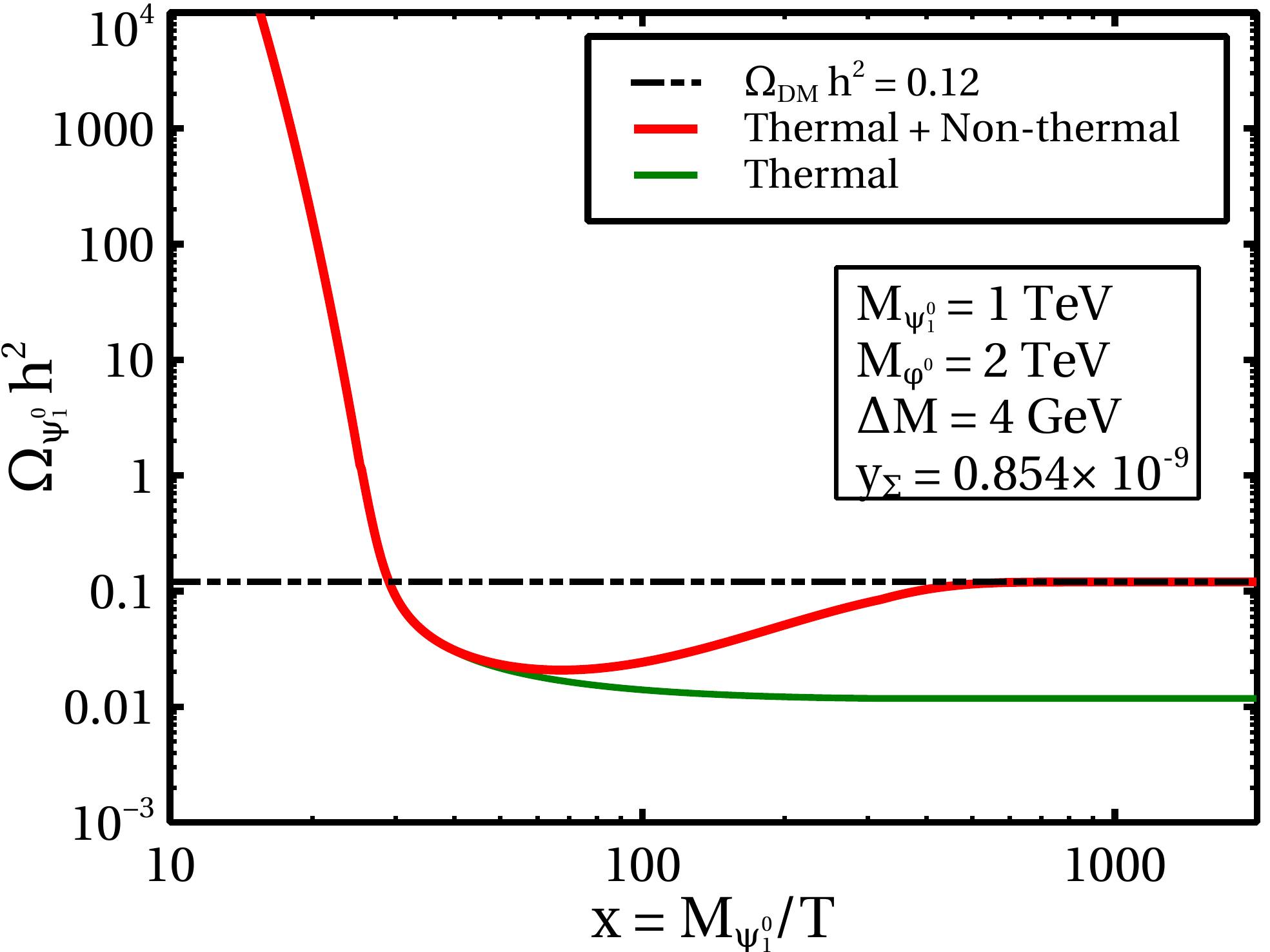}
\caption{Left panel: Solution of coupled Boltzmann equations given in
Eqs.\,\,\eqref{Eq:BEpsi0}-\eqref{Eq:BEpsi-} for a particular benchmark point
$M_{\dm}=1$ TeV, $M_{\phi^0}=2$ TeV, $M_s=200$ GeV, $\Delta{M}=4$ GeV,
$y_{\Sigma}=0.854\times10^{-9}$ and $y_s=1.5$. Right panel: Relic
density of $\dm$ for thermal interactions ($y_\Sigma=0$)
and thermal plus non-thermal interactions ($y_\Sigma=0.854\times10^{-9}$).}
\label{Fig:y-vs-x}
\end{figure}

In the left panel of Fig.\,\,\ref{Fig:y-vs-x}, we show the variation
of comoving number densities of $\phi^0$, $\dm$ and $\psi^{\pm}_1$ with
$x=\dfrac{M_{\dm}}{T}$. Here, the variation of $Y_{\phi^0}$ is
denoted by green solid line and as mentioned above due to the
finite decay width $\Gamma^{\rm Total}_{\phi^0}$, the comoving number
density of $\phi^0$, instead of becoming a constant with temperature $T$,
decreases sharply after its freeze-out. Due to this decrement of $Y_{\phi^0}$
there is an increment in $Y_{\dm}$ denoted by red solid line as more and more
$\dm$ are being produced from the decay of $\phi^0$ after the thermal freeze-out
of $\dm$. In this figure, the evolution of ${\psi_1^{\pm}}$
is denoted by blue solid line, which gets produced from the decay of $\phi^0$
at around $x\sim 25$ and eventually $Y_{\psi_1^{\pm}}$ becomes negligibly small after
having a minute contribution to $Y_{\dm}$. This plot has been drawn for $M_{\dm} = 1$ TeV,
$M_{\phi^0}=2$ TeV, $M_s=200$ GeV, $\Delta{M}=M_{A^0}-M_{\phi^0}
=M_{\phi^{\pm}}-M_{\phi^0}=4$ GeV
\footnote{In this work, we have assumed $M_{\phi^\pm}=M_{A^0}$,
which is possible if $\lambda_4=\lambda_5$.}, $y_s = 1.5$ and $y_{\Sigma}=0.854\times10^{-9}$
\footnote{For simplicity we have assumed Yukawa couplings $y^{11}_{\Sigma}
=y^{21}_{\Sigma}=y^{31}_{\Sigma}=y_{\Sigma}$.}.
In absence of any non-thermal contribution from the decay of $\phi^0$, the thermal
abundance of $\dm$ is denoted by a cyan dashed line, which clearly shows
the under abundance of $\dm$ for $M_{\dm}=1$ TeV due to its large annihilations
and co-annihilations. In the right panel of Fig.\,\,\ref{Fig:y-vs-x}, we plot
the variation of relic abundance of $\dm$ with $x$ for the same benchmark
point mentioned above. Here, the green solid line represents the
relic abundance of $\dm$ due to thermal freeze-out only  
for $M_{\dm} = 1$ TeV, i.e. the relic abundance of $\dm$
by considering its all possible annihilation and co-annihilation
channels as shown in Fig.\,\,\ref{Fig:feyn_dia}. From
this figure it is clearly seen that for this benchmark point
the thermal contribution, which is contributing only around
$\sim 10\%$ of the {\it canonical value} $\Omega_{\rm DM}h^2=0.12$,
is not enough to reproduce the correct dark matter relic abundance.
Hence there is need of an additional non-thermal contribution from the
decay of $\phi^0$ to compensate this deficit. Total
abundance of $\dm$ including contributions from both
thermal and non-thermal processes is shown by red
solid line.

\begin{figure}[h!]
\centering
\includegraphics[height=9cm,width=12cm]{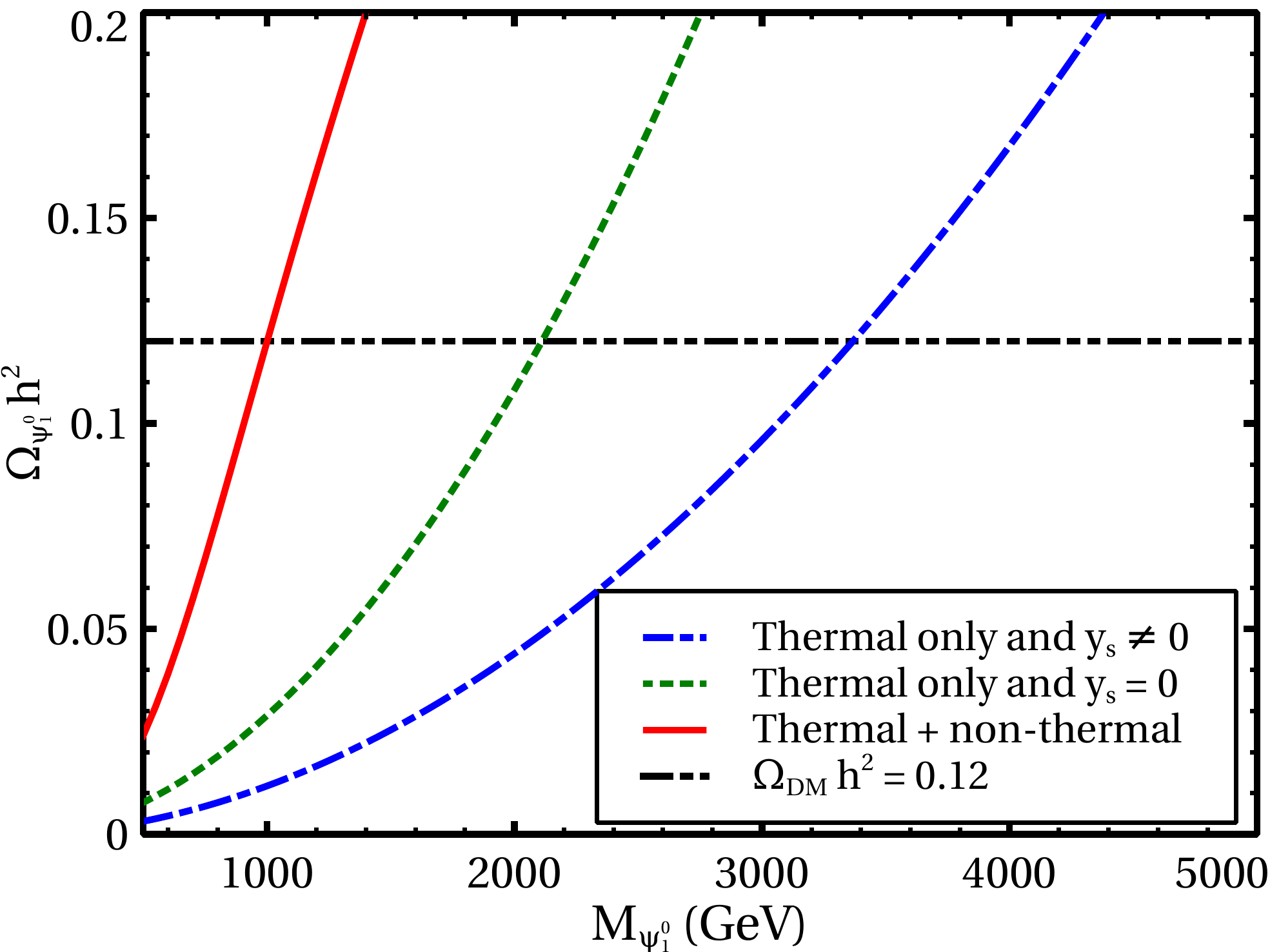}
\caption{Variation of $\Omega_{\dm}h^2$ with $M_{\dm}$.}
\label{Fig:omega-mdm}
\end{figure}

In Fig.\,\,\ref{Fig:omega-mdm} we show how $\Omega_{\dm}h^2$
varies with the mass of $\dm$. In this plot, green dotted
line is for the pure fermion triplet dark matter model
\cite{Ma:2008cu} (discussed in Section \ref{sec:ftdm}) and it is clearly seen that
in this model relic density of $\dm$ satisfies the Planck limit
for $M_{\dm} \sim 2.2$ TeV which is consistent with
the Ref. \cite{Ma:2008cu}. The blue dashed dotted line is for
the present model without any non-thermal contribution
to relic density i.e. Yukawa coupling $y_{\Sigma}=0$.
The difference between this case with the pure
fermion triplet dark matter is that here we have
extra annihilation and co-annihilation channels
involving one or two $S$ in the final states. In this case,
we have chosen $M_{s}=200$ GeV and $y_s=1.5$. Hence,
the relic density of $\dm$ for a particular mass
$M_{\dm}$ is further suppressed. This is
clearly evident from Fig.\,\,\ref{Fig:omega-mdm}.
Finally, the red solid line represents variation
of dark matter relic density, which has both
thermal as well as non-thermal contributions,
for $\Delta{M}=4$ GeV, $M_{\phi^0}=2$ TeV, $M_s=200$ GeV,
$y_{\Sigma}=0.854\times10^{-9}$ and $y_s=1.5$.
Moreover, it also shows that for the chosen
benchmark point the relic density of triplet
fermion $\dm$ satisfies the Planck limit 
for a mass as low as $M_{\dm}=1$ TeV. Please note that the plot shown in Fig. \ref{Fig:omega-mdm} is for illustrative purposes only in order to show the difference between possible scenarios discussed here. We have not taken non-perturbative effects into account in dark matter annihilations which can be significant if we go to high mass regime beyond 1 TeV. If we take them into account, as discussed by the authors of \cite{Hisano:2004ds, Hisano:2006nn}, the point in the green and the blue lines where correct relic is satisfied will shift further towards right.
\begin{figure}[h!]
\centering
\subfigure[Variation of $Y_{\dm}$ for different values of $\Delta{M}$]
{\includegraphics[height=5cm,width=7cm]{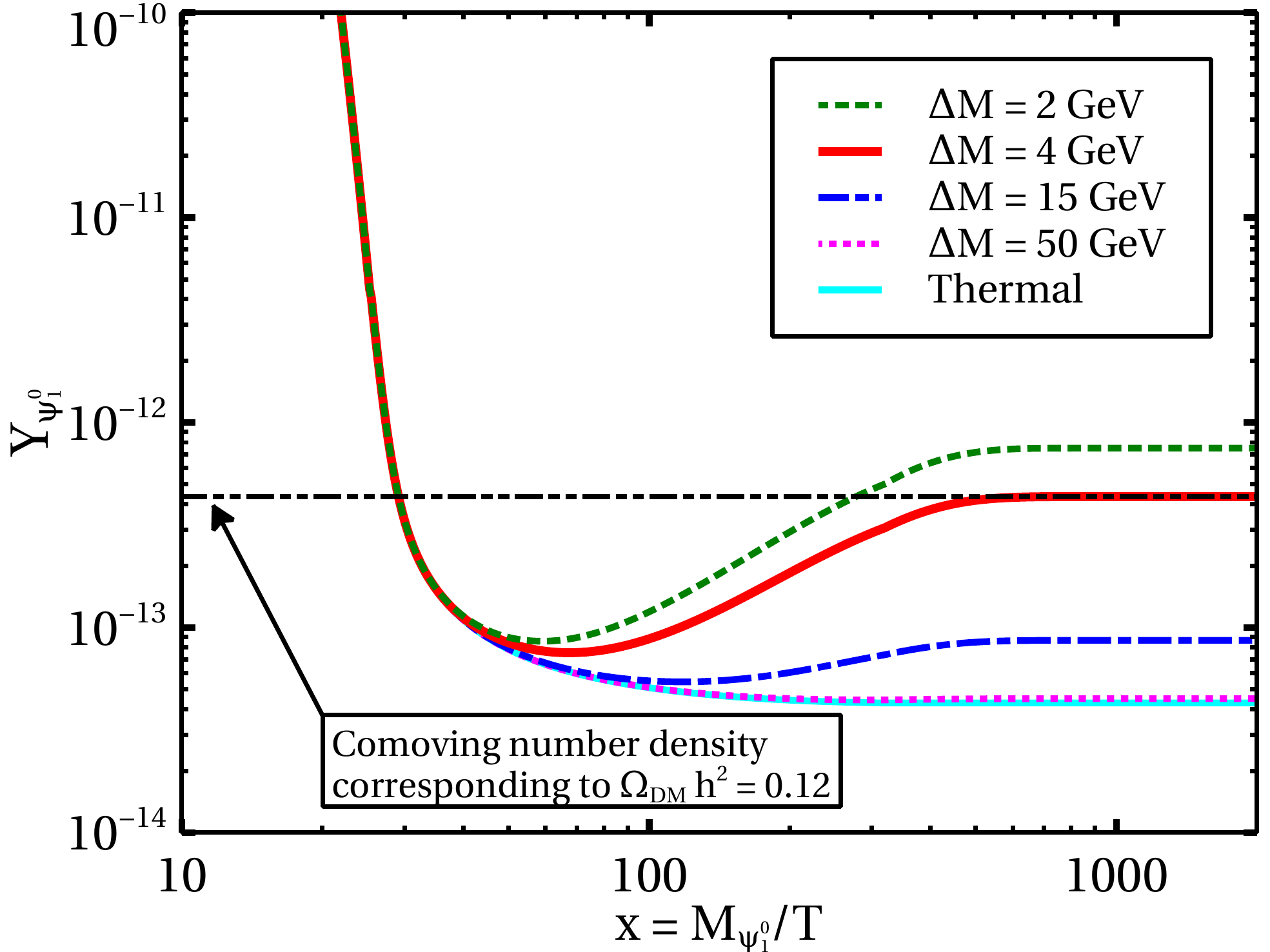}}
\,\,
\subfigure[Variation of $Y_{\dm}$ for different values of $M_{\phi^0}$]
{\includegraphics[height=5cm,width=7cm]{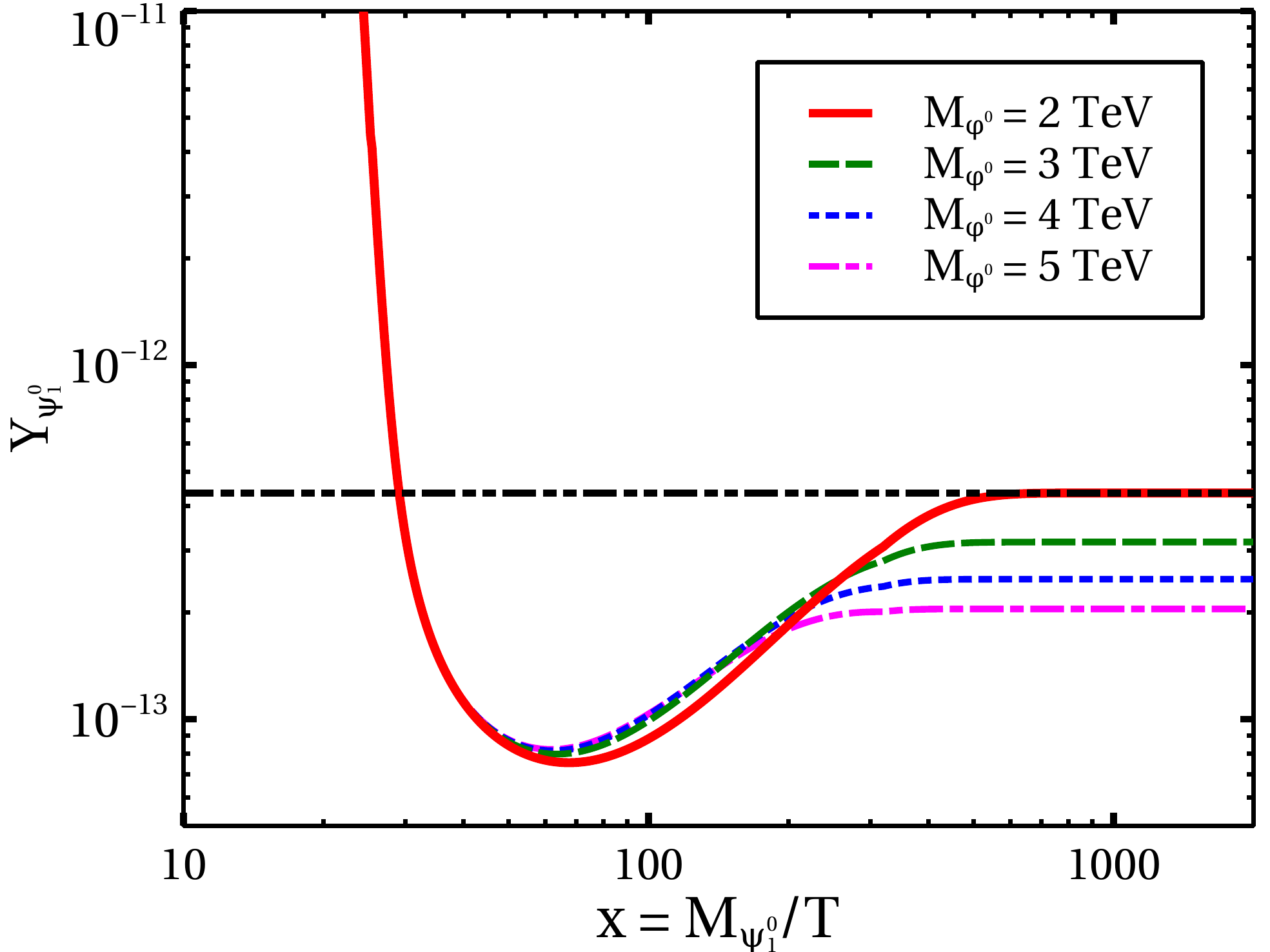}} \\
%\caption{}
%\end{figure}
%\begin{figure}[h!]
%\centering
\vskip 0.2in
\subfigure[Variation of $Y_{\dm}$ for different values of $M_{\dm}$]
{\includegraphics[height=5cm,width=7cm]{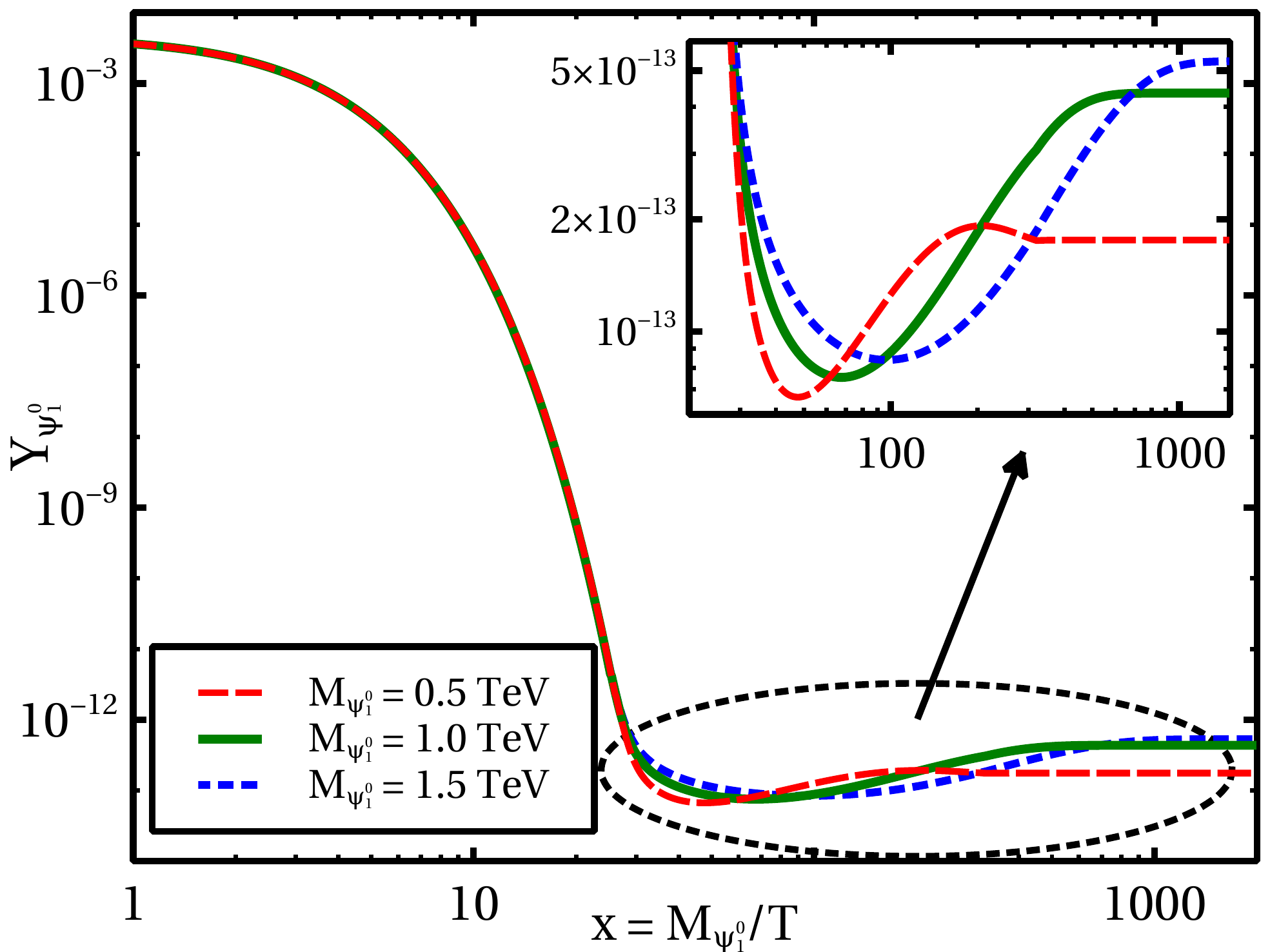}}
\,\,
\subfigure[Variation of $Y_{\dm}$ for different values of $y_{\Sigma}$]
{\includegraphics[height=5cm,width=7cm]{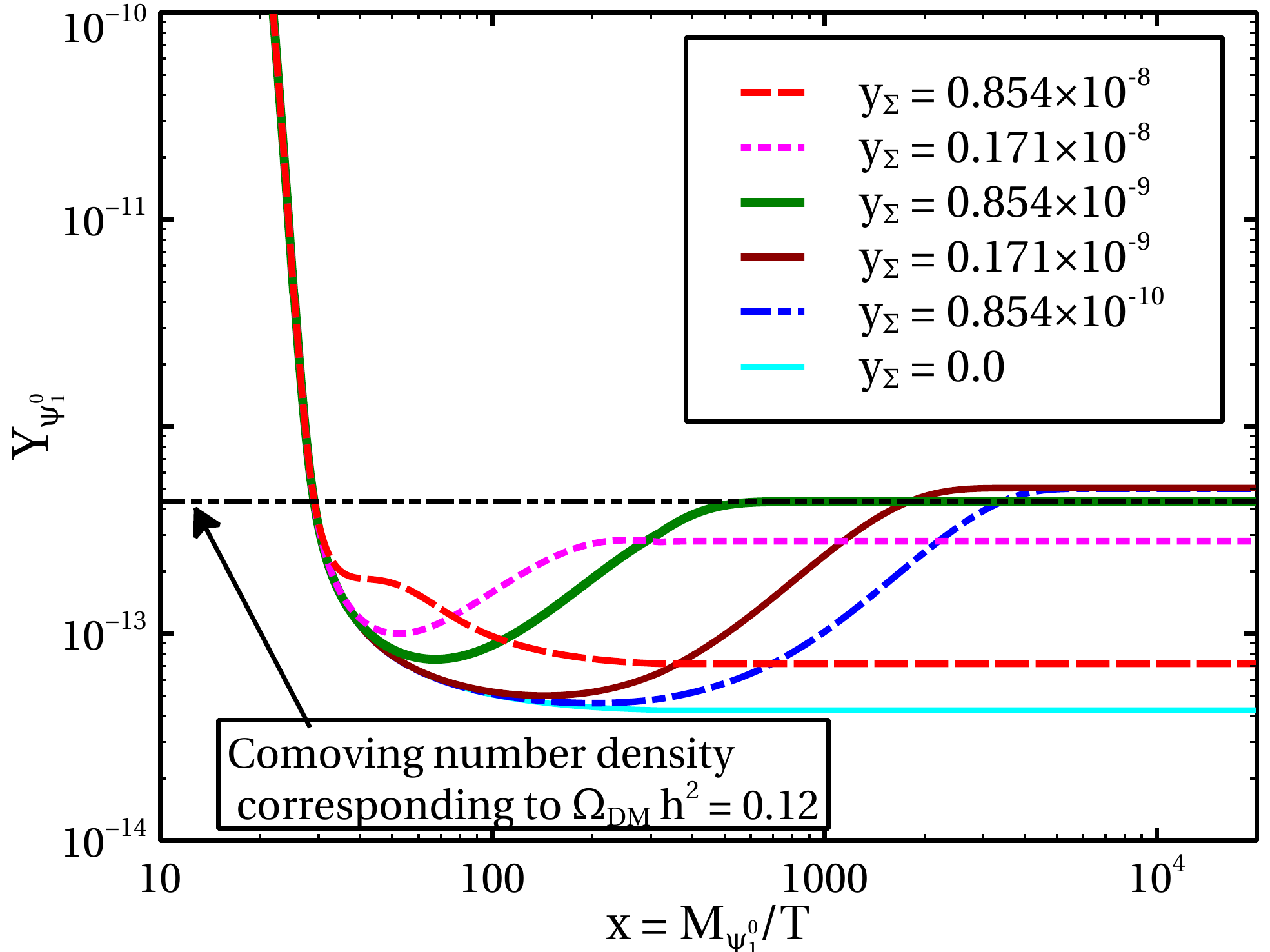}}
\caption{Comparison of $Y_{\dm}$ with respect to different model parameters.}
\label{Fig:compare-Y}
\end{figure}

Now, we will present the effect of four 
model parameters on $Y_{\dm}$.\,\,These four
model parameters have significant impacts
on the non-thermal contributions to $Y_{\dm}$.
While generating the four plots, we have
kept fixed the other parameters (expect
the particular one which has been varied) to the
same benchmark point we have used to generate Fig.\,\,\ref{Fig:y-vs-x}.
In Fig.\,\,\ref{Fig:compare-Y}(a), we show
the variation of $Y_{\dm}$ with $x$ for four different values
of $\Delta{M}=M_{\phi^{\pm}}-M_{\phi^0}=M_{A^{0}}-M_{\phi^0}$.
From this plot it is seen that the value of $Y_{\dm}$, after the thermal freeze-out of $\dm$,
increases as we decrease the mass splitting between $\phi^0$ and other components
of inert doublet $\Phi$ i.e. $\phi^{\pm}$ and $A^0$ \footnote{In this work,
for simplicity we choose $M_{\phi^{\pm}}=M_{A^0}$.}. This can be understood
as follows. From the inert doublet dark matter model
\cite{LopezHonorez:2006gr, LopezHonorez:2010tb, Honorez:2010re,
Arhrib:2013ela, Belyaev:2016lok, Borah:2017dqx, Borah:2017dfn}
we know that the abundance $\phi^0$ increases as $\Delta{M}$
decreases and it is due to the cancellation between four point
diagram and $t$, $u$ channel diagrams for annihilation channels
of $\phi^0$ into gauge boson final states $W^{+}W^{-}$, $ZZ$ and
this is also true for the co-annihilation channels of inert
scalars.
%and it is
%due to the effect of co-annihilation among
%$\phi^0$, $\phi^\pm$ and $A^0$, which enhances
%as the mass splitting decreases.
Now, similarly here also
$Y_{\phi^0}$ increases with decreasing $\Delta{M}$. Again, from Eq.\,\,\eqref{Eq:BEpsi0}
one can easily notice that the non-thermal contribution to $Y_{\dm}$
coming from the decay of $\phi^0$ is proportional to $Y_{\phi^0}$.
Hence $Y_{\dm}$ also increases as we decrease the mass splitting $\Delta{M}$.
The dependence of $Y_{\dm}$ on $M_{\phi^0}$ is shown in Fig.\,\,\ref{Fig:compare-Y}(b),
where we have considered four different values of $M_{\phi^0}$.
From this plot, we see that $Y_{\dm}$ keeps on increasing as we lower
the mass of $\phi^0$. This is due to the fact that the non-thermal
contribution to $Y_{\dm}$ is proportional to two quantities. One is
the abundance of $\phi^0$ during the decay to $\dm$, which occurs after
the freeze-out of $\phi^0$ (see left panel of Fig.\,\,\ref{Fig:y-vs-x})
and other one is the corresponding
decay width $\Gamma_{\phi^0\rightarrow\dm}$. The behaviour of
these two quantities are opposite with respect to $M_{\phi^0}$.
While decay width $\Gamma_{\phi^0\rightarrow\dm}$ is proportional to
$M_{\phi^0}$, the comoving number density of $\phi^0$ is less
and becomes more suppressed for heavier $\phi^0$ during
the period when maximum decay to $\dm$ occurs. This is because,
primarily the comoving number density of $\phi^0$  after its freeze-out
is less for higher value of $M_{\phi^0}$ (valid for a fixed value of $\Delta{M}$ and
$M_{\phi^0}\gtrsim 1$ TeV) and additionally the total decay width $\Gamma^{\rm Total}_{\phi^0}$ of $\phi^0$
which gets enhanced with $M_{\phi^0}$, has a negative impact on $Y_{\phi^0}$ (see Eq.\,\,\eqref{Eq:BEphi0}),
which further reduces $Y_{\phi^0}$ with respect to $M_{\phi^0}$
in the considered range.
Therefore, finally we get a combined effect of both the
terms $Y_{\phi^0}$ and $\Gamma_{\phi^0\rightarrow\dm}$
on $Y_{\dm}$, where $Y_{\dm}$ increases with decreasing
$M_{\phi^0}$. In Fig.\,\,\ref{Fig:compare-Y}(c), we demonstrate
how $Y_{\dm}$ depends on the mass of $\dm$. Here, we have
shown the variation of $Y_{\dm}$ with $x$ for three different
values of $M_{\dm}$ such as 500 GeV, 1000 GeV and 1500 GeV.
From this plot one can see that the final saturation
value of $Y_{\dm}$ is more for dark matter candidate
with heavier mass. This is due to the reason that the
thermal abundance of $\dm$, which is approximately inversely
proportional to its annihilation and co-annihilation cross sections,
is larger for heavier $\dm$. Also, we don't get same enhancement in $Y_{\dm}$
while going from 500 GeV to 1000 GeV and 1000 GeV to 1500 GeV. This
is because, the decay width $\Gamma_{\phi^0\rightarrow\dm}$ becomes
phase space suppressed as $M_{\dm} \rightarrow M_{\phi^0}$. As a result,
non-thermal contribution, which is proportional to $\Gamma_{\phi^0\rightarrow\dm}$,
decreases with $M_{\dm}$. Hence, we get the saturation values of $Y_{\dm}$
for 1000 GeV and 1500 GeV, which are not much different from each other.
Finally, we show the dependence of the evolution of $Y_{\dm}$ with $x$ on Yukawa coupling $y_{\Sigma}$, where we have 
considered six different values for the Yukawa coupling
$y_{\Sigma}$ and we find that the correct relic density is
achieved for $y_{\Sigma}=0.854\times10^{-9}$. As 
the total decay width $\Gamma^{\rm Total}_{\phi^0}$ of $\phi^0$
is proportional to $y^2_{\Sigma}$, the smaller $y_{\Sigma}$
results in a late decay of $\phi^0$ to $\dm$. Hence, the
increase of $Y_{\dm}$ due to the non-thermal decay
of $\phi^0\rightarrow \dm + \nu$ also occurs much
later for smaller value of $y_{\Sigma}$. This
effect can be understood by comparing the
magenta dashed curve and blue dashed-dotted curve
in Fig.\,\,\ref{Fig:compare-Y}(d), where
magenta and blue lines are for $y_{\Sigma}=0.171\times10^{-8}$
and $0.854\times10^{-10}$ respectively. Further, we have
seen an anomalous behaviour in the effect of Yukawa coupling $y_{\Sigma}$
on $Y_{\dm}$. Generally for the non-thermal dark matter (FIMP),
the relic abundance increases as we increase the coupling between mother particle
and dark matter. As a result, more dark matter
particles are produced from the decay of mother particle which is
assumed to be in thermal equilibrium and the equilibrium abundance
of mother particle does not depend on its decay width to FIMP
dark matter. However, in the present case the mother particle $\phi^0$ becomes out
of thermal equilibrium at the time of non-thermal production of $\dm$.
Now, from the Boltzmann equation of $\dm$ (Eq.\,\,\eqref{Eq:BEpsi0}), one can
easily see that the rate of increase of $Y_{\dm}$ due to the decay to $\phi^0$
is proportional to $Y_{\phi^0}$ and $\Gamma_{\phi^0\rightarrow\dm}$. The comoving
number density $Y_{\phi^0}$ can be obtained by solving the Boltzmann
equation for $\phi^0$ (Eq.\,\,\eqref{Eq:BEphi0}), where the last term
proportional to the total decay width of $\phi^0$ ($\Gamma^{\rm Total}_{\phi^0}$)
further decreases $Y_{\phi^0}$ from its freeze-out abundance. 
Now, if we increase the Yukawa coupling $y_{\Sigma}$, the total decay width $\Gamma^{\rm Total}_{\phi^0}$
which is proportional to $y^2_{\Sigma}$, also increases. This
in turn decreases $Y_{\phi^0}$. On the other hand, any increase of
$y_{\Sigma}$ is also accompanied by an enhancement of the decay width
$\Gamma_{\phi^0\rightarrow\dm}$. Therefore, in the Boltzmann equation
of $\dm$, there is a competition between the two quantities $Y_{\phi^0}$ and 
$\Gamma_{\phi^0\rightarrow\dm}$, which are behaving oppositely with
respect to the variation of $y_{\Sigma}$. The final abundance
of $\dm$ will follow the behaviour of that quantity which depends
more strongly on $y_{\Sigma}$.
From Fig.\,\,\ref{Fig:compare-Y}(d),
we find that the final abundance of $\dm$ actually decreases as we increase the
Yukawa coupling $y_{\Sigma}$ from $0.171\times 10^{-10}$ to $0.854\times10^{-8}$.
This makes the present scenario different from the usual FIMP scenario
\cite{Hall:2009bx, Biswas:2015sva, Biswas:2016yjr, Biswas:2018aib},
where the final abundance of dark matter always increases with
associated couplings.

\begin{figure}[h!]
\centering
\includegraphics[height=5cm,width=7.0cm]{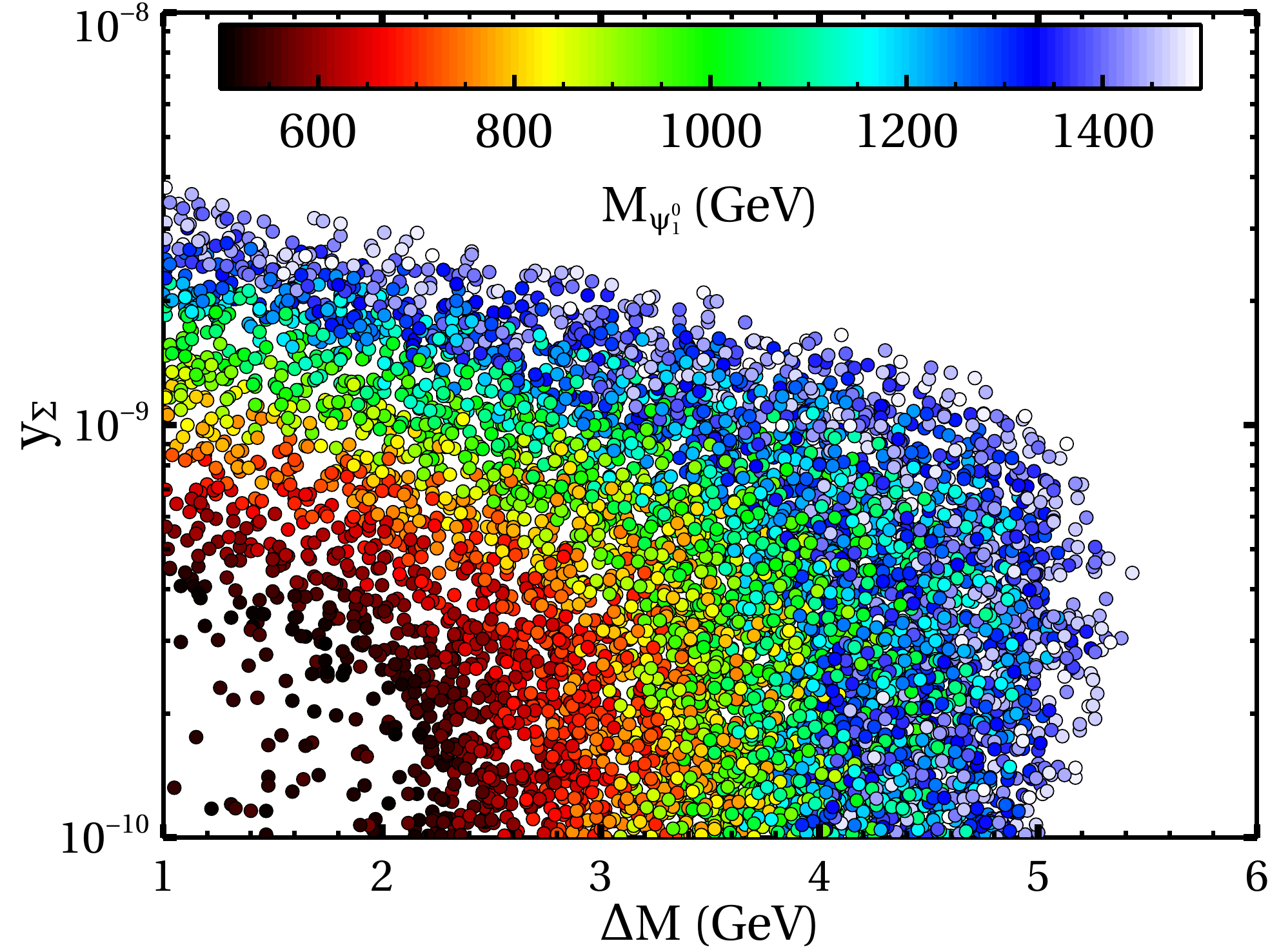}
\,\,
\includegraphics[height=5cm,width=7.0cm]{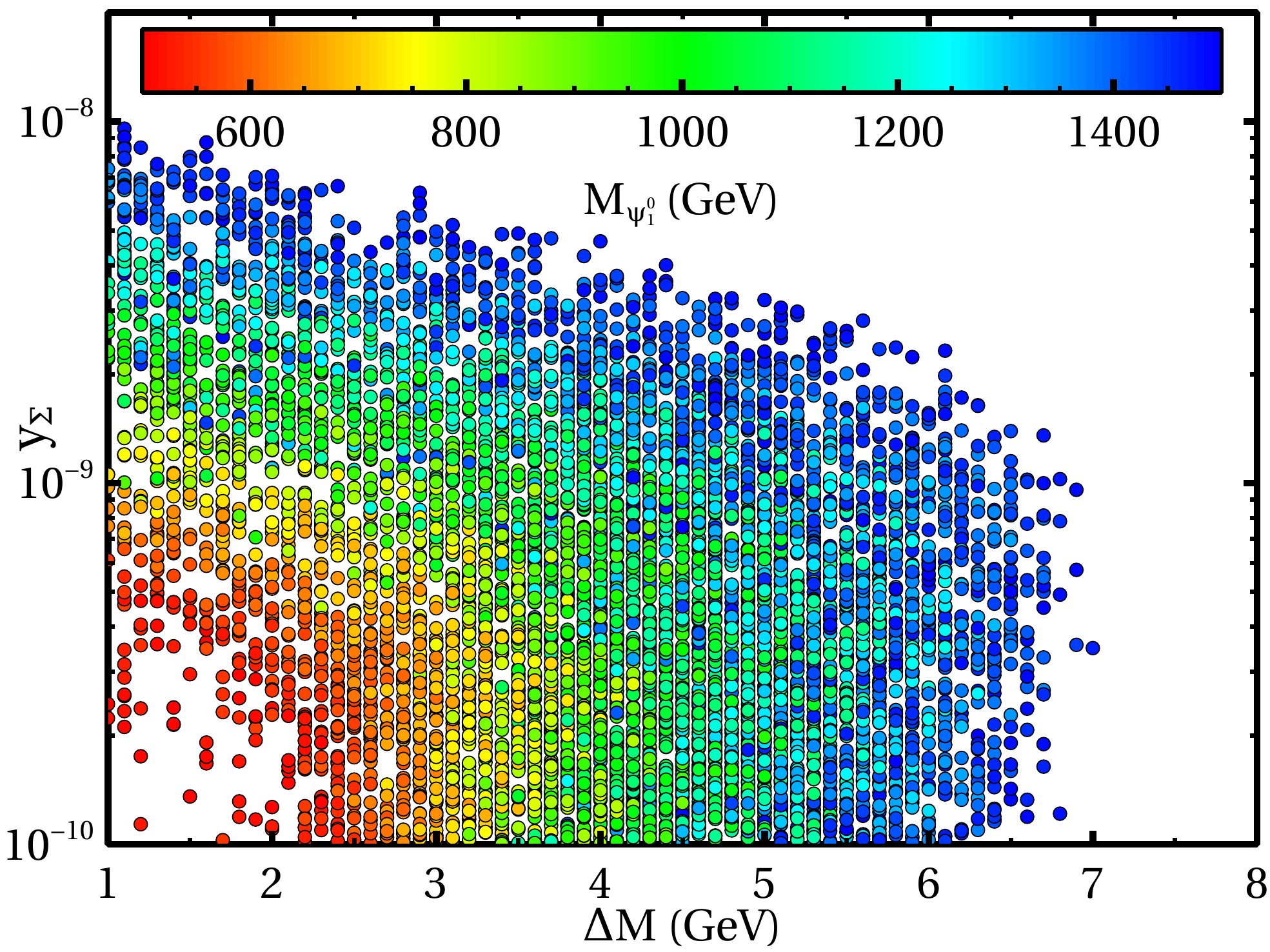}
\caption{Allowed parameter space in $y_{\Sigma}-\Delta{M}$ for $y_s=1.5$ (left panel)
and $0.1\leq y_s\leq 2.0$ (right panel). In both plots mass of $\phi^0$
has been varied between 1 TeV to 5 TeV and $M_s$ is kept fixed at 200 GeV.}
\label{Fig:scanplot}
\end{figure}

In Fig.\,\,\ref{Fig:scanplot}, we show our allowed parameter space
in $y_{\Sigma}-\Delta{M}$ plane which reproduces the correct dark matter
relic density. The colour code is indicating the mass
of our dark matter $\dm$ which we have varied between 500 GeV to 1500 GeV
while the corresponding mass of $\phi^0$ is scanned
over the following range: $2\,M_{\dm}\leq M_{\phi^0}\leq 2\,M_{\dm}+2000$ GeV.
The plot in left panel is for a fixed value of other Yukawa coupling
$y_s$ which we have kept fixed at 1.5 while for the plot in the right
panel, we have varied the Yukawa coupling $y_s$ between 0.1 to 2.
From the plot in the left panel, one can see that to produce
the dark matter relic density in the right ballpark, we need
$10^{-10}\leq y_{\Sigma}\leq 4\times 10^{-9}$ depending upon the 
mass splitting $\Delta{M}$ between the inert scalars, which is
also tightly constrained to be less than 5.5 GeV. However, from
the right panel it is also noticeable that these ranges of
Yukawa coupling $y_{\Sigma}$ and mass splitting
$\Delta{M}$ are slightly larger when we have also varied
the singlet Yukawa coupling $0.1\leq y_s\leq2$. This is because
the thermal contribution to relic density becomes fixed 
for a particular set of values of $M_{\dm}$, $M_{\phi^0}$, $M_s$, $\Delta{M}$
and $y_s$. In this case, the amount of deficit in relic density
which is compensated by the non-thermal production is also a definite
number as the total relic density should lie with the observed
band $0.1166\leq\Omega_{\rm DM} h^2\leq 0.1206$ in 68\% C.L.
However, if we vary the singlet Yukawa $y_s$ while keeping
others fixed at their respective values ($\Delta{M}, M_{\phi}$, $M_{\dm}$ and $M_s$) then
the thermal contribution to $\Omega_{\dm}h^2$ also
varies and hence we require different non-thermal contributions
to achieve correct dark matter relic density and consequently
more parameter space in $y_{\Sigma}-\Delta{M}$ plane become
allowed. Later when we will discuss the constrains coming from
the indirect detection we will see that for the considered mass
range of $M_{\dm}$ we need $y_s\gtrsim 1$ (see Fig.\,\,\ref{fig3}).
Moreover, from both these plots it is clearly evident the for
heavier dark matter mass we need larger values of $y_{\Sigma}$
and $\Delta{M}$ to achieve $\Omega_{\dm}h^2$ in the correct ballpark.
%%%%%%%%%%%%%%%%%%%%%%%%%%%%%%%%%%%%%%%%%%%%%%%%%%%%%%%
\section{LHC constraints on $\psi_1^{\pm}$}
\label{sec:lhc_cons}
\begin{figure}[h!]
\centering
\includegraphics[scale=0.45]{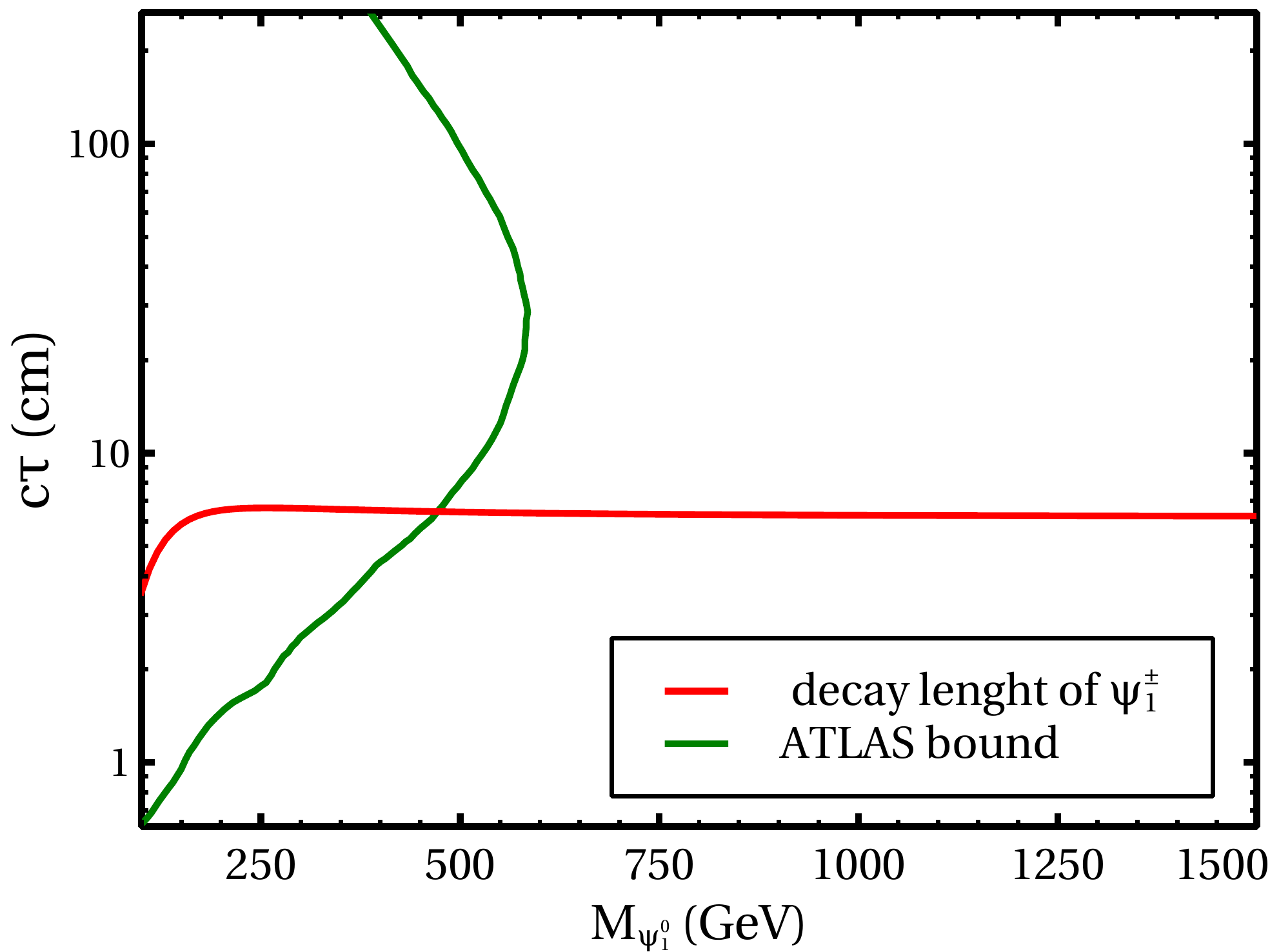}
\caption{Lifetime of $\psi_1^{\pm}$ versus DM mass compared with the ATLAS bound on disappearing charge track searches.}
\label{fig:lhc}
\end{figure}

In this section, we briefly discuss the testability of our model at the LHC experiment. There have been several dedicated searches for possible dark matter signatures at colliders, a recent summary of which can be found in \cite{Kahlhoefer:2017dnp, Penning:2017tmb}. Instead of usual missing transverse energy associated with dark matter production at colliders, in our model there exists a different (rather unique to a limited class of scenarios) signature that is within the reach of LHC. This is basically the collider production of different components of the fermion triplet 
$\Sigma_{1R}$ through gauge interactions and subsequent decay of the heavier components into the lighter one. Since dark matter candidate $\psi^0_1$ is the lightest component, the heavier component $\psi^{\pm}_1$ must decay into $\psi^0_1$ and other SM particles. This decay is actually interesting due to small mass splitting
$\Delta{M}_{\text{triplet}}\simeq 166$ MeV between $M_{\psi^{\pm}_1}$ and
$M_{\psi^0_1}$ for $M_{\Sigma}\gtrsim1\,{\rm TeV}$. For such a small mass difference, the dominant decay mode is $\psi_1^{\pm}\rightarrow \psi^0_1\,\pi^\pm$, the corresponding decay width of which is given by Eq. \eqref{psidecay}. Such tiny decay width keeps the lifetime of $\psi^{\pm}_1$ considerably long enough that it can reach the detector before decaying. In fact, the ATLAS experiment at the LHC has already searched for such long-lived charged particles with lifetime ranging from 10 ps to 10 ns, with maximum sensitivity around 1 ns \cite{Aaboud:2017mpt}. In the decay $\psi_1^{\pm}\rightarrow \psi^0_1\,\pi^\pm$, the final state pion typically has very low momentum and it is not reconstructed in the detector. On the other hand the dark matter particle in the final state $\psi^0_1$ is perfectly stable and leaves the detector without interacting. Therefore, it gives rise to a signature where a charged particle leaves a track in the inner parts of the detector and then disappears leaving no tracks in the portions of the detector at higher radii. The ATLAS constraints on such disappearing charged track signatures for a long lived chargino decaying into a pion and wino dark matter is shown as the solid green line in Fig. \ref{fig:lhc}. Since our dark matter multiplet is similar to the multiplet containing chargino, wino with similar production cross section at LHC as shown in Table \ref{Tab:LHC} for fermion triplets and in \cite{Fuks:2012qx,Fuks:2013vua} for gauginos, we compare our model predictions against this constraint from ATLAS. The lifetime predictions for $\psi_1^{\pm}$ as a function of DM mass is shown as the solid red line in Fig. \ref{fig:lhc}. It can be seen that the existing LHC constraint can already rule out DM masses below 500 GeV from its searches for disappearing charged tracks, keeping the DM parameter space considered in this study within near future sensitivity.
\begin{table}[h!]
\centering
\vspace{0.3cm}
\begin{tabular}{||l|l|l||}
\hline
$M_{\dm}$ (TeV)& $\sigma_{p\,p\,\rightarrow\psi_1^+\psi_1^-}$ (pb)
& $\sigma_{p\,p\,\rightarrow\psi_1^{\pm}\dm}$ (pb)\\
\hline 
\hline
0.5 & 1.493 $\times$ 10$^{-2}$ & 4.394 $\times$ 10$^{-2}$\\
\hline
0.6 & 6.338 $\times$ 10$^{-3}$ & 1.92 $\times$ 10$^{-2}$\\
\hline
0.7 & 2.938 $\times$ 10$^{-3}$ & 9.1 $\times$ 10$^{-3}$\\
\hline
0.8 & 1.464 $\times$ 10$^{-3}$ & 4.6 $\times$ 10$^{-3}$ \\
\hline
0.9 & 7.588 $\times$ 10$^{-4}$ & 2.43 $\times$ 10$^{-3}$ \\
\hline
1.0 & 4.103 $\times$ 10$^{-4}$ & 1.33 $\times$ 10$^{-3}$\\
\hline
1.1 & 2.305 $\times$ 10$^{-4}$ & 7.468 $\times$ 10$^{-4}$\\
\hline
1.2 & 1.305 $\times$ 10$^{-4}$ & 4.308 $\times$ 10$^{-4}$\\
\hline
1.3 & 7.600 $\times$ 10$^{-5}$ & 2.508 $\times$ 10$^{-4}$\\
\hline
1.4 & 4.490 $\times$ 10$^{-5}$ & 1.494 $\times$ 10$^{-4}$\\
\hline
1.5 & 2.687 $\times$ 10$^{-5}$ & 8.978 $\times$ 10$^{-5}$\\
\hline
\end{tabular}
\caption{Production cross sections of
$\psi_1^{+}\psi_1^{-}$ and $\psi_1^{\pm}\dm$
from $p\,p$ collisions 
at $\sqrt{s}=14$ TeV LHC.}
\label{Tab:LHC}
\end{table}
%%%%%%%%%%%%%%%%%%%%%%%%%%%%%%%%%%%%%%%%%%%%%%%%%%%%%%%
\section{Direct Detection}
\label{sec:dd}
\begin{figure}[h!]
\centering
\begin{tabular}{cc}
\epsfig{file=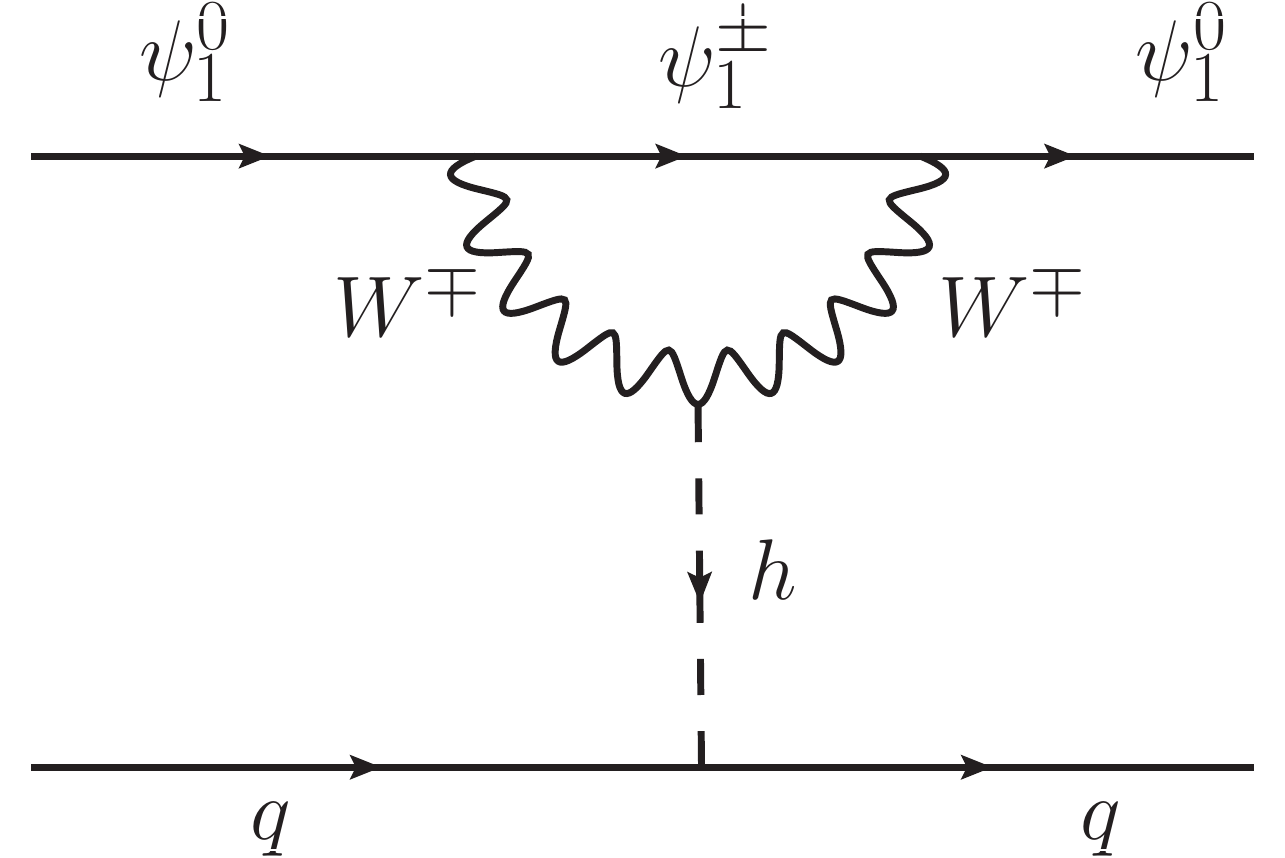,width=0.33\textwidth,clip=}
\epsfig{file=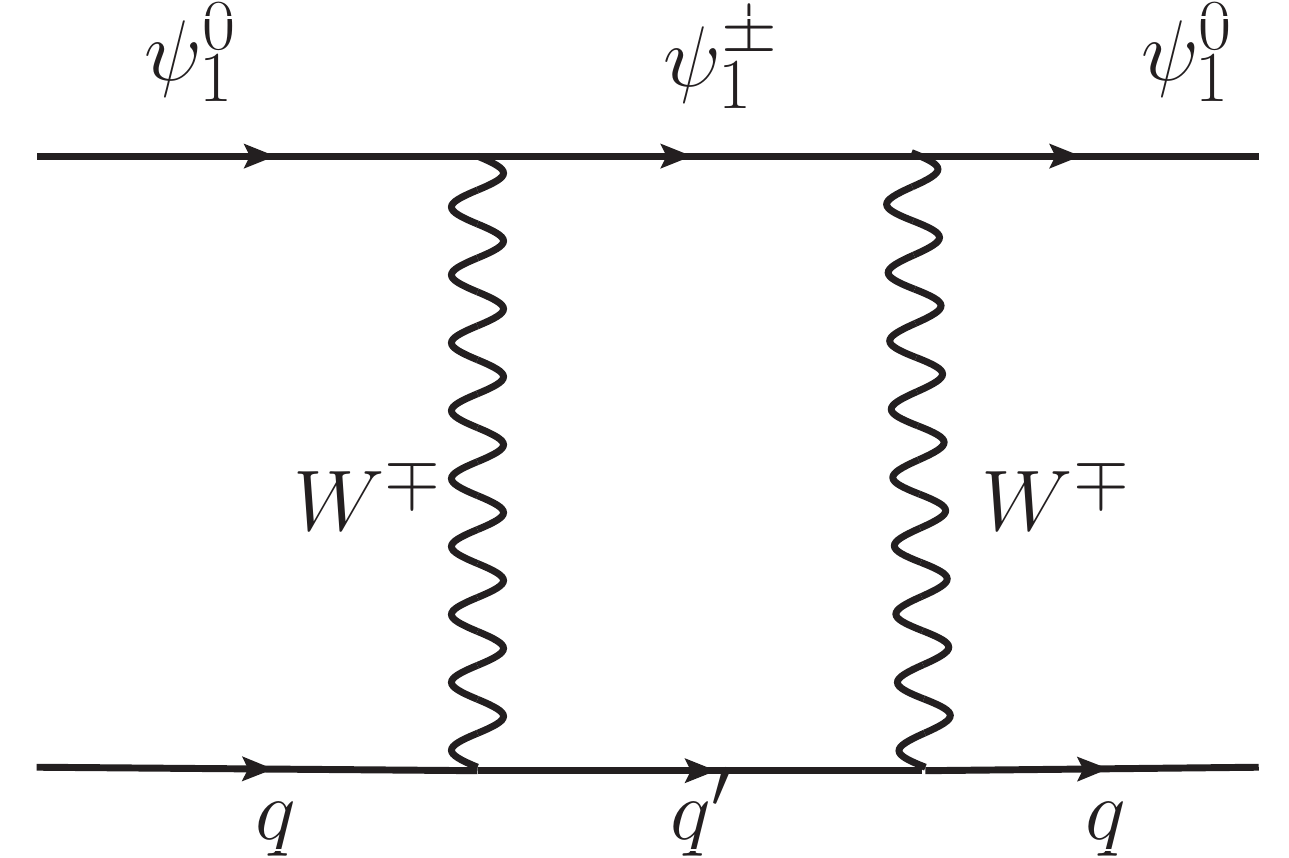,width=0.33\textwidth,clip=}
\epsfig{file=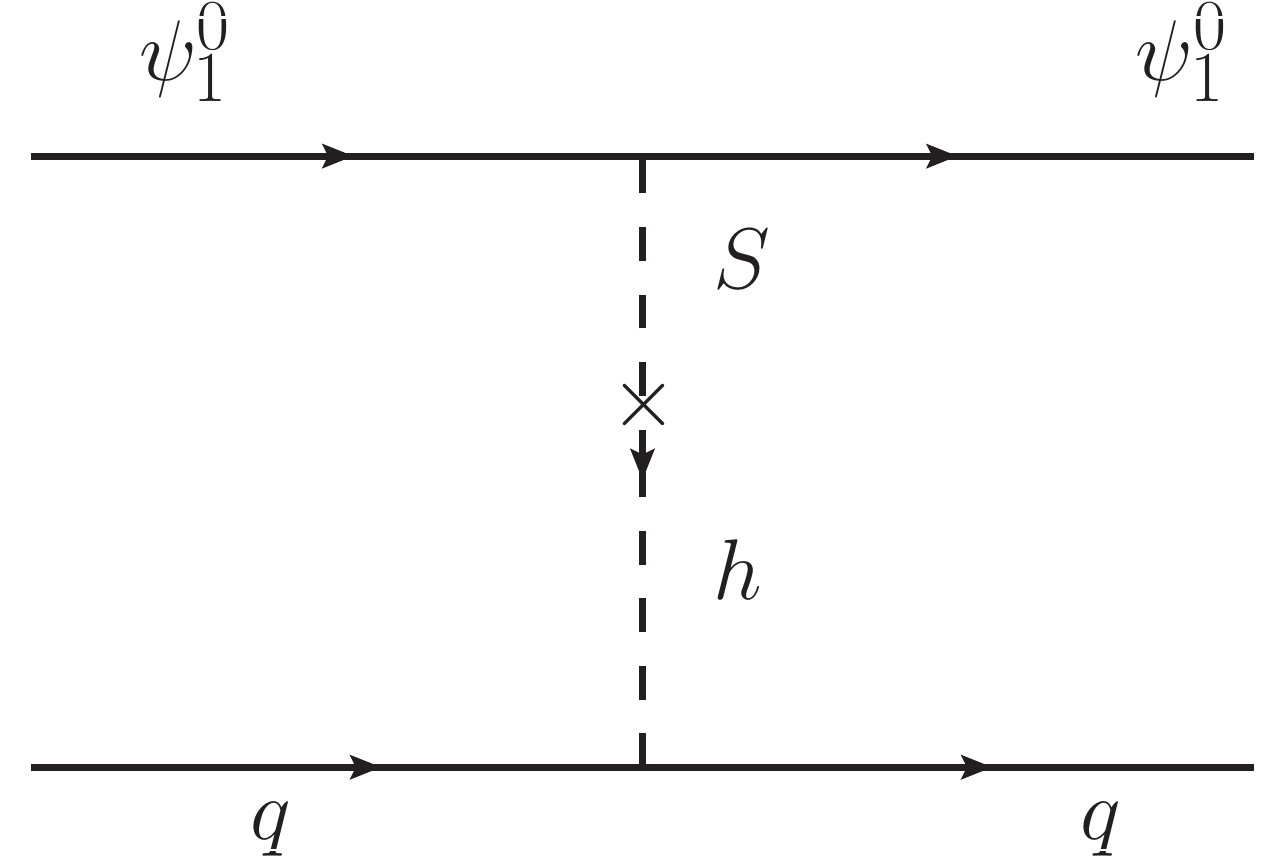,width=0.33\textwidth,clip=}
\end{tabular}
\caption{Spin-independent DD scattering processes in the model. The first two processes are for pure triplet fermion dark matter model while the last one arises additionally in the present model due to the introduction of the singlet scalar $S^{\prime}$.}
\label{fig1}
\end{figure}

\begin{figure}[h!]
\centering
\begin{tabular}{cc}
\epsfig{file=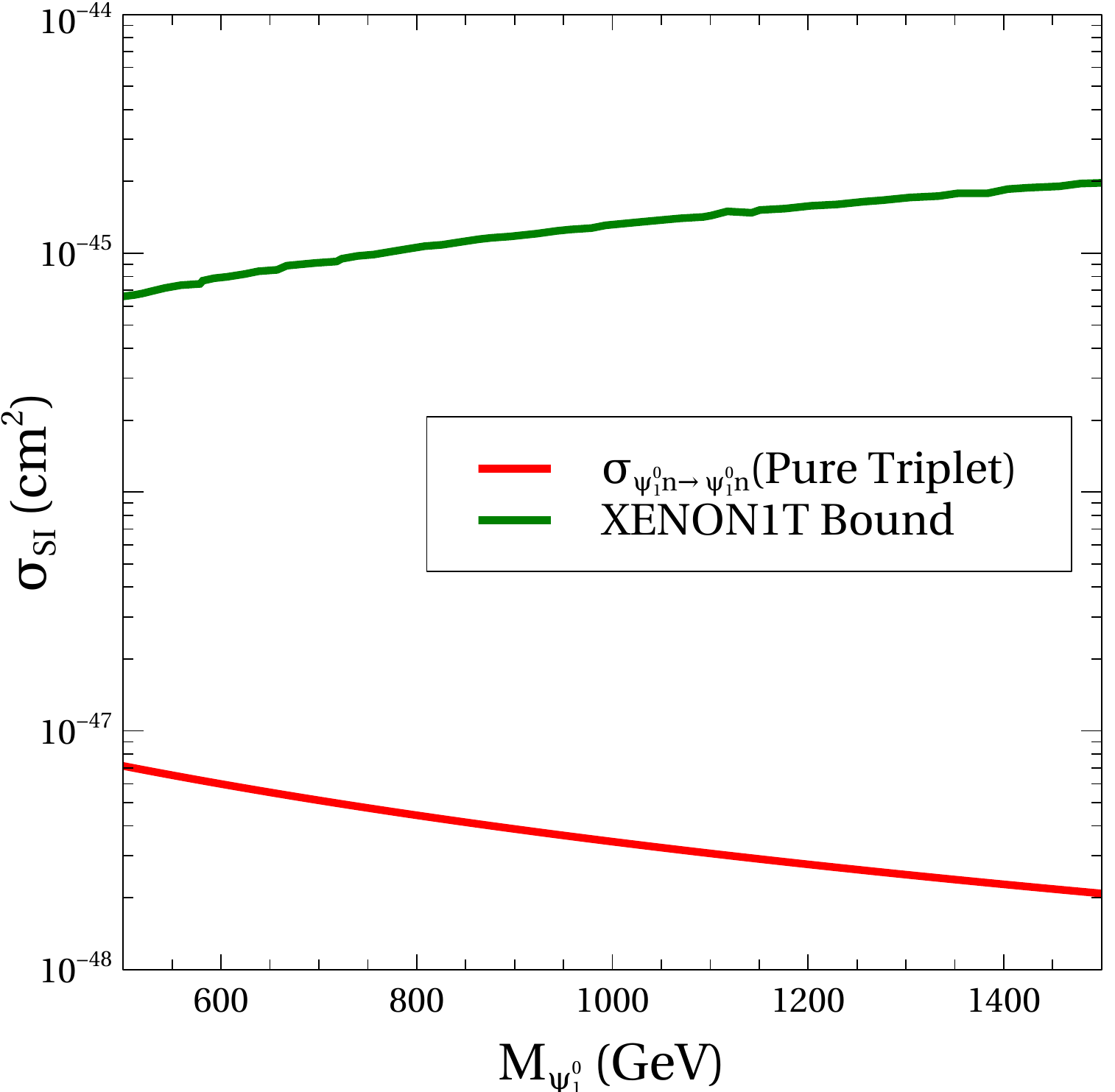,width=0.50\textwidth,clip=}
\epsfig{file=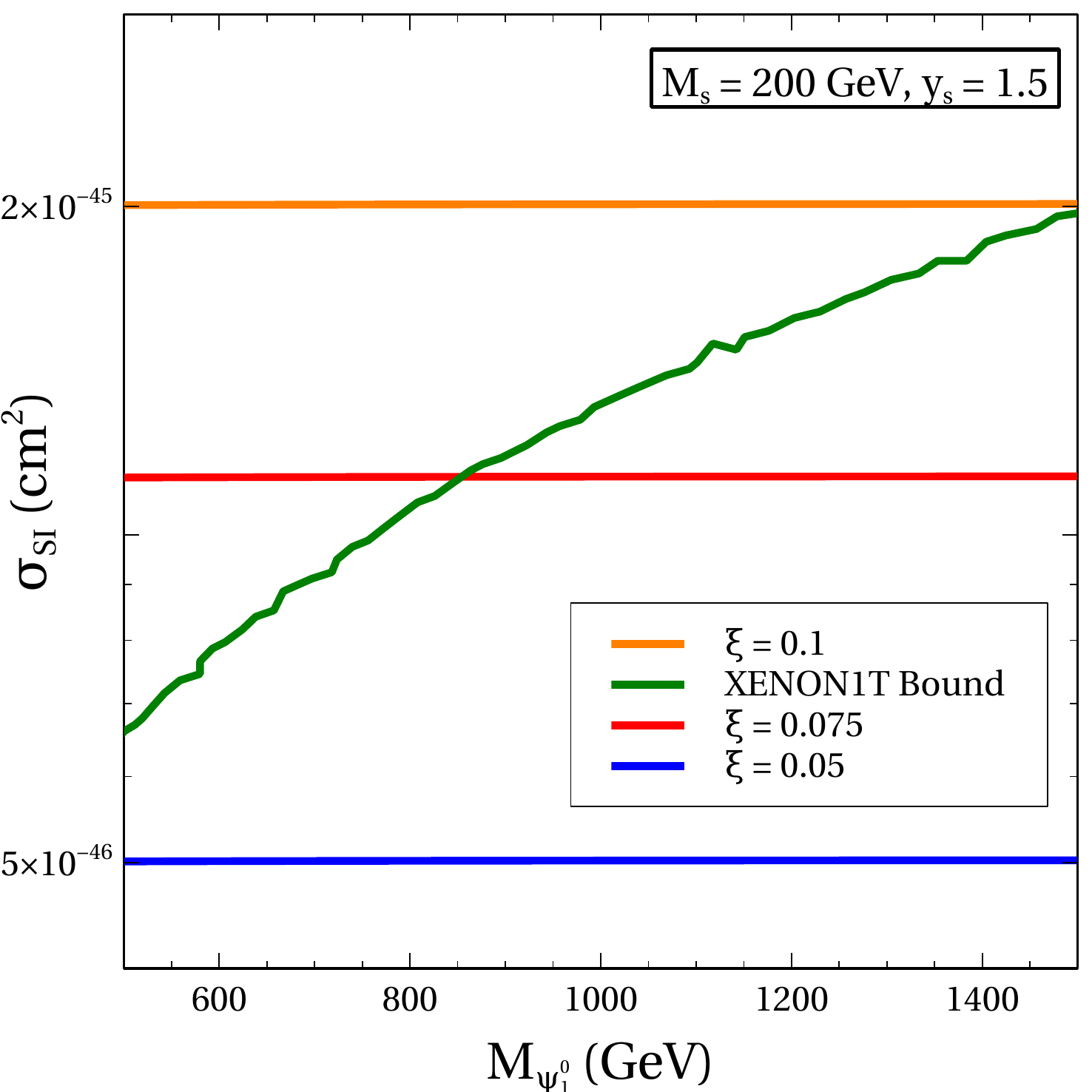,width=0.50\textwidth,clip=}
\end{tabular}
\caption{Left panel: DD cross-section for pure triplet fermion as a functions of DM mass. Right panel: DD cross-section for singlet scalar mediated diagram as a function of DM mass for different benchmark values of $\xi$ with fixed $M_{s}=$200 GeV and y$_{s}=$ 1.5.}
\label{fig2}
\end{figure}

%%%%%%%%%%%%%%%%%%%%%%%%%%%%%%%%%%%%%%%%%%%%%%%%%%%%%%%
The most convincing way to probe a particle dark matter candidate is the direct detection experiments. There have been serious efforts in this direction for last few decades where ground based detectors made up of different heavy nuclei have been used to probe processes where a DM particle passing through the detector can interact with a nuclei giving rise to some recoil. Such experiments have significantly improved their sensitivities over the years, with their present exclusion limits on spin independent DM-nucleon scattering cross section lying very close to the coherent neutrino-nucleon scattering rate \cite{Akerib:2016vxi, Tan:2016zwf, Cui:2017nnn, Aprile:2017iyp, Aprile:2018dbl}. Interestingly, our present model has good prospects for direct detection frontiers which is usually not there for purely freeze-in type dark matter candidates. Being part of an electroweak multiplet, the dark matter can interact with SM through gauge interactions and also through scalar interactions present due to the singlet scalar field $S^{\prime}$. Due to the absence of neutral current interactions of our dark matter candidate at tree level, the gauge boson mediated DM-nucleon scattering can arise only at radiative level. However, due to the existence of singlet scalar mixing with the SM like Higgs, such scalar mediated DM-nucleon scattering can occur at tree level as well. The Feynman diagrams corresponding to these processes are shown in Fig. \ref{fig1} all of which can lead to spin independent (SI) direct detection (DD) scattering processes. For pure triplet fermion case we have to exclude the last one as there exists no singlet scalar in that scenario. The dark matter candidate $\psi^0_{1}$ has two different one loop DD scattering: a Higgs penguin with $W^{\pm}$ loop and the box diagram in Fig. \ref{fig1}. The low energy effective Lagrangian for the $\psi^0_{1}$-quark one loop interaction have already been discussed in \cite{Cai:2015zza}. So the effective Lagrangian for the loop diagrams can be written as 
\begin{equation}
\mathcal{L}_{\psi^0_{1} q}=\sum_{i} \lambda_{q}^i \bar{\psi^0_{1}}\psi^0_{1}\bar{q}q 
\end{equation}

\begin{eqnarray}\nonumber
\lambda_{q}^i &=& -\alpha_2^2 \frac{m_{q_i}}{m_{\psi^0_{1}}m_h^2} \frac{1-4\eta_{W}+3\eta_{W}^2+(2-4\eta_{W})\log{\eta_{W}}}{(1-\eta_{W})^3}\\
&& + \alpha_2^2 \frac{m_{q_i}}{m_{\psi^0_{1}}m_W^2} \frac{2-3\eta_{W}+6\eta_{W}^2-5\eta_{W}^3+3\eta_{W}(1+\eta_{W}^2)\log{\eta_{W}}}{6(1-\eta_{W})^4}
\label{eqn2}
\end{eqnarray}
where $\eta_{W}\equiv \frac{m_{W}^2}{m_{\psi^0_{1}}^2}$ and $\alpha_2 = \frac{g^2}{4\pi}$. In Eq. \eqref{eqn2}, the first term is the effective coupling for the Higgs penguin diagram and the second term is the same for the box diagram. We have shown the corresponding spin-independent direct detection cross-section as a function of DM mass in the left panel of Fig. \ref{fig2}, which is about two orders of magnitude below the current XENON1T bound \cite{Aprile:2017iyp}. In spite of electroweak gauge interactions being involved in the scattering, this is in a way expected due to loop suppressions involved.

The other possible diagram coming from the introduction of the singlet scalar can, on the other hand, give rise to larger DD cross-section if the singlet scalar $S^{\prime}$ has sizeable mixing with the SM like Higgs. The relevant cross section can be written as 
\begin{equation}
\sigma_{\text{SI}}=\frac{ y_s^2 \ \mu_{\psi^0_{1} n}^2 \ m_n^2 \ }{\pi \ m_s^4 \ v^2}f_p^2 \xi^2
\end{equation}
where $y_{s}$, $f_{p}$, $\xi$ and $m_{n}$ are the new Yukawa coupling, form factor, $S^{\prime}$-Higgs mixing parameter (defined in Eq.\,\,(\ref{Eq:scalar_mixing_angle})), and nucleon mass respectively. We have also fixed the singlet mass (M$_{s}$) at 200 GeV, like in the calculation for relic abundance. As the DM-nucleon reduced mass $\mu_{\psi^0_{1} n}$ is nearly close to the mass of the nucleon for the chosen range of DM masses, this scattering cross section is almost independent of the DM mass. This in fact serves as another motivation of introducing the singlet scalar $S^{\prime}$ apart from being responsible for generating DM mass dynamically. This singlet scalar allows the model to be testable at ongoing and upcoming direct detection experiments for a wide range of DM masses. We show the contribution of this scalar mediated diagram to DD scattering in the right panel of Fig. \ref{fig2} for three different mixing parameter $\xi$ values ($\xi =$ 0.1, 0.075, 0.05). From the Fig. \ref{fig2} it is clear that the $\sigma_{\text{SI}}$ lies above the XENON1T bound for $\xi=0.1$, whereas for $\xi=0.075$ it is partially allowed and for $\xi=0.05$ it is just below the present bound for the chosen mass range of DM.

%%%%%%%%%%%%%%%%%%%%%%%%%%%%%%%%%%%%%%%%%%%%%%%%%%%%%%%%
\section{Indirect Detection}
\label{sec:id}
%%%%%%%%%%%%%%%%%%%%%%%%%%%%%%%%%%%%%%%%%%%%%%%%%%%%%%%%%
\begin{figure}[h!]
\centering
\begin{tabular}{cc}
\epsfig{file=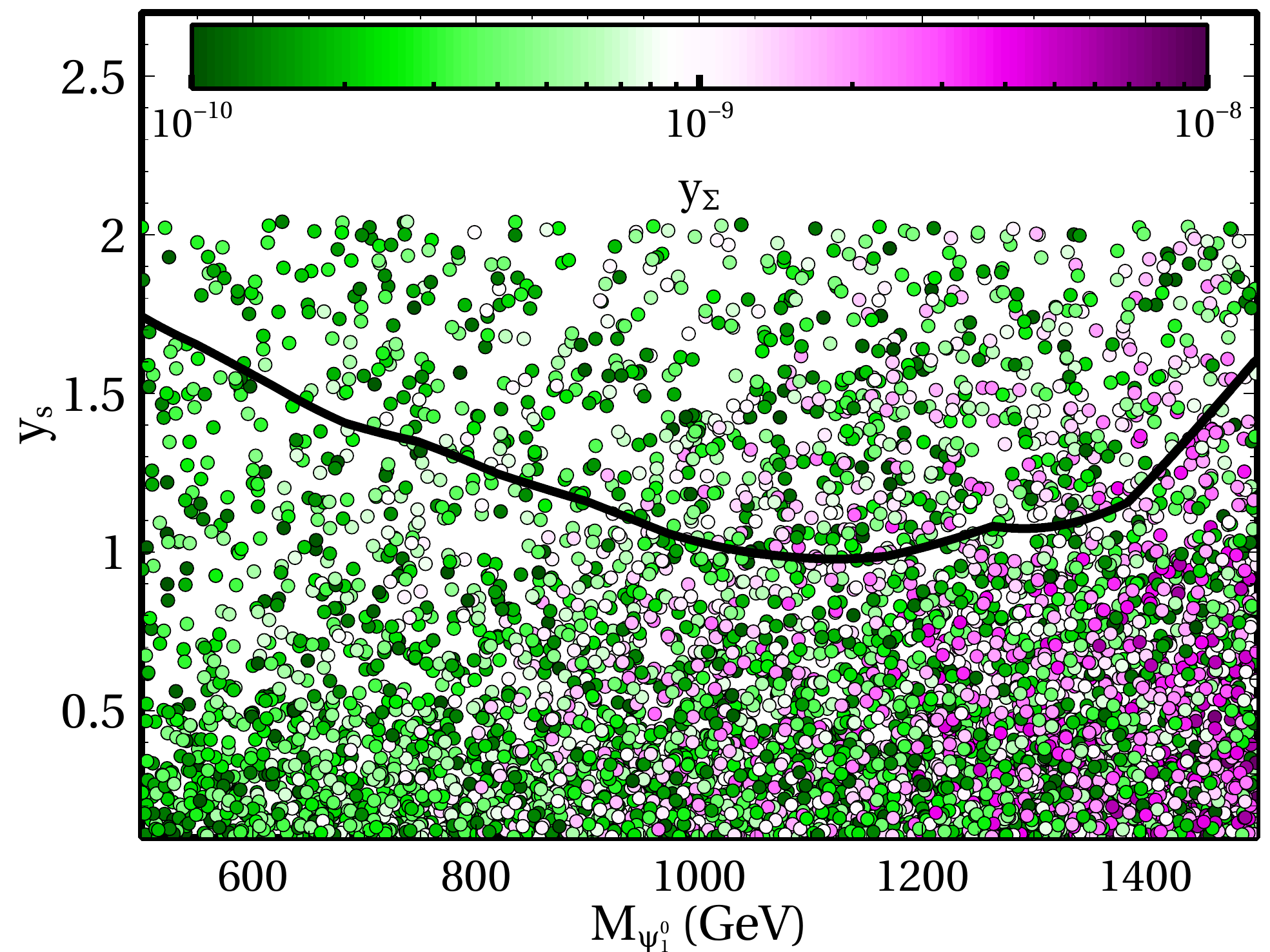,width=0.65\textwidth,clip=}
\end{tabular}
\caption{Parameter space of Yukawa coupling $y_{s}$ and $M_{\psi^0_{1}}$ that satisfy the relic density constraint. The color-bar indicates the variation of the another Yukawa $y_{\Sigma}$. The Black line corresponds to the Fermi-LAT bound, the points above that line are allowed. To generate this plot we have fixed the singlet mass ($M_s$) at 200 GeV.}
\label{fig3}
\end{figure}
Apart from direct detection experiments, DM parameter space in our model can also be probed at different indirect detection experiments (space as well as ground based) that are looking for SM particles produced either through DM annihilations or via DM decay in the local Universe. Among these final states, photon and neutrinos, being neutral and stable can reach the indirect detection experiments without getting affected much by intermediate regions. If the DM is of the type we have in our model, having TeV scale masses and sizeable interaction with the SM particles, these photons lie in the gamma ray regime that can be measured at space based telescopes like the Fermi-LAT or ground based telescopes like MAGIC. Here we constrain the DM parameter space from the indirect detection bounds arising from the global analysis of the Fermi-LAT and MAGIC observations of dSphs \cite{Ahnen:2016qkx}.

Since the heavier components of the fermion triplet are not there in the present Universe, we need to consider the DM self annihilations into the charged particles of the SM. The only possible process of this type is the DM annihilation into a pair of $W$ bosons which is indeed constrained tightly from gamma ray observations \cite{Ahnen:2016qkx}. Such bounds on $\langle \sigma v \rangle_{\rm DM DM \rightarrow W^+ W^-}$ are derived assuming $100\%$ annihilation of DM into these final states. Since we have another DM annihilation in this model namely, the one with a pair of singlet scalars $S^{\prime}$ in the final states which do not contribute to gamma rays due to neutral final states, we take into account the relative factor between these two annihilation rates while applying the indirect detection bound on $\langle \sigma v \rangle_{\rm DM DM \rightarrow W^+ W^-}$.\,\,Due to sizeable DM annihilation rates into neutral singlet scalars in our model, the gamma ray constraints on DM annihilations into W boson pairs get weaker in our model, saving the parameter space under study. It should be noted that in the pure fermion triplet dark matter model, gamma ray bound will completely rule out low mass region of DM mass we are studying \cite{Hisano:2004ds} as the annihilation rate for $\langle \sigma v \rangle_{\rm DM DM \rightarrow W^+ W^-}$ is $100\%$ in that model. Therefore, weakening the indirect detection bound arise as another motivation for the singlet scalar $S^{\prime}$ apart from generating DM mass dynamically and giving rise to tree level direct detection rates as discussed before. In Fig. \ref{fig3} we have shown allowed parameter space in the $y_{s}$ versus $M_{\psi^0_{1}}$ plane where we have also varied Yukawa of the inert doublet $y_{\Sigma}$ while fixing the singlet scalar mass at 200 GeV. All the points shown there are satisfying the relic density constraint and the black solid line corresponds to the present Fermi-LAT plus MAGIC bound on DM annihilation to $W^\pm$. The points below the black line are ruled out from these bounds which shows that we need large $y_{s}$ to satisfy the present bound. As the x-axis goes all the way upto DM mass of 1.5 TeV where non-perturbative effects (Sommerfeld enhancement) on DM annihilations can be sizeable, we use the results of \cite{Hisano:2004ds} to derive the exclusion line shown in Fig.\,\,\ref{fig3}. As we go into higher mass values, the Sommerfeld enhancement becomes more and more efficient which can be seen by the sharp rise in the black solid line towards the right end of the plot, requiring stronger Yukawa coupling $y_{s}$ to keep the DM self annihilations to neutral particles more dominant over $\langle \sigma v \rangle_{\rm DM DM \rightarrow W^+ W^-}$. Future data from gamma ray telescopes should be able to probe more parts of this region of dark matter masses as well as Yukawa coupling $y_{s}$, keeping the indirect detection prospects of the model very promising.
%%%%%%%%%%%%%%%%%%%%%%%%%%%%%%%%%%%%%%%%%%%%%%%%%%%%%%
\section{Summary and Conclusion}
\label{sec:conclu}
%%%%%%%%%%%%%%%%%%%%%%%%%%%%%%%%%%%%%%%%%%%%%%%%%%%%%%%
We have studied a minimal model for dark matter and radiative neutrino mass where dark matter relic abundance is generated from a hybrid setup consisting of both freeze-out and freeze-in scenarios. Considering a radiative type III seesaw scenario with the lightest fermion triplet being the dark matter candidate stabilised by an in built $\mathbb{Z}_2$ symmetry, we first show how the neutral component of this fermion triplet having mass around 1 TeV remains under-abundant from thermal freeze-out mechanism. We then consider a non-thermal (freeze-in) contribution from the late decay of $\mathbb{Z}_2$-odd scalar doublet to fill the deficit created during thermal freeze-out. We solve the coupled Boltzmann equations involving the mother particles and the dark matter candidate and find the parameter space that can lead to the correct relic abundance of dark matter. Due to the coupled nature of these equations, we can simultaneously constrain both dark matter as well as mother particle parameter space. The mother particle, being heavy, freezes out first followed by the freeze-out of dark matter and then the late decay of the mother particle into the dark matter particles. Such a hybrid setup has some interesting differences from the usual freeze-in dark matter scenarios. One notable difference we found was the dependence of non-thermal contribution on the corresponding Yukawa coupling between mother particle and dark matter. In the usual FIMP scenario, the non-thermal contribution always increases with such Yukawa, while in the present case it is not necessarily so. This is due to the role of the same Yukawa coupling in deciding the freeze-out abundance of the mother particle which later gets converted into non-thermal part of DM through late decay.

Apart from satisfying the correct relic abundance of dark matter within this hybrid setup having distinguishable features from pure WIMP and pure FIMP scenarios, the model also has promising detection prospects at collider, direct and indirect detection experiments. The presence of a scalar singlet which generates the dark matter mass dynamically in this model, plays a non-trivial role in keeping the model parameter space not only allowed from current experimental constraints but also very close to present as well as upcoming experimental sensitivity leaving a very promising scope for verifiability. This is in contrast to the typical freeze-in dark matter models which have very limited detection prospects due to tiny couplings. In principle, any mass of dark matter in the thermally under-abundant regime $(M_{\psi^0_1} \leq 2.2 \;\text{TeV})$ can be revived by adding a non-thermal contribution. However, due to tight constraints from LHC (evident from Fig. \ref{fig:lhc}), we get a lower bound of around 500 GeV. Also, as we go above 1 TeV mass, the indirect detection constraints become severe (evident from Fig. \ref{fig3}), requiring large couplings of dark with the singlet scalar. Such large couplings will get large corrections under renormalisation group evolutions, making the model non-perturbative at a low scale. Therefore we stick to a mass window near the 1 TeV.

In the neutrino sector also this model remains predictive as the lightest neutrino mass is vanishingly small due to tiny Yukawa couplings of the lightest fermion triplet, required for its non-thermal production at late epochs. Such vanishing lightest neutrino mass can have interesting consequences at experiments sensitive to the absolute neutrino mass scale. The model however suffers from the fine-tuning issue related to the above mentioned Yukawa coupling of the lightest fermion triplet to the leptons. Such tiny couplings can be generated within the framework of more general scenarios some of which also provide a UV completion \cite{Biswas:2018aib, Borah:2018gjk, Bhattacharyya:2018evo}.

%%%%%%%%%%%%%%%%%%%%%%%%%%%%%%%%%%%%%%%%%%%%%%%%%%%%%%%%%%%%%%%%%%%%
\acknowledgments
One of the authors AB would like
to acknowledge financial support from SERB, Govt.\,\,of INDIA
through NPDF fellowship under the project id. PDF/2017/000490.
He also thanks Alexander Pukhov for a few email conversation
regarding package \texttt{micrOMEGAs}. 
Moreover, AB gratefully acknowledges the cluster computing facility
at Harish-Chandra Research Institute, Allahabad (http://www.hri.res.in/cluster/).
AB and DN would like to thank Amit Dutta Banik for various discussions.
DB acknowledges the support from IIT Guwahati start-up grant
(reference number: xPHYSUGIITG01152xxDB001) and Associateship
Programme of IUCAA, Pune. He also acknowledges the hospitality and
facilities provided by School of Liberal Arts, Seoul-Tech,
Seoul 139-743, Korea where part of this work was completed.
%%%%%%%%%%%%%%%%%%%%%%%%%%%%%%%%%%%%%%%%%%%%%%%%%%%%%%%%%%%%%%%%%%%%
\appendix
%%%%%%%%%%%%%%%%%%%%%%%%%%%%%%%%%%%%%%%%%%%%%%%%%%%%%%%%%%%%%%%%%%%%
\section{Full calculation of the Lagrangian of fermionic triplet $\Sigma$}
\label{App:lag}
%%%%%%%%%%%%%%%%%%%%%%%%%%%%%%%%%%%%%%%%%%%%%%%%%%%%%%%%%%%%%%%%%%%%
We have constructed a fermionic triplet $\Sigma$ using $\Sigma_R$ and
its CP conjugate ${\Sigma_R}^c=C\,\overline{\Sigma_R}^T$ as $\Sigma = \Sigma_R + {\Sigma_R}^c$.
The Lagrangian of $\Sigma$ can be written as
\begin{eqnarray}
\mathcal{L}_{triplet} &=& \dfrac{i}{2}{\rm Tr}
[\overline{\Sigma}\,\slashed{D} \Sigma] -
\dfrac{1}{2}\,{\rm Tr}[\overline{{\Sigma}}\,M_{\Sigma}\,\Sigma] \,,\\
%\dfrac{1}{2}\,{\rm Tr}[\overline{\Sigma^c}\, M^{\star}_{\Sigma}\,\Sigma]\,,
&&=\dfrac{i}{2}{\rm Tr}[\overline{\Sigma_R}\,\slashed{D}\,{\Sigma_R}]
+\dfrac{i}{2}{\rm Tr}[\overline{{\Sigma_R}^c}\,\slashed{D}\,{\Sigma_R}^c]-
\left(\dfrac{1}{2}\,{\rm Tr}[\overline{{\Sigma_R}^c}\,
M_{\Sigma}\,\Sigma_R]+h.c.\right) \,.
\label{Eq:lag}
\end{eqnarray}
The covariant derivative of $\Sigma_R$ is defined as
\begin{eqnarray}
D_{\mu} \Sigma_R &=& \partial_{\mu} \Sigma_R + i\,g
\left[\sum_{a=1}^{3}\dfrac{\sigma^a}{2}\,W^a_{\mu},
\Sigma_R\right]\,,
\label{Eq:dmu}   
\end{eqnarray}
where, $g$ is the SU(2)$_{\rm L}$ gauge coupling
and $W_{\mu}^a$ ($a=1$ to 3) are three corresponding gauge bosons.
Now, using the expression of $\Sigma_R$ given in Eq.\,\,(\ref{sigmar}),
the covariant derivative of $\Sigma_R$ can be further expressed in terms
of the components of $\Sigma_R$ as
\begin{eqnarray}
D_{\mu} \Sigma_R &=& \partial_{\mu} \Sigma_R + \dfrac{i\,g}{2}
\left(\begin{array}{cc}
   \sqrt{2}(W^{+}_{\mu}\,\Sigma^{-}_R-W_{\mu}^{-}\,\Sigma_R^{+}) &
   2\,(W^{3}_{\mu}\,\Sigma^{+}_R-W_{\mu}^{+}\,\Sigma_R^{0}) \\
    2\,(W^{-}_{\mu}\,\Sigma^{0}_R-W_{\mu}^{3}\,\Sigma_R^{-}) &
    \sqrt{2}(W^{-}_{\mu}\,\Sigma^{+}_R-W_{\mu}^{+}\,\Sigma_R^{-}) \\
  \end{array} \right)\,\,.
  \label{Eq:dmu_sigr_comp}
\end{eqnarray}
Moreover, using Eq.\,\,(\ref{sigmar}) and the definition of $\Sigma_R^c$,
the CP conjugate of $\Sigma_R$ can also be expressed in a $2\times2$ matrix
notation as
\begin{eqnarray}
\Sigma_R^c~=~\left(\begin{array}{cc}
   {{\Sigma_R^0}^c}/{\sqrt{2}} & {\Sigma^{-}_R}^c \\
   {\Sigma^{+}_R}^c  & -\,{{\Sigma_R^0}^c}/{\sqrt{2}} \\
  \end{array} \right)\,,
\label{Eq:sigmarc}  
\end{eqnarray}
Similarly, using Eq.\,\,(\ref{Eq:dmu}) and Eq.\,\,(\ref{Eq:sigmarc}),
the covariant derivative of $\Sigma^c_R$ is given by
\begin{eqnarray}
D_{\mu} \Sigma_R^c &=& \partial_{\mu} \Sigma_R^c + \dfrac{i\,g}{2}
\left(\begin{array}{cc}
   \sqrt{2}(W^{+}_{\mu}\,{\Sigma^{+}_R}^c-W_{\mu}^{-}\,{\Sigma_R^{-}}^c) &
   2\,(W^{3}_{\mu}\,{\Sigma^{-}_R}^c-W_{\mu}^{+}\,{\Sigma_R^{0}}^c) \\
    2\,(W^{-}_{\mu}\,{\Sigma^{0}_R}^c-W_{\mu}^{3}\,{\Sigma_R^{+}}^c) &
    \sqrt{2}(W^{-}_{\mu}\,{\Sigma^{-}_R}^c-W_{\mu}^{+}\,{\Sigma_R^{+}}^c) \\
  \end{array} \right)\,\,.
  \label{Eq:dmu_sigrc_comp}
\end{eqnarray}
Now, using Eq.\,\,(\ref{Eq:dmu_sigr_comp}), let us simplify the first term
of Eq.\,\,(\ref{Eq:lag}):
\begin{eqnarray}
&&\dfrac{i}{2}{\rm Tr}[\overline{\Sigma_R}\,\slashed{D}\,{\Sigma_R}]\nonumber \\
&=&\dfrac{i}{2}\,\left(\overline{\Sigma^0_R}\,\slashed{\partial}\Sigma^0_R
+ \overline{\Sigma^{-}_R}\,\slashed{\partial}\Sigma^{-}_R
+ \overline{\Sigma^{+}_R}\,\slashed{\partial}\Sigma^{+}_R
\right)-\dfrac{g}{4}\left(2\,\overline{\Sigma^0_R}\gamma^{\mu}\Sigma^{-}_R\,W^{+}_{\mu}
-2\,\overline{\Sigma^0_R}\gamma^{\mu}\Sigma^{+}_R\,W^{-}_{\mu}
\right.\nonumber\\&&\left.
+2\,\overline{\Sigma^{-}_R}\gamma^{\mu}\Sigma^{0}_R\,W^{-}_{\mu}
-2\,\overline{\Sigma^+_R}\gamma^{\mu}\Sigma^{0}_R\,W^{+}_{\mu}
+2\,\overline{\Sigma^+_R}\gamma^{\mu}\Sigma^{+}_R\,W^{3}_{\mu}
-2\,\overline{\Sigma^-_R}\gamma^{\mu}\Sigma^{-}_R\,W^{3}_{\mu}
\right)\,.
\end{eqnarray}
Moreover, using the relations ${\Sigma_R^{+}}^c = C\,\overline{\Sigma_R^{+}}^T$
and ${\Sigma_R^{0}}^c = C\,\overline{\Sigma_R^{0}}^T$, 
one can further rewrite the above equation in terms of ${\Sigma_R^{+}}^c$
instead of $\Sigma^{+}_R$ as
\begin{eqnarray}
&&\dfrac{i}{2}{\rm Tr}[\overline{\Sigma_R}\,\slashed{D}\,{\Sigma_R}]\nonumber \\
&=&\dfrac{i}{2}\,\left(\overline{\Sigma^0_R}\,\slashed{\partial}\Sigma^0_R
+ \overline{\Sigma^{-}_R}\,\slashed{\partial}\Sigma^{-}_R
+ \overline{{\Sigma^{+}_R}^c}\,\slashed{\partial}{\Sigma^{+}_R}^c
\right)-\dfrac{g}{4}\left(2\,\overline{\Sigma^0_R}\gamma_{\mu}\Sigma^{-}_R\,W^{+}_{\mu}
+2\,\overline{{\Sigma^{+}_R}^c}\gamma^{\mu}{\Sigma^{0}_R}^c\,W^{-}_{\mu}
\right.\nonumber\\&&\left.
+2\,\overline{\Sigma^{-}_R}\gamma^{\mu}\Sigma^{0}_R\,W^{-}_{\mu}
+2\,\overline{{\Sigma^0_R}^c}\gamma^{\mu}{\Sigma^{+}_R}^c\,W^{+}_{\mu}
-2\,\overline{{\Sigma^{+}_R}^c}\gamma^{\mu}{\Sigma^{+}_R}^c\,W^{3}_{\mu}
-2\,\overline{\Sigma^-_R}\gamma^{\mu}\Sigma^{-}_R\,W^{3}_{\mu}
\right)\,.
\label{Eq:simplified1term}
\end{eqnarray}
In the above equation, we have used the following properties of charge
conjugation operator:
\begin{eqnarray}
C^{-1}\,\gamma_{\mu}\,C&=&-\gamma_{\mu}^T\,\,,\nonumber \\
C^{-1}\,\gamma_{5}\,C&=&\gamma_{5}^T\,\,,\nonumber \\
C^{T} &=& -C\,\,,\nonumber \\
C^{\dagger} &=& C^{-1}\,.
\label{Eq:Cprop}
\end{eqnarray}

Similarly, let us simplify the second term of Eq.\,\,(\ref{Eq:lag})
using Eq.\,\,(\ref{Eq:dmu_sigrc_comp}):
\begin{eqnarray}
&&\dfrac{i}{2}{\rm Tr}[\overline{{\Sigma_R}^c}\,\slashed{D}\,{{\Sigma_R}^c}]\nonumber \\
&=&\dfrac{i}{2}\,\left(\overline{{\Sigma^0_R}^c}\,\slashed{\partial}{\Sigma^0_R}^c
+ \overline{{\Sigma^{-}_R}^c}\,\slashed{\partial}{\Sigma^{-}_R}^c
+ \overline{{\Sigma^{+}_R}^c}\,\slashed{\partial}{\Sigma^{+}_R}^c
\right)-\dfrac{g}{4}\left(2\,\overline{{\Sigma^0_R}^c}\gamma^{\mu}{\Sigma^{+}_R}^c\,W^{+}_{\mu}
-2\,\overline{{\Sigma^{0}_R}^c}\gamma^{\mu}{\Sigma^{-}_R}^c\,W^{-}_{\mu}
\right.\nonumber\\&&\left.
+2\,\overline{{\Sigma^{+}_R}^c}\gamma^{\mu}{\Sigma^{0}_R}^c\,W^{-}_{\mu}
-2\,\overline{{\Sigma^{-}_R}^c}\gamma^{\mu}{\Sigma^{0}_R}^c\,W^{+}_{\mu}
+2\,\overline{{\Sigma^{-}_R}^c}\gamma^{\mu}{\Sigma^{-}_R}^c\,W^{3}_{\mu}
-2\,\overline{{\Sigma^{+}_R}^c}\gamma^{\mu}{\Sigma^{+}_R}^c\,W^{3}_{\mu}
\right)\,.
%\label{Eq:simplified2term}
\end{eqnarray}
One can change the field ${\Sigma^{-}_R}^c$ by $\Sigma^{-}_R$ in the 
above equation using the relation ${\Sigma^{-}_R}^c=C\,\overline{\Sigma^{-}_R}^T$
and the properties of $C$ operator mentioned in Eq.\,\,(\ref{Eq:Cprop}) as
\begin{eqnarray}
&&\dfrac{i}{2}{\rm Tr}[\overline{{\Sigma_R}^c}\,\slashed{D}\,{{\Sigma_R}^c}]\nonumber \\
&=&\dfrac{i}{2}\,\left(\overline{{\Sigma^0_R}^c}\,\slashed{\partial}{\Sigma^0_R}^c
+ \overline{{\Sigma^{-}_R}}\,\slashed{\partial}{\Sigma^{-}_R}
+ \overline{{\Sigma^{+}_R}^c}\,\slashed{\partial}{\Sigma^{+}_R}^c
\right)-\dfrac{g}{4}\left(2\,\overline{{\Sigma^0_R}^c}\gamma^{\mu}{\Sigma^{+}_R}^c\,W^{+}_{\mu}
+2\,\overline{{\Sigma^{-}_R}}\gamma^{\mu}{\Sigma^{0}_R}\,W^{-}_{\mu}
\right.\nonumber\\&&\left.
+2\,\overline{{\Sigma^{+}_R}^c}\gamma^{\mu}{\Sigma^{0}_R}^c\,W^{-}_{\mu}
+2\,\overline{{\Sigma^{0}_R}}\gamma^{\mu}{\Sigma^{-}_R}\,W^{+}_{\mu}
-2\,\overline{{\Sigma^{-}_R}}\gamma^{\mu}{\Sigma^{-}_R}\,W^{3}_{\mu}
-2\,\overline{{\Sigma^{+}_R}^c}\gamma^{\mu}{\Sigma^{+}_R}^c\,W^{3}_{\mu}
\right)\,.
\label{Eq:simplified2term}
\end{eqnarray}
Now, adding Eq.\,\,(\ref{Eq:simplified1term}) and Eq.\,\,(\ref{Eq:simplified2term}),
we get
\begin{eqnarray}
&&\dfrac{i}{2}{\rm Tr}[\overline{\Sigma_R}\,\slashed{D}\,{\Sigma_R}]+
\dfrac{i}{2}{\rm Tr}[\overline{{\Sigma_R}^c}\,\slashed{D}\,{{\Sigma_R}^c}]\,, \nonumber \\
&=& \dfrac{i}{2}\,\left(\overline{\Sigma^0_R}\,\slashed{\partial} \Sigma^0_R
+ \overline{{\Sigma^{0}_R}^c}\,\slashed{\partial}{\Sigma^{0}_R}^c
\right)
+ i\,\left(\overline{{\Sigma^{-}_R}}\,\slashed{\partial}{\Sigma^{-}_R}
+ \overline{{\Sigma^{+}_R}^c}\,\slashed{\partial}{\Sigma^{+}_R}^c\right)
-{g}\left(
\overline{\Sigma^{-}_R}\gamma^{\mu}\Sigma^{0}_R\,W^{-}_{\mu}
+\overline{{\Sigma^{+}_R}^c}\gamma^{\mu}{\Sigma^{0}_R}^c\,W^{-}_{\mu}
\right.\nonumber\\&&\left.
+\overline{\Sigma^0_R}\gamma^{\mu}\Sigma^{-}_R\,W^{+}_{\mu}
+\overline{{\Sigma^0_R}^c}\gamma^{\mu}{\Sigma^{+}_R}^c\,W^{+}_{\mu}
-\overline{{\Sigma^{+}_R}^c}\gamma^{\mu}{\Sigma^{+}_R}^c\,W^{3}_{\mu}
-\overline{\Sigma^-_R}\gamma^{\mu}\Sigma^{-}_R\,W^{3}_{\mu}
\right)\,.
\label{Eq:term1+2}
\end{eqnarray}
Let us define two fermionic state as
\begin{eqnarray}
\psi^0 &=& \Sigma^0_R + {\Sigma^0_R}^c \,\,,
\label{Eq:psi0}\\
\psi^{-} &=& \Sigma^{-}_R + {\Sigma^{+}_R}^c \,\,.
\label{Eq:psi-}
\end{eqnarray}
Finally, we can write the Eq.\,\,(\ref{Eq:term1+2}) in terms of $\psi^0$
and $\psi^-$ as
\begin{eqnarray}
&&\dfrac{i}{2}{\rm Tr}[\overline{\Sigma_R}\,\slashed{D}\,{\Sigma_R}]+
\dfrac{i}{2}{\rm Tr}[\overline{{\Sigma_R}^c}\,\slashed{D}\,{{\Sigma_R}^c}]\,, \nonumber \\
&=& \dfrac{i}{2}\,\overline{\psi^0}\,\slashed{\partial} \psi^0
+ i\,\overline{\psi^{-}}\,\slashed{\partial}{\psi^-} 
-g\,\left(\overline{\psi^{-}}\gamma^{\mu}\psi^0\,W^{-} + h.c.\right)
+g\,\overline{\psi^{-}}\gamma^{\mu}\psi^{-}W^3_{\mu}\,.
\label{Eq:simplifiedKEterm}
\end{eqnarray}

Furthermore, we can also simplify the mass terms in Eq.\,\,(\ref{Eq:lag}) as well
using the expressions of $\Sigma_R$ and ${\Sigma_R}^c$
(Eqs.\,\,(\ref{sigmar} and \ref{Eq:sigmarc}))
and considering $M_{\Sigma}$ real, i.e.
\begin{eqnarray}
&&-\dfrac{1}{2}\,{\rm Tr}[\overline{{\Sigma_R}^c}\,
M_{\Sigma}\,\Sigma_R]- \dfrac{1}{2}\,{\rm Tr}[\overline{{\Sigma_R}}\,
M_{\Sigma}\,{\Sigma_R}^c]\nonumber \\
&=&-\dfrac{M_{\Sigma}}{2}\left(\overline{\Sigma^0_R}{\Sigma^0_R}^c
+\overline{\Sigma^{-}_R}{\Sigma^{+}_R}^c
+\overline{\Sigma^{+}_R}{\Sigma^{-}_R}^c\right)
-\dfrac{M_{\Sigma}}{2}\left(\overline{{\Sigma^0_R}^c}{\Sigma^0_R}
+\overline{{\Sigma^{-}_R}^c}{\Sigma^{+}_R}
+\overline{{\Sigma^{+}_R}^c}{\Sigma^{-}_R}\right)\,,
\end{eqnarray}
Now, using the relation ${\Sigma^{\pm}_R}=C\,\overline{\Sigma^{\pm}_R}^T$,
one can have the following identities
\begin{eqnarray}
\overline{\Sigma^{+}_R}{\Sigma^{-}_R}^c &=& \overline{\Sigma^{-}_R}{\Sigma^{+}_R}^c\,, \nonumber \\
\overline{{\Sigma^{-}_R}^c}{\Sigma^{+}_R} &=& \overline{{\Sigma^{+}_R}^c}{\Sigma^{-}_R}\,.
\label{Eq:massidentities}
\end{eqnarray}
Therefore, using Eq.\,\,(\ref{Eq:massidentities}), we can further
simplify the mass terms of triplet $\Sigma$ as
\begin{eqnarray}
&&-\dfrac{1}{2}\,{\rm Tr}[\overline{{\Sigma_R}^c}\,
M_{\Sigma}\,\Sigma_R]- \dfrac{1}{2}\,{\rm Tr}[\overline{{\Sigma_R}}\,
M_{\Sigma}\,{\Sigma_R}^c]\nonumber \\
&=&-\dfrac{M_{\Sigma}}{2}
\left(
\overline{\Sigma^0_R}{\Sigma^0_R}^c
+\overline{{\Sigma^0_R}^c}{\Sigma^0_R}
\right)
-M_{\Sigma}
\left(
\overline{\Sigma^{-}_R}{\Sigma^{+}_R}^c
+\overline{{\Sigma^{+}_R}^c}{\Sigma^{-}_R}
\right)\,,\nonumber \\
&=&-\dfrac{M_{\Sigma}}{2} \overline{{\psi^0}}\psi^0
-M_{\Sigma} \overline{\psi^-}{\psi^{-}}\,,
\label{Eq:simplifiedmassterm}
\end{eqnarray}
while deriving in the last line we have used the definitions of $\psi^0$
and $\psi^{-}$ fields given in Eq.\,\,(\ref{Eq:psi0}) and Eq.\,\,(\ref{Eq:psi-}).
Finally, using Eq.\,\,(\ref{Eq:simplifiedKEterm}) and Eq.\,\,(\ref{Eq:simplifiedmassterm}),
we can write the Lagrangian of the triplet field $\Sigma$ in terms of newly defined
two fields $\psi^0$ and $\psi^{-}$ as
\begin{eqnarray}
\mathcal{L}_{triplet} &=& \dfrac{i}{2}{\rm Tr}
[\overline{\Sigma}\,\slashed{D} \Sigma] -
\dfrac{1}{2}\,{\rm Tr}[\overline{{\Sigma}}\,M_{\Sigma}\,\Sigma] \,,\\
&&=\dfrac{i}{2}{\rm Tr}[\overline{\Sigma_R}\,\slashed{D}\,{\Sigma_R}]
+\dfrac{i}{2}{\rm Tr}[\overline{{\Sigma_R}^c}\,\slashed{D}\,{\Sigma_R}^c]-
\left(\dfrac{1}{2}\,{\rm Tr}[\overline{{\Sigma_R}^c}\,
M_{\Sigma}\,\Sigma_R]+h.c.\right) \,,\nonumber \\
&&= \dfrac{i}{2}\,\overline{\psi^0}\,\slashed{\partial} \psi^0
+ i\,\overline{\psi^{-}}\,\slashed{\partial}{\psi^-} 
-\dfrac{M_{\Sigma}}{2} \overline{{\psi^0}}\psi^0
-M_{\Sigma} \overline{\psi^-}{\psi^{-}}
\nonumber \\ &&
-g\,\left(\overline{\psi^{-}}\gamma^{\mu}\psi^0\,W^{-} + h.c.\right)
+g\,\overline{\psi^{-}}\gamma^{\mu}\psi^{-}W^3_{\mu}\,.
\end{eqnarray} 

\bibliographystyle{JHEP}
%\bibliography{ref_ftdm.bib}

\begin{thebibliography}{10}

\bibitem{Zwicky:1933gu}
F.~Zwicky, \emph{{Die Rotverschiebung von extragalaktischen Nebeln}},
  \href{https://doi.org/10.1007/s10714-008-0707-4}{\emph{Helv. Phys. Acta}
  {\bfseries 6} (1933) 110}.

\bibitem{Rubin:1970zza}
V.~C. Rubin and W.~K. Ford, Jr., \emph{{Rotation of the Andromeda Nebula from a
  Spectroscopic Survey of Emission Regions}},
  \href{https://doi.org/10.1086/150317}{\emph{Astrophys. J.} {\bfseries 159}
  (1970) 379}.

\bibitem{Clowe:2006eq}
D.~Clowe, M.~Bradac, A.~H. Gonzalez, M.~Markevitch, S.~W. Randall, C.~Jones
  et~al., \emph{{A direct empirical proof of the existence of dark matter}},
  \href{https://doi.org/10.1086/508162}{\emph{Astrophys. J.} {\bfseries 648}
  (2006) L109} [\href{https://arxiv.org/abs/astro-ph/0608407}{{\ttfamily
  astro-ph/0608407}}].

\bibitem{Ade:2015xua}
{\scshape Planck} collaboration, P.~A.~R. Ade et~al., \emph{{Planck 2015
  results. XIII. Cosmological parameters}},
  \href{https://doi.org/10.1051/0004-6361/201525830}{\emph{Astron. Astrophys.}
  {\bfseries 594} (2016) A13}
  [\href{https://arxiv.org/abs/1502.01589}{{\ttfamily 1502.01589}}].

\bibitem{Taoso:2007qk}
M.~Taoso, G.~Bertone and A.~Masiero, \emph{{Dark Matter Candidates: A Ten-Point
  Test}}, \href{https://doi.org/10.1088/1475-7516/2008/03/022}{\emph{JCAP}
  {\bfseries 0803} (2008) 022}
  [\href{https://arxiv.org/abs/0711.4996}{{\ttfamily 0711.4996}}].

\bibitem{Akerib:2016vxi}
{\scshape LUX} collaboration, D.~S. Akerib et~al., \emph{{Results from a search
  for dark matter in the complete LUX exposure}},
  \href{https://doi.org/10.1103/PhysRevLett.118.021303}{\emph{Phys. Rev. Lett.}
  {\bfseries 118} (2017) 021303}
  [\href{https://arxiv.org/abs/1608.07648}{{\ttfamily 1608.07648}}].

\bibitem{Tan:2016zwf}
{\scshape PandaX-II} collaboration, A.~Tan et~al., \emph{{Dark Matter Results
  from First 98.7 Days of Data from the PandaX-II Experiment}},
  \href{https://doi.org/10.1103/PhysRevLett.117.121303}{\emph{Phys. Rev. Lett.}
  {\bfseries 117} (2016) 121303}
  [\href{https://arxiv.org/abs/1607.07400}{{\ttfamily 1607.07400}}].

\bibitem{Aprile:2017iyp}
{\scshape XENON} collaboration, E.~Aprile et~al., \emph{{First Dark Matter
  Search Results from the XENON1T Experiment}},
  \href{https://doi.org/10.1103/PhysRevLett.119.181301}{\emph{Phys. Rev. Lett.}
  {\bfseries 119} (2017) 181301}
  [\href{https://arxiv.org/abs/1705.06655}{{\ttfamily 1705.06655}}].

\bibitem{Cui:2017nnn}
{\scshape PandaX-II} collaboration, X.~Cui et~al., \emph{{Dark Matter Results
  From 54-Ton-Day Exposure of PandaX-II Experiment}},
  \href{https://doi.org/10.1103/PhysRevLett.119.181302}{\emph{Phys. Rev. Lett.}
  {\bfseries 119} (2017) 181302}
  [\href{https://arxiv.org/abs/1708.06917}{{\ttfamily 1708.06917}}].

\bibitem{Aprile:2018dbl}
E.~Aprile et~al., \emph{{Dark Matter Search Results from a One
  Tonne$\times$Year Exposure of XENON1T}},
  \href{https://arxiv.org/abs/1805.12562}{{\ttfamily 1805.12562}}.

\bibitem{McDonald:2001vt}
J.~McDonald, \emph{{Thermally generated gauge singlet scalars as
  selfinteracting dark matter}},
  \href{https://doi.org/10.1103/PhysRevLett.88.091304}{\emph{Phys. Rev. Lett.}
  {\bfseries 88} (2002) 091304}
  [\href{https://arxiv.org/abs/hep-ph/0106249}{{\ttfamily hep-ph/0106249}}].

\bibitem{Hall:2009bx}
L.~J. Hall, K.~Jedamzik, J.~March-Russell and S.~M. West, \emph{{Freeze-In
  Production of FIMP Dark Matter}},
  \href{https://doi.org/10.1007/JHEP03(2010)080}{\emph{JHEP} {\bfseries 03}
  (2010) 080} [\href{https://arxiv.org/abs/0911.1120}{{\ttfamily 0911.1120}}].

\bibitem{Konig:2016dzg}
J.~König, A.~Merle and M.~Totzauer, \emph{{keV Sterile Neutrino Dark Matter
  from Singlet Scalar Decays: The Most General Case}},
  \href{https://doi.org/10.1088/1475-7516/2016/11/038}{\emph{JCAP} {\bfseries
  1611} (2016) 038} [\href{https://arxiv.org/abs/1609.01289}{{\ttfamily
  1609.01289}}].

\bibitem{Biswas:2016bfo}
A.~Biswas and A.~Gupta, \emph{{Freeze-in Production of Sterile Neutrino Dark
  Matter in U(1)$_{\rm B-L}$ Model}},
  \href{https://doi.org/10.1088/1475-7516/2017/05/A01,
  10.1088/1475-7516/2016/09/044}{\emph{JCAP} {\bfseries 1609} (2016) 044}
  [\href{https://arxiv.org/abs/1607.01469}{{\ttfamily 1607.01469}}].

\bibitem{Biswas:2016iyh}
A.~Biswas and A.~Gupta, \emph{{Calculation of Momentum Distribution Function of
  a Non-thermal Fermionic Dark Matter}},
  \href{https://doi.org/10.1088/1475-7516/2017/03/033,
  10.1088/1475-7516/2017/05/A02}{\emph{JCAP} {\bfseries 1703} (2017) 033}
  [\href{https://arxiv.org/abs/1612.02793}{{\ttfamily 1612.02793}}].

\bibitem{Bernal:2017kxu}
N.~Bernal, M.~Heikinheimo, T.~Tenkanen, K.~Tuominen and V.~Vaskonen, \emph{{The
  Dawn of FIMP Dark Matter: A Review of Models and Constraints}},
  \href{https://doi.org/10.1142/S0217751X1730023X}{\emph{Int. J. Mod. Phys.}
  {\bfseries A32} (2017) 1730023}
  [\href{https://arxiv.org/abs/1706.07442}{{\ttfamily 1706.07442}}].

\bibitem{Fairbairn:2008fb}
M.~Fairbairn and J.~Zupan, \emph{{Dark matter with a late decaying dark
  partner}}, \href{https://doi.org/10.1088/1475-7516/2009/07/001}{\emph{JCAP}
  {\bfseries 0907} (2009) 001}
  [\href{https://arxiv.org/abs/0810.4147}{{\ttfamily 0810.4147}}].

\bibitem{Cheung:2010gj}
C.~Cheung, G.~Elor, L.~J. Hall and P.~Kumar, \emph{{Origins of Hidden Sector
  Dark Matter I: Cosmology}},
  \href{https://doi.org/10.1007/JHEP03(2011)042}{\emph{JHEP} {\bfseries 03}
  (2011) 042} [\href{https://arxiv.org/abs/1010.0022}{{\ttfamily 1010.0022}}].

\bibitem{Medina:2014bga}
A.~D. Medina, \emph{{Higgsino-like Dark Matter From Sneutrino Late Decays}},
  \href{https://doi.org/10.1016/j.physletb.2017.04.054}{\emph{Phys. Lett.}
  {\bfseries B770} (2017) 161}
  [\href{https://arxiv.org/abs/1409.2560}{{\ttfamily 1409.2560}}].

\bibitem{Gherghetta:2015ysa}
T.~Gherghetta, B.~von Harling, A.~D. Medina, M.~A. Schmidt and T.~Trott,
  \emph{{SUSY implications from WIMP annihilation into scalars at the Galactic
  Center}}, \href{https://doi.org/10.1103/PhysRevD.91.105004}{\emph{Phys. Rev.}
  {\bfseries D91} (2015) 105004}
  [\href{https://arxiv.org/abs/1502.07173}{{\ttfamily 1502.07173}}].

\bibitem{Borah:2017dfn}
D.~Borah and A.~Gupta, \emph{{New viable region of an inert Higgs doublet dark
  matter model with scotogenic extension}},
  \href{https://doi.org/10.1103/PhysRevD.96.115012}{\emph{Phys. Rev.}
  {\bfseries D96} (2017) 115012}
  [\href{https://arxiv.org/abs/1706.05034}{{\ttfamily 1706.05034}}].

\bibitem{Ma:2006km}
E.~Ma, \emph{{Verifiable radiative seesaw mechanism of neutrino mass and dark
  matter}}, \href{https://doi.org/10.1103/PhysRevD.73.077301}{\emph{Phys. Rev.}
  {\bfseries D73} (2006) 077301}
  [\href{https://arxiv.org/abs/hep-ph/0601225}{{\ttfamily hep-ph/0601225}}].

\bibitem{Olive:2016xmw}
{\scshape Particle Data Group} collaboration, C.~Patrignani et~al.,
  \emph{{Review of Particle Physics}},
  \href{https://doi.org/10.1088/1674-1137/40/10/100001}{\emph{Chin. Phys.}
  {\bfseries C40} (2016) 100001}.

\bibitem{Minkowski:1977sc}
P.~Minkowski, \emph{{$\mu \to e\gamma$ at a Rate of One Out of $10^{9}$ Muon
  Decays?}}, \href{https://doi.org/10.1016/0370-2693(77)90435-X}{\emph{Phys.
  Lett.} {\bfseries B67} (1977) 421}.

\bibitem{Mohapatra:1979ia}
R.~N. Mohapatra and G.~Senjanovic, \emph{{Neutrino Mass and Spontaneous Parity
  Violation}}, \href{https://doi.org/10.1103/PhysRevLett.44.912}{\emph{Phys.
  Rev. Lett.} {\bfseries 44} (1980) 912}.

\bibitem{Schechter:1980gr}
J.~Schechter and J.~W.~F. Valle, \emph{{Neutrino Masses in SU(2) x U(1)
  Theories}}, \href{https://doi.org/10.1103/PhysRevD.22.2227}{\emph{Phys. Rev.}
  {\bfseries D22} (1980) 2227}.

\bibitem{Ma:2008cu}
E.~Ma and D.~Suematsu, \emph{{Fermion Triplet Dark Matter and Radiative
  Neutrino Mass}}, \href{https://doi.org/10.1142/S021773230903059X}{\emph{Mod.
  Phys. Lett.} {\bfseries A24} (2009) 583}
  [\href{https://arxiv.org/abs/0809.0942}{{\ttfamily 0809.0942}}].

\bibitem{Hisano:2004ds}
J.~Hisano, S.~Matsumoto, M.~M. Nojiri and O.~Saito, \emph{{Non-perturbative
  effect on dark matter annihilation and gamma ray signature from galactic
  center}}, \href{https://doi.org/10.1103/PhysRevD.71.063528}{\emph{Phys. Rev.}
  {\bfseries D71} (2005) 063528}
  [\href{https://arxiv.org/abs/hep-ph/0412403}{{\ttfamily hep-ph/0412403}}].

\bibitem{Hisano:2006nn}
J.~Hisano, S.~Matsumoto, M.~Nagai, O.~Saito and M.~Senami,
  \emph{{Non-perturbative effect on thermal relic abundance of dark matter}},
  \href{https://doi.org/10.1016/j.physletb.2007.01.012}{\emph{Phys. Lett.}
  {\bfseries B646} (2007) 34}
  [\href{https://arxiv.org/abs/hep-ph/0610249}{{\ttfamily hep-ph/0610249}}].

\bibitem{Garcia-Cely:2015quu}
C.~Garcia-Cely and J.~Heeck, \emph{{Phenomenology of left-right symmetric dark
  matter}},  \href{https://arxiv.org/abs/1512.03332}{{\ttfamily 1512.03332}}.

\bibitem{Cirelli:2005uq}
M.~Cirelli, N.~Fornengo and A.~Strumia, \emph{{Minimal dark matter}},
  \href{https://doi.org/10.1016/j.nuclphysb.2006.07.012}{\emph{Nucl. Phys.}
  {\bfseries B753} (2006) 178}
  [\href{https://arxiv.org/abs/hep-ph/0512090}{{\ttfamily hep-ph/0512090}}].

\bibitem{Choubey:2017yyn}
S.~Choubey, S.~Khan, M.~Mitra and S.~Mondal, \emph{{Singlet-Triplet Fermionic
  Dark Matter and LHC Phenomenology}},
  \href{https://doi.org/10.1140/epjc/s10052-018-5785-1}{\emph{Eur. Phys. J.}
  {\bfseries C78} (2018) 302}
  [\href{https://arxiv.org/abs/1711.08888}{{\ttfamily 1711.08888}}].

\bibitem{Chardonnet:1993wd}
P.~Chardonnet, P.~Salati and P.~Fayet, \emph{{Heavy triplet neutrinos as a new
  dark matter option}},
  \href{https://doi.org/10.1016/0550-3213(93)90101-T}{\emph{Nucl. Phys.}
  {\bfseries B394} (1993) 35}.

\bibitem{Foot:1988aq}
R.~Foot, H.~Lew, X.~G. He and G.~C. Joshi, \emph{{Seesaw Neutrino Masses
  Induced by a Triplet of Leptons}},
  \href{https://doi.org/10.1007/BF01415558}{\emph{Z. Phys.} {\bfseries C44}
  (1989) 441}.

\bibitem{Ma:1998dn}
E.~Ma, \emph{{Pathways to naturally small neutrino masses}},
  \href{https://doi.org/10.1103/PhysRevLett.81.1171}{\emph{Phys. Rev. Lett.}
  {\bfseries 81} (1998) 1171}
  [\href{https://arxiv.org/abs/hep-ph/9805219}{{\ttfamily hep-ph/9805219}}].

\bibitem{Ma:2002pf}
E.~Ma and D.~P. Roy, \emph{{Heavy triplet leptons and new gauge boson}},
  \href{https://doi.org/10.1016/S0550-3213(02)00815-5}{\emph{Nucl. Phys.}
  {\bfseries B644} (2002) 290}
  [\href{https://arxiv.org/abs/hep-ph/0206150}{{\ttfamily hep-ph/0206150}}].

\bibitem{Chao:2012sz}
W.~Chao, \emph{{Dark matter, LFV and neutrino magnetic moment in the radiative
  seesaw model with fermion triplet}},
  \href{https://doi.org/10.1142/S0217751X15500074}{\emph{Int. J. Mod. Phys.}
  {\bfseries A30} (2015) 1550007}
  [\href{https://arxiv.org/abs/1202.6394}{{\ttfamily 1202.6394}}].

\bibitem{vonderPahlen:2016cbw}
F.~von~der Pahlen, G.~Palacio, D.~Restrepo and O.~Zapata, \emph{{Radiative Type
  III Seesaw Model and its collider phenomenology}},
  \href{https://doi.org/10.1103/PhysRevD.94.033005}{\emph{Phys. Rev.}
  {\bfseries D94} (2016) 033005}
  [\href{https://arxiv.org/abs/1605.01129}{{\ttfamily 1605.01129}}].

\bibitem{Ahnen:2016qkx}
{\scshape Fermi-LAT, MAGIC} collaboration, M.~L. Ahnen et~al., \emph{{Limits to
  dark matter annihilation cross-section from a combined analysis of MAGIC and
  Fermi-LAT observations of dwarf satellite galaxies}},
  \href{https://doi.org/10.1088/1475-7516/2016/02/039}{\emph{JCAP} {\bfseries
  1602} (2016) 039} [\href{https://arxiv.org/abs/1601.06590}{{\ttfamily
  1601.06590}}].

\bibitem{Aad:2012tfa}
{\scshape ATLAS} collaboration, G.~Aad et~al., \emph{{Observation of a new
  particle in the search for the Standard Model Higgs boson with the ATLAS
  detector at the LHC}},
  \href{https://doi.org/10.1016/j.physletb.2012.08.020}{\emph{Phys. Lett.}
  {\bfseries B716} (2012) 1} [\href{https://arxiv.org/abs/1207.7214}{{\ttfamily
  1207.7214}}].

\bibitem{Chatrchyan:2012xdj}
{\scshape CMS} collaboration, S.~Chatrchyan et~al., \emph{{Observation of a new
  boson at a mass of 125 GeV with the CMS experiment at the LHC}},
  \href{https://doi.org/10.1016/j.physletb.2012.08.021}{\emph{Phys. Lett.}
  {\bfseries B716} (2012) 30}
  [\href{https://arxiv.org/abs/1207.7235}{{\ttfamily 1207.7235}}].

\bibitem{Griest:1990kh}
K.~Griest and D.~Seckel, \emph{{Three exceptions in the calculation of relic
  abundances}}, \href{https://doi.org/10.1103/PhysRevD.43.3191}{\emph{Phys.
  Rev.} {\bfseries D43} (1991) 3191}.

\bibitem{Edsjo:1997bg}
J.~Edsjo and P.~Gondolo, \emph{{Neutralino relic density including
  coannihilations}},
  \href{https://doi.org/10.1103/PhysRevD.56.1879}{\emph{Phys. Rev.} {\bfseries
  D56} (1997) 1879} [\href{https://arxiv.org/abs/hep-ph/9704361}{{\ttfamily
  hep-ph/9704361}}].

\bibitem{Belanger:2013oya}
G.~Belanger, F.~Boudjema, A.~Pukhov and A.~Semenov, \emph{{micrOMEGAs 3: A
  program for calculating dark matter observables}},
  \href{https://doi.org/10.1016/j.cpc.2013.10.016}{\emph{Comput. Phys. Commun.}
  {\bfseries 185} (2014) 960}
  [\href{https://arxiv.org/abs/1305.0237}{{\ttfamily 1305.0237}}].

\bibitem{LopezHonorez:2006gr}
L.~Lopez~Honorez, E.~Nezri, J.~F. Oliver and M.~H.~G. Tytgat, \emph{{The Inert
  Doublet Model: An Archetype for Dark Matter}},
  \href{https://doi.org/10.1088/1475-7516/2007/02/028}{\emph{JCAP} {\bfseries
  0702} (2007) 028} [\href{https://arxiv.org/abs/hep-ph/0612275}{{\ttfamily
  hep-ph/0612275}}].

\bibitem{LopezHonorez:2010tb}
L.~Lopez~Honorez and C.~E. Yaguna, \emph{{A new viable region of the inert
  doublet model}},
  \href{https://doi.org/10.1088/1475-7516/2011/01/002}{\emph{JCAP} {\bfseries
  1101} (2011) 002} [\href{https://arxiv.org/abs/1011.1411}{{\ttfamily
  1011.1411}}].

\bibitem{Honorez:2010re}
L.~Lopez~Honorez and C.~E. Yaguna, \emph{{The inert doublet model of dark
  matter revisited}},
  \href{https://doi.org/10.1007/JHEP09(2010)046}{\emph{JHEP} {\bfseries 09}
  (2010) 046} [\href{https://arxiv.org/abs/1003.3125}{{\ttfamily 1003.3125}}].

\bibitem{Arhrib:2013ela}
A.~Arhrib, Y.-L.~S. Tsai, Q.~Yuan and T.-C. Yuan, \emph{{An Updated Analysis of
  Inert Higgs Doublet Model in light of the Recent Results from LUX, PLANCK,
  AMS-02 and LHC}},
  \href{https://doi.org/10.1088/1475-7516/2014/06/030}{\emph{JCAP} {\bfseries
  1406} (2014) 030} [\href{https://arxiv.org/abs/1310.0358}{{\ttfamily
  1310.0358}}].

\bibitem{Belyaev:2016lok}
A.~Belyaev, G.~Cacciapaglia, I.~P. Ivanov, F.~Rojas and M.~Thomas,
  \emph{{Anatomy of the Inert Two Higgs Doublet Model in the light of the LHC
  and non-LHC Dark Matter Searches}},
  \href{https://arxiv.org/abs/1612.00511}{{\ttfamily 1612.00511}}.

\bibitem{Borah:2017dqx}
D.~Borah, S.~Sadhukhan and S.~Sahoo, \emph{{Lepton Portal Limit of Inert Higgs
  Doublet Dark Matter with Radiative Neutrino Mass}},
  \href{https://arxiv.org/abs/1703.08674}{{\ttfamily 1703.08674}}.

\bibitem{Biswas:2015sva}
A.~Biswas, D.~Majumdar and P.~Roy, \emph{{Nonthermal two component dark matter
  model for Fermi-LAT ?-ray excess and 3.55 keV X-ray line}},
  \href{https://doi.org/10.1007/JHEP04(2015)065}{\emph{JHEP} {\bfseries 04}
  (2015) 065} [\href{https://arxiv.org/abs/1501.02666}{{\ttfamily
  1501.02666}}].

\bibitem{Biswas:2016yjr}
A.~Biswas, S.~Choubey and S.~Khan, \emph{{FIMP and Muon ($g-2$) in a
  U$(1)_{L_{\mu}-L_{\tau}}$ Model}},
  \href{https://doi.org/10.1007/JHEP02(2017)123}{\emph{JHEP} {\bfseries 02}
  (2017) 123} [\href{https://arxiv.org/abs/1612.03067}{{\ttfamily
  1612.03067}}].

\bibitem{Biswas:2018aib}
A.~Biswas, D.~Borah and A.~Dasgupta, \emph{{A UV Complete Framework of
  Freeze-in Massive Particle Dark Matter}},
  \href{https://arxiv.org/abs/1805.06903}{{\ttfamily 1805.06903}}.

\bibitem{Kahlhoefer:2017dnp}
F.~Kahlhoefer, \emph{{Review of LHC Dark Matter Searches}},
  \href{https://doi.org/10.1142/S0217751X1730006X}{\emph{Int. J. Mod. Phys.}
  {\bfseries A32} (2017) 1730006}
  [\href{https://arxiv.org/abs/1702.02430}{{\ttfamily 1702.02430}}].

\bibitem{Penning:2017tmb}
B.~Penning, \emph{{The Pursuit of Dark Matter at Colliders - An Overview}},
  \href{https://doi.org/10.1088/1361-6471/aabea7}{\emph{J. Phys.} {\bfseries
  G45} (2018) 063001} [\href{https://arxiv.org/abs/1712.01391}{{\ttfamily
  1712.01391}}].

\bibitem{Aaboud:2017mpt}
{\scshape ATLAS} collaboration, M.~Aaboud et~al., \emph{{Search for long-lived
  charginos based on a disappearing-track signature in $pp$ collisions at
  $\sqrt{s}$ = 13 TeV with the ATLAS detector}},
  \href{https://arxiv.org/abs/1712.02118}{{\ttfamily 1712.02118}}.

\bibitem{Fuks:2012qx}
B.~Fuks, M.~Klasen, D.~R. Lamprea and M.~Rothering, \emph{{Gaugino production
  in proton-proton collisions at a center-of-mass energy of 8 TeV}},
  \href{https://doi.org/10.1007/JHEP10(2012)081}{\emph{JHEP} {\bfseries 10}
  (2012) 081} [\href{https://arxiv.org/abs/1207.2159}{{\ttfamily 1207.2159}}].

\bibitem{Fuks:2013vua}
B.~Fuks, M.~Klasen, D.~R. Lamprea and M.~Rothering, \emph{{Precision
  predictions for electroweak superpartner production at hadron colliders with
  {\sc Resummino}}},
  \href{https://doi.org/10.1140/epjc/s10052-013-2480-0}{\emph{Eur. Phys. J. C}
  {\bfseries 73} (2013) 2480}
  [\href{https://arxiv.org/abs/1304.0790}{{\ttfamily 1304.0790}}].

\bibitem{Cai:2015zza}
Y.~Cai and A.~P. Spray, \emph{{Fermionic Semi-Annihilating Dark Matter}},
  \href{https://doi.org/10.1007/JHEP01(2016)087}{\emph{JHEP} {\bfseries 01}
  (2016) 087} [\href{https://arxiv.org/abs/1509.08481}{{\ttfamily
  1509.08481}}].

\bibitem{Borah:2018gjk}
D.~Borah, B.~Karmakar and D.~Nanda, \emph{{Common Origin of Dirac Neutrino Mass
  and Freeze-in Massive Particle Dark Matter}},
  \href{https://arxiv.org/abs/1805.11115}{{\ttfamily 1805.11115}}.

\bibitem{Bhattacharyya:2018evo}
G.~Bhattacharyya, M.~Dutra, Y.~Mambrini and M.~Pierre, \emph{{Freezing-in dark
  matter through a heavy invisible $Z'$}},
  \href{https://arxiv.org/abs/1806.00016}{{\ttfamily 1806.00016}}.

\end{thebibliography}

\providecommand{\href}[2]{#2}\begingroup\raggedright\endgroup

\end{document}